\documentclass[aps,pra,twocolumn,showpacs,superscriptaddress,floatfix,nofootinbib]{revtex4}
\usepackage{graphicx}
\usepackage{amsmath}
\usepackage{epsfig}
\usepackage{helvet}
\usepackage{amssymb}

\newcommand{\be}{\begin{equation}}
\newcommand{\ee}{\end{equation}}
\newcommand{\bea}{\begin{eqnarray}}
\newcommand{\eea}{\end{eqnarray}}

\newcommand{\tr}{\mbox{tr}}
\newcommand{\bra}[1]{\mbox{$\langle #1 |$}}
\newcommand{\ket}[1]{\mbox{$| #1 \rangle$}}

\def\tr{ \mbox{tr}}

\begin{document}

\title{Algorithms for entanglement renormalization}
\author{G. Evenbly}
\author{G. Vidal}
\affiliation{School of Physical Sciences, the University of
Queensland, QLD 4072, Australia} 
\date{\today}

\begin{abstract}
We describe an iterative method to optimize the multi-scale entanglement renormalization ansatz (MERA) for the low-energy subspace of local Hamiltonians on a $D$-dimensional lattice. For translation invariant systems the cost of this optimization is logarithmic in the linear system size. Specialized algorithms for the treatment of infinite systems are also described. Benchmark simulation results are presented for a variety of $1D$ systems, namely Ising, Potts, XX and Heisenberg models. The potential to compute expected values of local observables, energy gaps and correlators is investigated.
\end{abstract}


\maketitle


\tableofcontents

\section{Introduction}
\label{sect:Intro}

Entanglement renormalization \cite{ER} is a numerical technique based on locally reorganizing the Hilbert space of a quantum many-body system with the aim to reduce the amount of entanglement in its wave function. It was introduced to address a major computational obstacle in real space renormalization group (RG) methods \cite{Wilson,DMRG,CORE}, responsible for limitations in their performance and range of applicability, namely the proliferation of degrees of freedom that occurs under successive applications of a RG transformation.

Entanglement renormalization is built around the assumption that, as a result of the local character of physical interactions, some of the relevant degrees of freedom in the ground state of a many-body system can be decoupled from the rest by unitarily transforming small regions of space. Accordingly, unitary transformations known as \emph{disentanglers} are applied locally to the system in order to identify and decouple such degrees of freedom, which are then safely removed and therefore do no longer appear in any subsequent coarse-grained description. This prevents the harmful accumulation of degrees of freedom and thus leads to a sustainable real space RG transformation, able to explore arbitrarily large $1D$ and $2D$ lattice systems, even at a quantum critical point. It also leads to the \emph{multi-scale entanglement renormalization ansatz} (MERA), a variational ansatz for many-body states \cite{MERA}. 

The MERA, based in turn on a class of quantum circuits, is particularly successful at describing ground states at quantum criticality \cite{ER,MERA,FreeFermions,FreeBosons,Finite2D, Transfer,CFT,Scalable2D} or with topological order \cite{QuantumDouble,StringNet}. From the computational viewpoint, the key property of the MERA is that it can be manipulated efficiently, due to the causal structure of the underlying quantum circuit \cite{MERA}. As a result, it is possible to efficiently evaluate the expected value of local observables, and to efficiently optimize its tensors. Thus, well-established simulation techniques for matrix product states, such as energy minimization \cite{MPSvariational} or simulation of time evolution \cite{TEBD}, can be readily generalized to the MERA \cite{FlowEq,TimeEvolution,OldAlgorithm}.

In this paper we describe a simple algorithm (and several variations thereof) to compute an approximation of the low energy subspace of a local Hamiltonian with the MERA, and present benchmark calculations for 1D lattice systems. 

Our goal is to provide the interested reader with a rather self-contained explanation of the algorithm, with enough information to implement it. Sect. \ref{sect:MERA} and \ref{sect:local} review and elaborate on the theoretical foundations of the MERA \cite{ER,MERA}, and establish the notation and nomenclature used in the rest of the paper. Specifically, Sect. \ref{sect:MERA} introduces the MERA, both from the perspective of quantum circuits and of the renormalization group, and describes several realizations in 1D and 2D lattices. Then Sect. \ref{sect:local} explains how to compute expected values of local observables and two-point correlators. Central to this discussion is the \emph{past causal cone} of a small block of lattice sites and the \emph{ascending} and \emph{descending} superoperators, which can be used to move local observables and density matrices up and down the causal cone. 

Sect. \ref{sect:optim} considers how to optimize a single tensor of the MERA during an energy minimization. This optimization involves linearizing a quadratic cost function for the (isometric) tensor, and computing its \emph{environment}. In Sect. \ref{sect:algorithm} we describe algorithms to minimize the energy of the state/subspace represented by a MERA. A highlight of the algorithms is their computational cost. For an inhomogeneous lattice with $N$ sites, the cost scales as $O(N)$, whereas for translation invariant systems it drops to just $O(\log N)$. Other variations of the algorithm allow us to address infinite systems, and scale invariant systems (e.g. quantum critical systems), at a cost independent of $N$. 

Sect. \ref{sect:benchmark} presents benchmark calculations for different 1D quantum lattice models, namely Ising, 3-level Potts, XX and Heisenberg models. We compute ground state energies, magnetizations and two-point correlators throughout the phase diagram, which includes a second order quantum phase transition. We find that, at the critical point of an infinite system, the error in the ground state energy decays exponentially with the refinement parameter $\chi$, whereas the two-point correlators remain accurate even at distances of millions of lattice sites. We then extract critical exponents from the order parameter and from two-point correlators. Finally, we also compute a MERA that includes the first excited state, from which the energy gap can be obtained and seen to vanish as $1/N$ at criticality.

This paper replaces similar notes on MERA algorithms presented in Ref. \cite{OldAlgorithm}. For the sake of concreteness, we have not included several of the algorithms of Ref. \cite{OldAlgorithm}, which nevertheless remain valid proposals. We have also focused the discussion on a ternary MERA for 1D lattices (instead of the binary MERA used in all previous references) because it is somewhat computationally advantageous (e.g. see computation of two-point correlators) and also leads to a much more convenient generalization in 2D. 

\begin{figure}[!htb]
\begin{center}
\includegraphics[width=6cm]{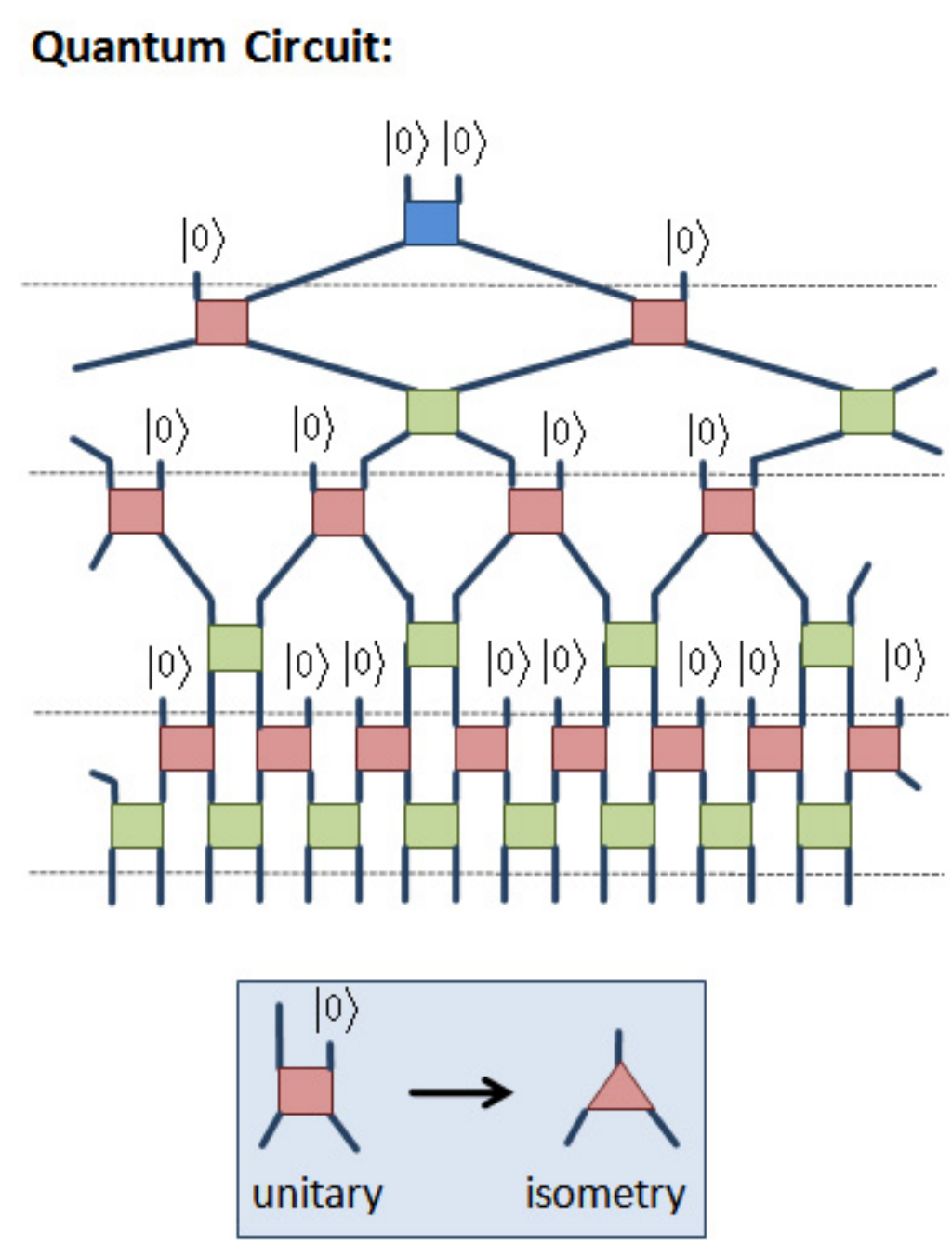}
\caption{(Colour online) Quantum circuit $\mathcal{C}$ corresponding to a specific realization of the MERA, namely the binary 1D MERA of Fig. \ref{fig:2MERA}. In this particular example, circuit $\mathcal{C}$ is made of gates involving two incoming wires and two outgoing wires, $p=p_{in}=p_{out}=2$. Some of the unitary gates in this circuit have one incoming wire in the fixed state $\ket{0}$ and can be replaced with an isometry $w$ of type (1,2). By making this replacement, we obtain the isometric circuit of Fig. \ref{fig:2MERA}.} 
\label{fig:QuantCirc}
\end{center}
\end{figure}

\section{The MERA}
\label{sect:MERA}

Let $\mathcal{L}$ denote a $D$-dimensional lattice made of $N$ sites, where each site is described by a Hilbert space $\mathbb{V}$ of finite dimension $d$, so that $\mathbb{V}_{\mathcal{L}} \cong \mathbb{V}^{\otimes N}$. The MERA is an ansatz to describe certain pure states $\ket{\Psi}\in \mathbb{V}_{\mathcal{L}}$ of the lattice or, more generally, subspaces $\mathbb{V}_{U} \subseteq \mathbb{V}_{\mathcal{L}}$.

There are two useful ways of thinking about the MERA that can be used to motivate its specific structure as a tensor network, and also help understand its properties and how the algorithms ultimately work. One way is to regard the MERA as a quantum circuit $\mathcal{C}$ whose output wires correspond to the sites of the lattice $\mathcal{L}$ \cite{MERA}. Alternatively, we can think of the MERA as defining a coarse-graining transformation that maps $\mathcal{L}$ into a sequence of increasingly coarser lattices, thus leading to a renormalization group transformation \cite{ER}. Next we briefly review these two complementary interpretations. Then we compare several MERA schemes and discuss how to exploit space symmetries.

\begin{figure}[!tb]
\begin{center}
\includegraphics[width=8cm]{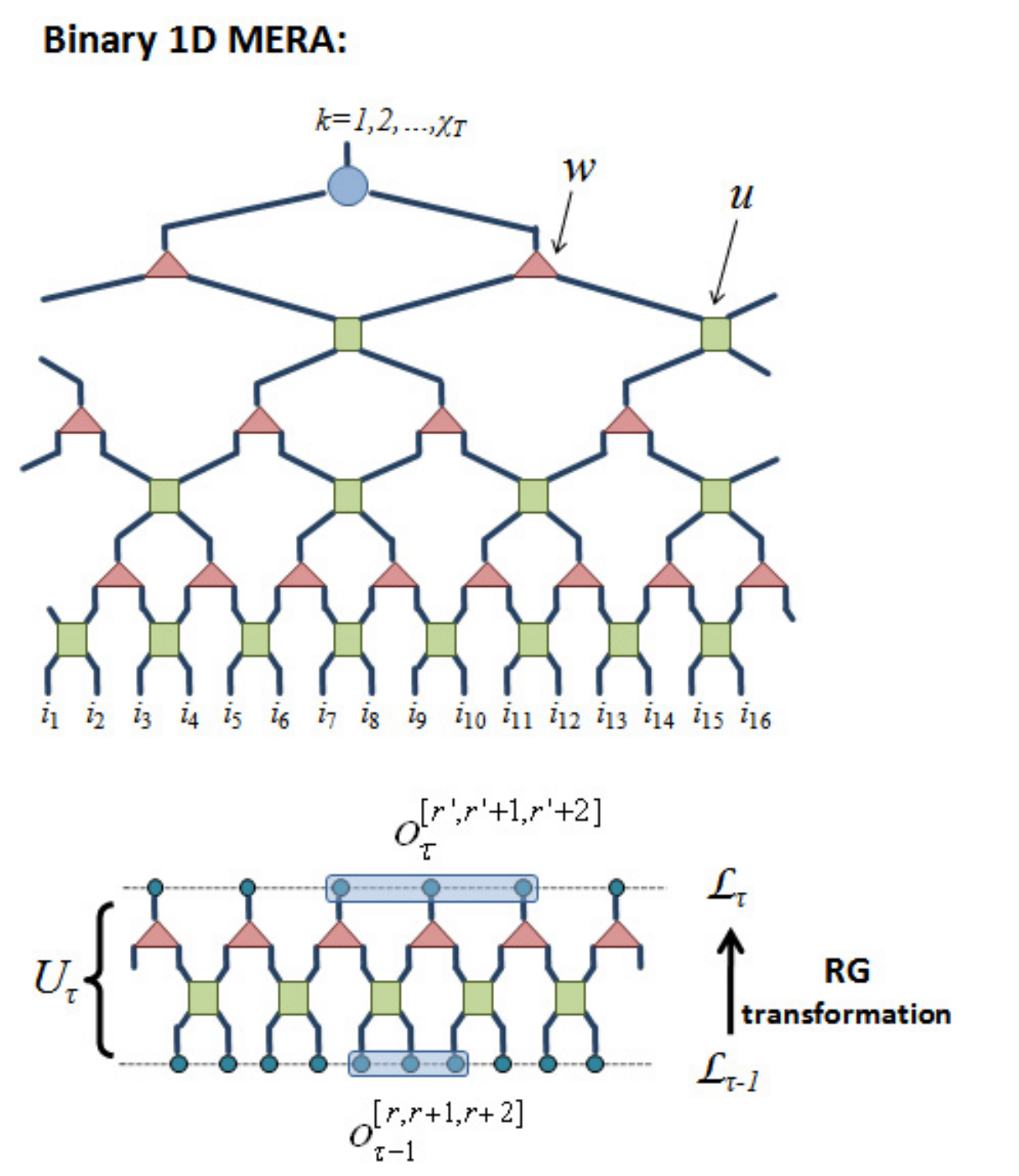}
\caption{(Colour online) \emph{Top:} Example of a binary 1D MERA for a lattice $\mathcal{L}$ with $N=16$ sites. It contains two types of isometric tensors, organized in $T=4$ layers. The input (output) wires of a tensor are those that enter it from the top (leave it from the bottom). The \emph{top tensor} is of type $(1,2)$ and the rank $\chi_T$ of its upper index determines the dimension of the subspace $\mathbb{V}_{U}\subseteq \mathbb{V}_{\mathcal{L}}$ represented by the MERA. The \emph{isometries} $w$ are of type $(1,2)$ and are used to replace each block of two sites with a single effective site. Finally, the \emph{disentanglers} $u$ are of type (2,2) and are used to disentangle the blocks of sites before coarse-graining. \emph{Bottom:} Under the renormalization group transformation induced by the binary 1D MERA, three-site operators are mapped into three-site operators.} 
\label{fig:2MERA}
\end{center}
\end{figure}

\subsection{Quantum circuit}

As a quantum circuit $\mathcal{C}$, the MERA for a pure state $\ket{\Psi} \in \mathbb{V}_{\mathcal{L}}$ is made of $N$ quantum wires, each one described by a Hilbert space $\mathbb{V}$, and unitary gates $u$ that transform the unentangled state $\ket{0}^{\otimes N}$ into $\ket{\Psi}$ (see Fig. \ref{fig:QuantCirc}).  

In a generic case, each unitary gate $u$ in the circuit $\mathcal{C}$ involves some small number $p$ of wires,
\begin{equation}
	u:\mathbb{V}^{\otimes p} \rightarrow\mathbb{V}^{\otimes p},~~~~~ u^\dagger u = u u^{\dagger} = \mathbb{I},
	\label{eq:unitary}
\end{equation}
where $I$ is the identity operator in $\mathbb{V}^{\otimes p}$. For some gates, however, one or several of the input wires are in a fixed state $\ket{0}$. In this case we can replace the unitary gate $u$ with an isometric gate $w$
\begin{equation}
	w:\mathbb{V}_{in}\rightarrow\mathbb{V}_{out},~~~~~ w^\dagger w = \mathbb{I}_{\mathbb{V}_{in}}, ~~~
\label{eq:isometry}
\end{equation}
where $\mathbb{V}_{in}\cong \mathbb{V}^{\otimes p_{in}}$ is the space of the $p_{in}$ input wires that are not in a fixed state $\ket{0}$ and $\mathbb{V}_{out}\cong \mathbb{V}^{\otimes p_{out}}$ is the space of the $p_{out}=p$ output wires. We refer to $w$ as a $(p_{in}, p_{out})$ gate or tensor. 

\begin{figure}[!tb]
\begin{center}
\includegraphics[width=8cm]{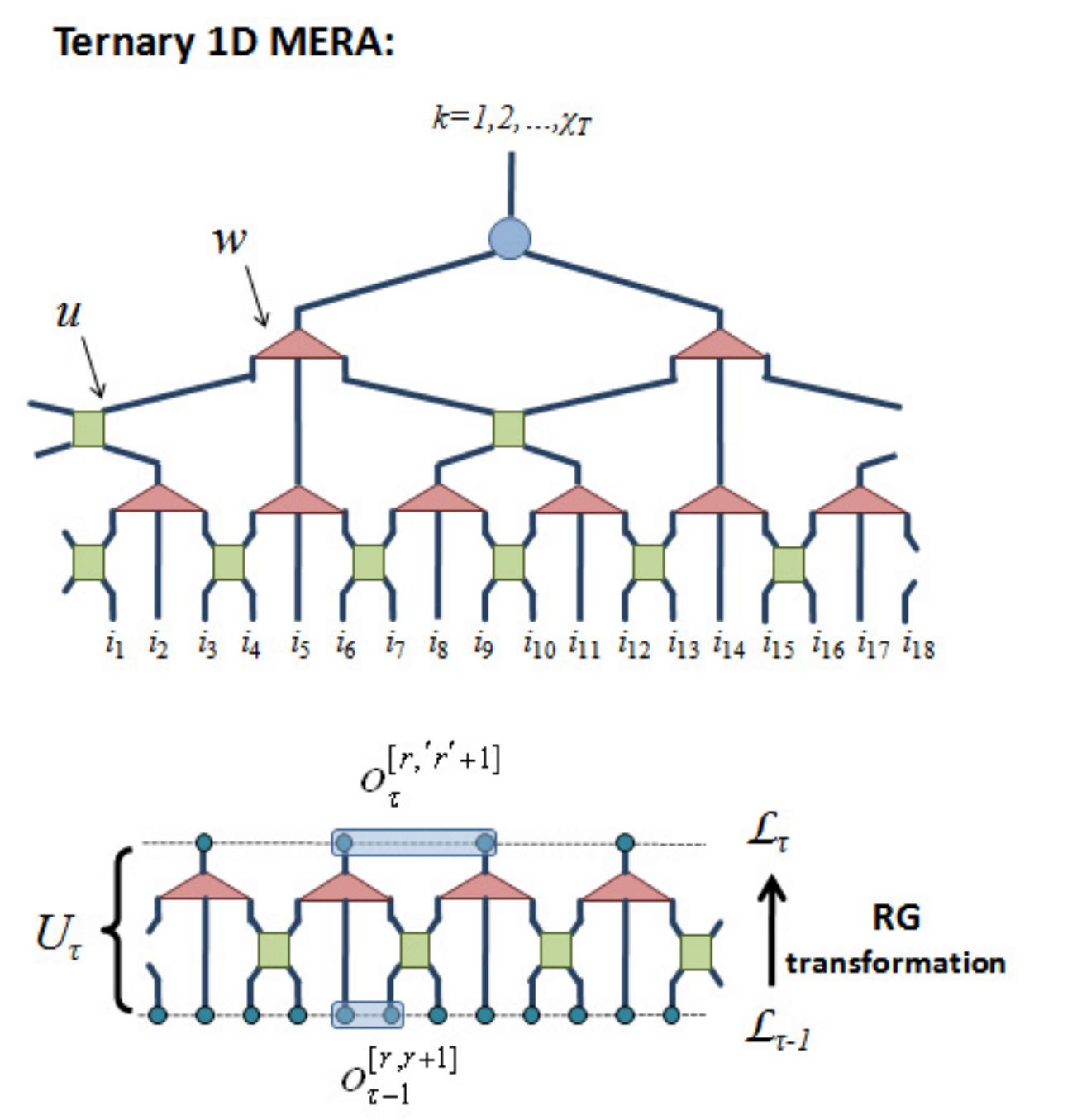}
\caption{(Colour online) \emph{Top:} Example of ternary 1D MERA (rank $\chi_T$, $T=3$) for a lattice of 18 sites. It differs from the binary 1D MERA of Fig. \ref{fig:2MERA} in that the \emph{isometries} are of type $(1,3)$, so that blocks of three sites are replaced with one effective site. \emph{Bottom:} As a result, two-site operators are mapped into two-site operators during the coarse-graining.} 
\label{fig:3MERA}
\end{center}
\end{figure}

\begin{figure}[!hbt]
\begin{center}
\includegraphics[width=6cm]{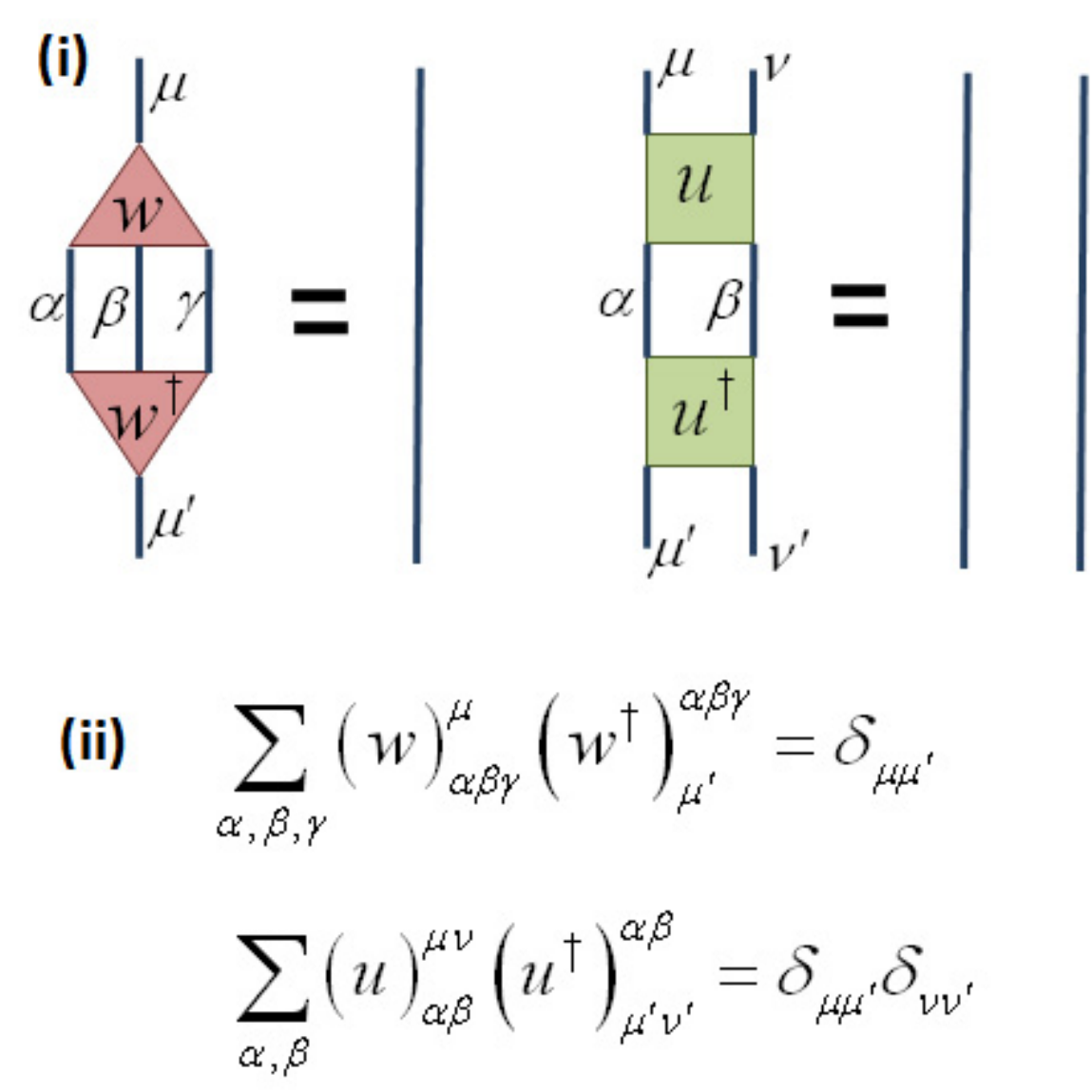}
\caption{(Colour online) The tensors which comprise a MERA are constrained to be isometric, cf. Eq. \ref{eq:isometry}. The constraints for the isometries $w$ and disentanglers $u$ of the ternary MERA can be equivalently expressed (i) diagramatically or (ii) with equations. In this paper we will mostly use the diagramatic notation, which remains simple for complicated tensor networks. } 
\label{fig:NotationCompare}
\end{center}
\end{figure}

Fig. \ref{fig:2MERA} shows an example of a MERA for a 1D lattice $\mathcal{L}$ made of $N=16$ sites. Its tensors are of types $(1,2)$ and $(2,2)$. We call the $(1,2)$ tensors \emph{isometries} $w$ and the $(2,2)$ tensors \emph{disentanglers} $u$ for reasons that will be explained shortly, and refer to Fig. \ref{fig:2MERA} as a binary 1D MERA, since it becomes a binary tree when we remove the disentanglers. Most of the previous work for 1D lattices \cite{ER,MERA,FreeFermions,FreeBosons,FlowEq,TimeEvolution,OldAlgorithm} has been done using the binary 1D MERA. However, there are many other possible choices. In this paper, for instance, we will mostly use the ternary 1D MERA of Fig. \ref{fig:3MERA}, where the isometries $w$ are of type $(1,3)$ and the disentanglers $u$ remain of type $(2,2)$. Fig. \ref{fig:NotationCompare} makes more explicit the meaning of Eq. \ref{eq:isometry} for these tensors. Notice that describing tensors and their manipulations by means of diagrams is fully equivalent to using equations and often much more clear.

\begin{figure}[!tb]
\begin{center}
\includegraphics[width=8cm]{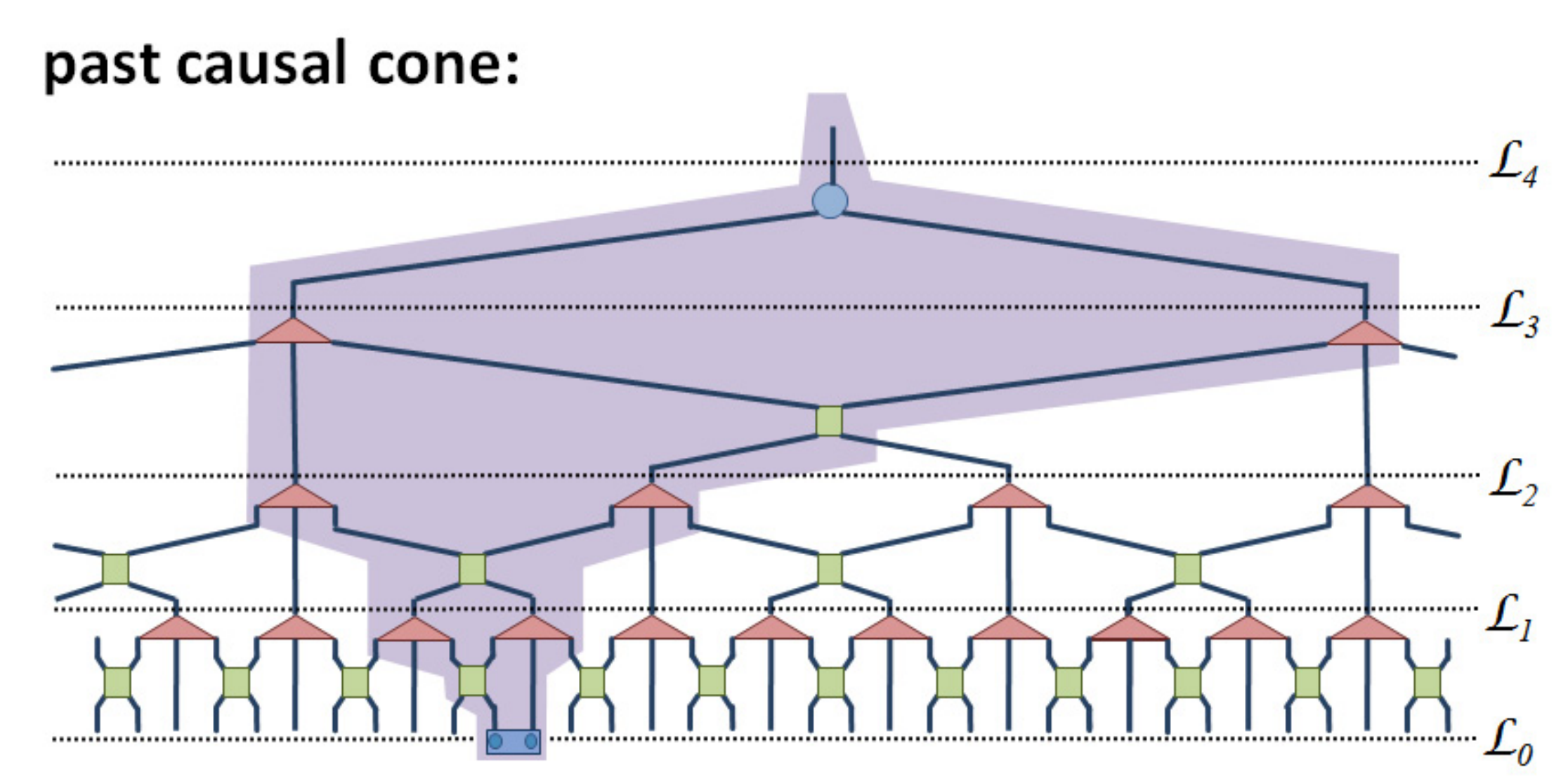}
\caption{(Colour online) The past causal cone of a group of sites in $\mathcal{L}_0 \equiv \mathcal{L}$ is the subset of wires and gates that can affect the state of those sites. The example shows the causal cone of a pair of nearest neighbor sites of $\mathcal{L}_0$ for the ternary 1D MERA. Notice that for each lattice $\mathcal{L}_{\tau}$, $\tau=0,1,2,3,4$, the causal cone involves at most 2 sites. This can be seen to be the case for any pair of contiguous sites of $\mathcal{L}_0$. We refer to this property by saying that the causal cones of the MERA have bounded width. } 
\label{fig:CausalCone}
\end{center}
\end{figure}

Eq. \ref{eq:isometry} encapsulates a distinctive property of the MERA as a tensor network: each of its tensors is isometric (notice that Eq. \ref{eq:unitary} is a particular case of Eq. \ref{eq:isometry}). A second key feature of the MERA refers to its causal structure. We define the past \emph{causal cone} of an outgoing wire of circuit $\mathcal{C}$ as the set of wires and gates that can affect the state on that wire. A quantum circuit $\mathcal{C}$ leads to a MERA only when the causal cone of an outgoing wire involves just a constant (that is, independent of $N$) number of wires at any fixed past time (Fig. \ref{fig:CausalCone}). We refer to this property by saying that the causal cone has a bounded 'width'. 

The usefulness of the quantum circuit interpretation of the MERA will become clear in the next section, in the context of computing expected values for local observables. There, the two defining properties, namely Eq. \ref{eq:isometry} and the peculiar structure of the causal cones of $\mathcal{C}$, will be the key to making such calculations efficient. 

\subsection{Renormalization group transformation}

Let us now review how the MERA defines a coarse-graining transformation for lattice systems that leads to a real-space renormalization group scheme, known as entanglement renormalization \cite{ER}. 

We start by grouping the tensors in the MERA into $T\approx \log N$ different layers, where each layer contains a row of isometries $w$ and a row of disentanglers $u$. We label these layers with an integer $\tau=1,2,\cdots T$, with $\tau=1$ for the lowest layer and with increasing values of $\tau$ as we climb up the tensor network, and denote by $U_{\tau}$ the isometric transformation implemented by all tensors in layer $\tau$, see Figs. \ref{fig:2MERA} and \ref{fig:3MERA}. Notice that the incoming wires of each $U_{\tau}$ define the Hilbert space of a lattice $\mathcal{L}_{\tau}$ with a number of sites $N_{\tau}$ that decreases exponentially with $\tau$ (specifically, as $N2^{-\tau}$ and $N3^{-\tau}$ for the binary and ternary 1D MERA). That is, the MERA implicitly defines a sequence of lattices 
\begin{equation}
	\mathcal{L}_0 \rightarrow \mathcal{L}_1 \rightarrow \cdots \rightarrow \mathcal{L}_T,
\end{equation}
where $\mathcal{L}_0\equiv \mathcal{L}$ is the original lattice, and where we can think of lattice $\mathcal{L}_\tau$ as the result of coarse-graining lattice $\mathcal{L}_{\tau-1}$.

\begin{figure}[!tb]
\begin{center}
\includegraphics[width=8cm]{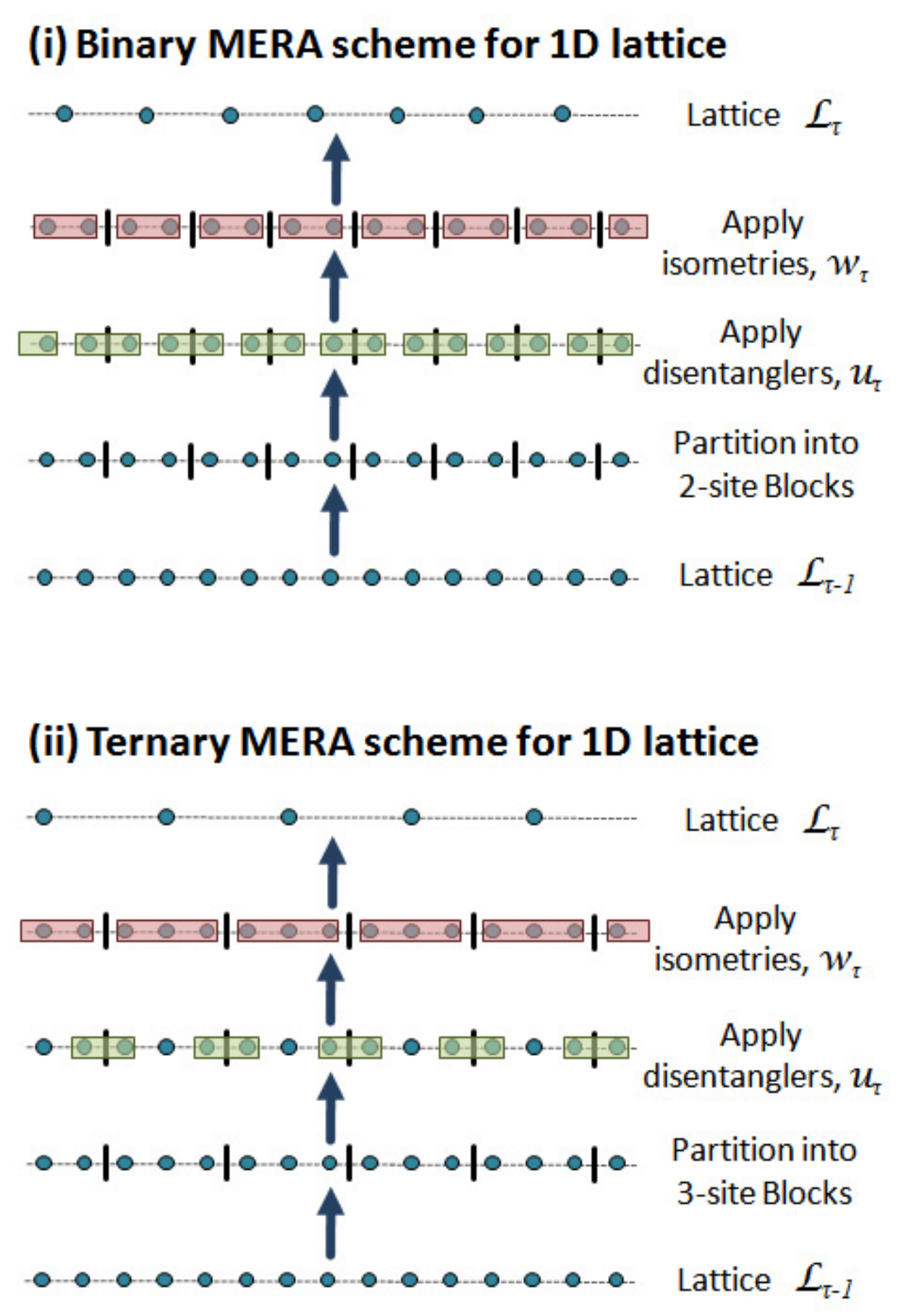}
\caption{(Colour online) Detailed description of the real-space renormalization group transformation for 1D lattices induced by ($i$) the binary 1D MERA and ($ii$) the ternary 1D MERA.} 
\label{fig:1DAltSchemes}
\end{center}
\end{figure}

Specifically, as illustrated in Figs. \ref{fig:2MERA} and (\ref{fig:3MERA}, this coarse-graining transformation is implemented by the operator $U_{\tau}^{\dagger}$ that maps pure states of the lattice $\mathcal{L}_{\tau-1}$ into pure states of the lattice $\mathcal{L}_{\tau}$,
\begin{equation}
	U_{\tau}^{\dagger}:\mathbb{V}_{\mathcal{L}_{\tau-1}} \rightarrow \mathbb{V}_{\mathcal{L}_{\tau}},
	\label{eq:Utau}
\end{equation}
and that proceeds in two steps (Fig. \ref{fig:1DAltSchemes}). Let us partition the lattice $\mathcal{L}_{\tau-1}$ into blocks of neighboring sites. The first step consists of applying the disentanglers $u$ on the boundaries of the blocks, aiming to reduce the amount of short range entanglement in the system. Once (part of) the short-range entanglement between neighboring blocks has been removed, the isometries $w$ are used in the second step to map each block of sites of lattice $\mathcal{L}_{\tau-1}$ into a single effective site of lattice $\mathcal{L}_{\tau}$.

By composition, we obtain a sequence of increasingly coarse-grained states, 
\begin{equation}
	\ket{\Psi_0} \rightarrow \ket{\Psi_1} \rightarrow \cdots \rightarrow \ket{\Psi_{T}},
\label{eq:sequePsi}
\end{equation}
for the lattices $\{\mathcal{L}_0, \mathcal{L}_1, \cdots, \mathcal{L}_T\}$, where $\ket{\Psi_{\tau}} \equiv U_{\tau}^{\dagger} \ket{\Psi_{\tau-1}}$ and $\ket{\Psi_0}\equiv \ket{\Psi}$ is the original state. Overall the MERA corresponds to the transformation $U\equiv U_1 U_2 \cdots U_T$,
\begin{equation}
	U :\mathbb{V}_{\mathcal{L}_{T}}\rightarrow \mathbb{V}_{\mathcal{L}_0},
\label{eq:U}
\end{equation}
with $\ket{\Psi_0} = U\ket{\Psi_T}$. 

\begin{figure}[!tb]
\begin{center}
\includegraphics[width=8cm]{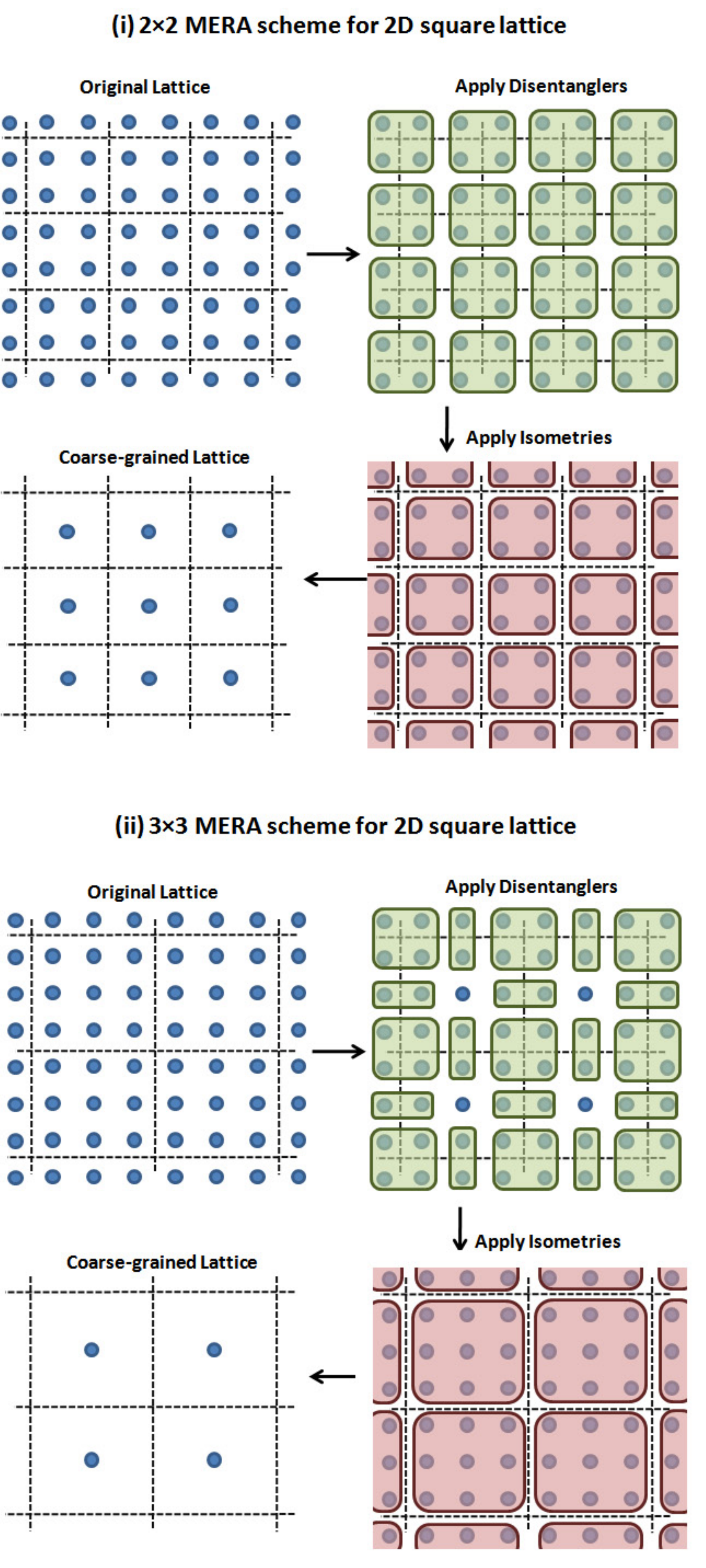}
\caption{(Colour online) Detailed description of the real-space renormalization group transformation for a 2D square lattice induced by two possible realizations of the MERA,  generalizing the 1D schemes of Fig. \ref{fig:1DAltSchemes}. In the first case the isometries map a block of $2\times 2$ sites into a single site, which can be seen to imply that the natural size of a local operator, equivalently the causal width of the scheme, is $3\times 3$ sites. In the second case the isometries map a block of $3\times 3$ sites into a single site and the natural size of a local operator is $2\times 2$. As a result, the computational cost in the second scheme is much smaller than in the first scheme.} 
\label{fig:2DAltSchemes}
\end{center}
\end{figure}

Regarding the MERA from the perspective of the renormalization group is quite instructive. It tells us that this ansatz is likely to describe states with a specific structure of internal correlations, namely, states in which the entanglement is organized in different length scales. Let us briefly explain what we mean by this. 

We say that the state $\ket{\Psi}$ contains entanglement at a given length scale $\lambda$ if by applying a unitary operation (i.e. a disentangler) on a region $R$ of linear size $\lambda$, we are able to decouple (i.e. disentangle) some of the local degrees of freedom, that is, if we are able to convert the state $\ket{\Psi}$ into a product state $\ket{\Psi'}\otimes \ket{0}$, where $\ket{0}$ is the state of the local degrees of freedom that have been decoupled and $\ket{\Psi'}$ is the state of the rest of the system. [Here we assumed, of course, that the decoupling is not possible with a unitary operation that acts on a subregion $R'$ of the region $R$, where the size $\lambda'$ of $R'$ is smaller than the size of $R$, $\lambda' < \lambda$].

What makes the MERA useful is that the entanglement in most ground states of local Hamiltonians seems to decompose into moderate contributions corresponding to different length scales. We can identify two behaviors, depending on whether the system is in a phase characterized by symmetry-breaking order or by topological order (see \cite{LevinWen} and references therein). In systems with symmetry-breaking order, ground-state entanglement spans all length scales $\lambda$ smaller than the correlation length $\xi$ in the system --- and, consequently, at a quantum critical point, where the correlation length $\xi$ diverges, entanglement is present at all length scales \cite{ER}. In a system with topological order, instead, the ground state displays some form of (topological) entanglement affecting all length scales even when the correlation length vanishes \cite{QuantumDouble,StringNet}.

\subsection{Choose your MERA}
 
We have introduced the MERA as a tensor network originating in a quantum circuit. Its tensors have incoming and outgoing wires/indices according to a well-defined direction of time in the circuit. Therefore, a MERA can be regarded as a tensor network equipped with a (fictitious) time direction and with two properties:
\begin{itemize}
\item Its tensors are isometric (Eq. \ref{eq:isometry}). 
\item Past causal cones have bounded width (Fig. \ref{fig:CausalCone}). 
\end{itemize}

From a computational perspective, these are the only properties that we need to retain. In particular, there is no need to keep the vector space dimension of the quantum wires (equivalently, of the sites in the coarse-grained lattice) constant throughout the tensor network. Accordingly, we will consider a MERA where the vector space dimension of a site of lattice $\mathcal{L}_{\tau}$, denoted $\chi_{\tau}$, may depend on the layer $\tau$ (this dimension could also be different for each individual site of layer $\tau$, but for simplicity we will not consider this case here). Notice that $\chi_0=d$ corresponds to the sites of the original lattice $\mathcal{L}$. 

\textbf{Bond dimension.---} Often, however, the sites in most layers will have the same vector space dimension (except, for instance, the sites of the original lattice $\mathcal{L}$, with $\chi_0=d$, or the single site of the top lattice $\mathcal{L}_{T}$, with dimension $\chi_T$). In this case we denote the dominant dimension simply by $\chi$, and we refer to the MERA as having \emph{bond dimension} $\chi$. The computational cost of the algorithms described in subsequent sections is often expressed as a power of the bond dimension $\chi$.

\textbf{Rank.---} We refer to the dimension $\chi_T$ of the space $\mathbb{V}_{\mathcal{L}_{T}}$ (corresponding to the single site of the uppermost lattice $\mathcal{L}_T$) as the rank of the MERA. For $\chi_T=1$, the MERA represents a pure state $\ket{\Psi}\in\mathbb{V}_{\mathcal{L}}$. More generally, a rank $\chi_T$ MERA encodes a $\chi_T$-dimensional subspace $\mathbb{V}_{U} \subseteq \mathbb{V}_{\mathcal{L}}$. For instance, given a Hamiltonian $H$ on the lattice $\mathcal{L}$, we could use a rank $\chi_T$ MERA to describe the ground subspace of $H$ (assuming it had dimension $\chi_T$); or the ground state of $H$ (if it was not degenerate) and its $\chi_T-1$ excitations with lowest energy. The isometric transformation $U$ in Eq. \ref{eq:U} can be used to build a projector $P \equiv UU^{\dagger}$,
\begin{equation}
	P :\mathbb{V}_{\mathcal{L}} \rightarrow \mathbb{V}_{\mathcal{L}},~~~~~~~~P^2 = P,~~~ \tr(P) = \chi_T,
	\label{eq:P}
\end{equation}
onto the subspace $\mathbb{V}_{U} \subseteq \mathbb{V}_{\mathcal{L}}$.

Given the above definition of the MERA, many different realizations are possible depending on what kind of isometric tensors are used and how they are interconnected. We have already met two examples for a 1D lattice, based on a binary and ternary underlying tree. Fig. \ref{fig:2DAltSchemes} shows two schemes for a 2D square lattice. It is natural to ask, given a lattice geometry, what realization of the MERA is the most convenient from a computational point of view. A definitive answer to this question does not seem simple. An important factor, however, is given by the fixed-point size of the support of local observables under successive RG transformations---which corresponds to the width of the past causal cones. 

\textbf{Support of local observables.---} In each MERA scheme, under successive coarse-graining transformations a local operator eventually becomes supported in a characteristic number of sites. This is the result of two competing effects: disentanglers $u$ tend to extend the support of the local observable (by adding new sites at its boundary), whereas the isometries $w$ tend to reduce it (by transforming blocks of sites into single sites). For instance, in the binary 1D MERA, local observables end up supported in three contiguous sites (Fig. \ref{fig:2MERA}), whereas in the ternary 1D MERA local observables become supported in two contiguous sites (Fig. \ref{fig:3MERA}).

Therefore, an important difference between the binary and ternary 1D schemes is in the natural support of local observables. This can be seen to imply that the cost of a computation scales as a larger power of the bond dimension $\chi$ for the binary scheme than for the ternary scheme, namely as $O(\chi^9)$ compared to $O(\chi^8)$. However, it turns out that the binary scheme is more effective at removing entanglement, and as a result a smaller $\chi$ is already sufficient in order to achieve the same degree of accuracy in the computation of, say, a ground state energy. In the end, we find that for the 1D systems analyzed in Sect. \ref{sect:benchmark}, the two effects compensate and the cost required in both schemes in order to achieve the same accuracy is comparable. On the other hand, in the ternary 1D MERA, two-point correlators between selected sites can be computed at a cost $O(\chi^8)$, whereas analogous calculations in the binary 1D MERA are much more expensive. Therefore in any context where the calculation of two-point correlators is important, the ternary 1D MERA is a better choice.

The number of possible realizations of the MERA for 2D lattices is greater than for 1D lattices. For a square lattice, the two schemes of Fig. \ref{fig:2DAltSchemes} are obvious generalizations of the above ones for 1D lattices. The first scheme, proposed in \cite{FreeFermions} (see also \cite{Finite2D}), involves isometries of type $(1,4)$ and the natural support of local observables is a block of $3\times 3$ sites. The second scheme involves isometries of type $(1,9)$ and local observables end up supported in blocks of $2\times 2$ sites. Here, the much narrower causal cones of the second scheme leads to a much better scaling of the computational cost with $\chi$, only $O(\chi^{16})$ compared to $O(\chi^{28})$ for the first scheme.

Another remark in relation to possible realizations concerns the type of tensors we use. So far we have insisted in distinguishing between disentanglers $u$ (unitary tensors of type $p\rightarrow p$) and isometries $w$ (isometric tensors of type $1\rightarrow p'$). We will continue to use this terminology throughout this paper, but we emphasize that a more general form of isometric tensor, e.g. of type $(2,4)$, that both disentangles the system and coarse-grains sites, is possible and is actually used in some realizations \cite{Scalable2D}.

\begin{figure}[!tb]
\begin{center}
\includegraphics[width=8cm]{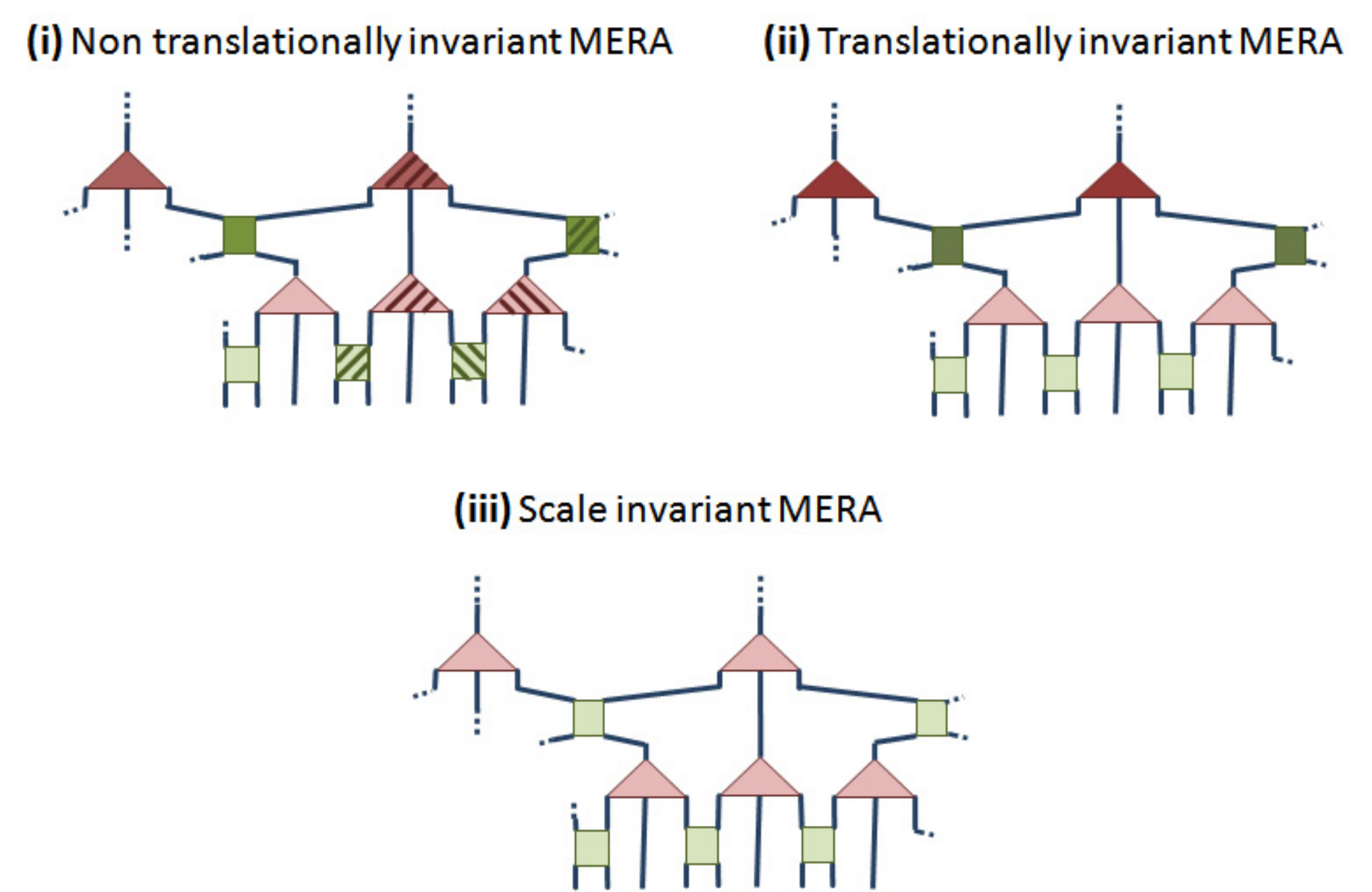}
\caption{(Colour online) Ternary 1D MERA in the presence of space symmetries. (i) In order to represent an inhomogeneous state/subspace, all disentanglers $u$ and isometries $w$ are different (denoted by different colouring). Notice that there are $N/3$ disentanglers (isometries) in the first layer, $N/9$ in the second, and more generally $N/3^{\tau}$ in layer $\tau$, so that the total number of tensors is $2N \sum_{\tau=1}^{\log N} 1/3^{\tau} < 2N$. Therefore the total number of parameters required to specify the MERA is proportional to the size $N$ of the lattice $\mathcal{L}$. (ii) In order to represent a state/subspace that is invariant under translations, we choose all disentanglers and isometries on a given layer of the MERA to be the same. In this case the MERA is completely specified by $O(\log N)$ disentanglers and isometries. (iii) In a scale invariant MERA, the same disentangler and isometry is in addition used in all layers.} 
\label{fig:MERAtypes}
\end{center}
\end{figure}

\subsection{Exploiting symmetries}

Symmetries have a direct impact on the efficiency of computations, because they can be used to drastically reduce the number of parameters in the MERA. Important examples are given by space symmetries, such as translation and scale invariance, see Fig. \ref{fig:MERAtypes}.

The MERA is made of $O(N)$ disentanglers and isometries. In order to describe an inhomogeneous state $\ket{\Psi}\in \mathbb{V}_{\mathcal{L}}$ or subspace $\mathbb{V}_{U} \subseteq \mathbb{V}_{\mathcal{L}}$, all these tensors are chosen to be different. Therefore, for fixed $\chi$ the number of parameters in the MERA scales linearly in $N$.

However, in the presence of translation invariance, one can use a \emph{translation invariant} MERA, where we choose all the disentanglers $u$ and isometries $w$ of any given layer $\tau$ to be the same, thus reducing the number of parameters to $O(\log N)$ (if there are $T\approx \log N$ layers). We emphasize that a translation invariant MERA, as just defined, does not necessarily represent a translation invariant state $\ket{\Psi}\in \mathbb{V}_{\mathcal{L}}$ or subspace $\mathbb{V}_{U} \subseteq \mathbb{V}_{\mathcal{L}}$. The reason is that different sites of $\mathcal{L}$ are placed in inequivalent positions with respect to the MERA. As a result, often the MERA can only approximately reproduce translation invariant states/subspaces, although the departure from translation invariance is seen to typically decrease fast with increasing $\chi$. In order to further mitigate inhomogeneities, we often consider an average of local observables/reduced density matrices over all possible sites, as will be discussed in the next section.

In systems that are invariant under changes of scale, we will use a \emph{scale invariant} MERA, where all the disentanglers and isometries can be chosen to be the same and we only need to store a constant number of parameters. The scale invariant MERA is useful to represent the ground state of some quantum critical systems \cite{ER} and the ground subspace of systems with topological order at the infrared limit of the RG flow \cite{QuantumDouble,StringNet}.

A reduction in parameters (as a function of $\chi$) is also possible in the presence of internal symmetries, such as $U(1)$ (e.g. particle conservation) or $SU(2)$ (e.g. spin isotropy). We defer their analysis to Ref. \cite{Sukhi}.

For the sake of concreteness, the explanations in the rest of this manuscript refer to the ternary 1D scheme of Fig. \ref{fig:3MERA}. However, analogous considerations also apply to any other realization of the MERA. 

\section{Computation of expected values of local observables and correlators}
\label{sect:local}

Let $o^{[r,r+1]}$ denote a local observable defined on two contiguous sites $[r,r\!+\!1]$ of $\mathcal{L}$. In this section we explain how to compute the expected value 
\begin{equation}
	\langle o^{[r,r+1]} \rangle_{\mathbb{V}_{U}} \equiv \tr (o^{[r,r+1]}P).
\label{eq:ev_o}
\end{equation}
Here $P$ is a projector (see Eq. \ref{eq:P}) onto the $\chi_T$-dimensional subspace $\mathbb{V}_{U}\subseteq \mathbb{V}_{\mathcal{L}}$ represented by the MERA. For a rank $\chi_T=1$ MERA, representing a pure state $\ket{\Psi}\in  \mathbb{V}_{\mathcal{L}}$, the above expression reduces to 
\begin{equation}
	\langle o^{[r,r+1]} \rangle_{\Psi} \equiv \bra{\Psi}o^{[r,r+1]}\ket{\Psi}. 
\end{equation}
Evaluating Eq. \ref{eq:ev_o} is necessary in order to extract physically relevant information from the MERA, such as e.g. the energy and magnetization in a spin system. In addition, the manipulations involved are also required as a central part of the optimization algorithms described in Sects. \ref{sect:optim} and \ref{sect:algorithm}. The results of this section remain relevant even in cases where no optimization algorithm is required (for instance when an exact expression of the MERA is known \cite{QuantumDouble,StringNet}).

As explained below, the expected value of Eq. (\ref{eq:ev_o}) can be computed in a number of ways:
\begin{itemize}
\item By repeated use of the \emph{ascending superoperator} $\mathcal{A}$, the local operator $o^{[r,r+1]}$ is mapped onto a coarse-grained operator $o_T$ on lattice $\mathcal{L}_T$. Eq. (\ref{eq:ev_o}) can then be evaluated as the trace of the coarse-grained operator $o_T$, $\tr (o^{[r,r+1]}P) = \tr(o_T)$.
\item Alternatively, by repeated use of the \emph{descending superoperator} $\mathcal{D}$, a two-site reduced density matrix $\rho^{[r,r+1]}$ for lattice $\mathcal{L}$ is obtained. Eq. (\ref{eq:ev_o}) can then evaluated as $\tr (o^{[r,r+1]}P) = \tr (o^{[r,r+1]}\rho^{[r,r+1]})$.
\item More generally, the ascending and descending superoperators $\mathcal{A}$ and $\mathcal{D}$ can be used to compute an operator $o^{[r',r'+1]}_{\tau}$ and density matrix $\rho^{[r',r'+1]}_{\tau}$ for the coarse-grained lattice $\mathcal{L}_{\tau}$. Eq. (\ref{eq:ev_o}) can then be evaluated as  $\tr (o^{[r,r+1]}P) = \tr (o^{[r',r'+1]}_{\tau}\rho^{[r',r'+1]}_{\tau})$. 
\end{itemize}
First we introduce the ascending and descending superoperators $\mathcal{A}$ and $\mathcal{D}$ and explain in detail how to perform the computation of the expected value of Eq. (\ref{eq:ev_o}). Then we address also the computation of the expected value
\begin{equation}
		\langle O \rangle_{\mathbb{V}_{U}} \equiv \tr (OP),~~~~~~~~	O \equiv \sum_{r} o^{[r,r+1]},
\label{eq:ev_O}
\end{equation}
where $O$ is an operator that decomposes as a sum of local operators in $\mathcal{L}$;
as well as the computation of two-point correlators. Finally, we revisit these tasks in the presence of translation invariance and scale invariance.

The ascending and descending superoperators are an essential part of the MERA formalism that was introduced in Ref. \cite{MERA} (see e.g. Fig. 4 of Ref. \cite{MERA} for an explicit representation of the descending superoperator $\mathcal{D}$). These superoperators have been also called MERA quantum channel/MERA transfer matrix in Ref. \cite{Transfer}.

\subsection{Ascending and descending superoperators}

\begin{figure}[!tb]
\begin{center}
\includegraphics[width=8cm]{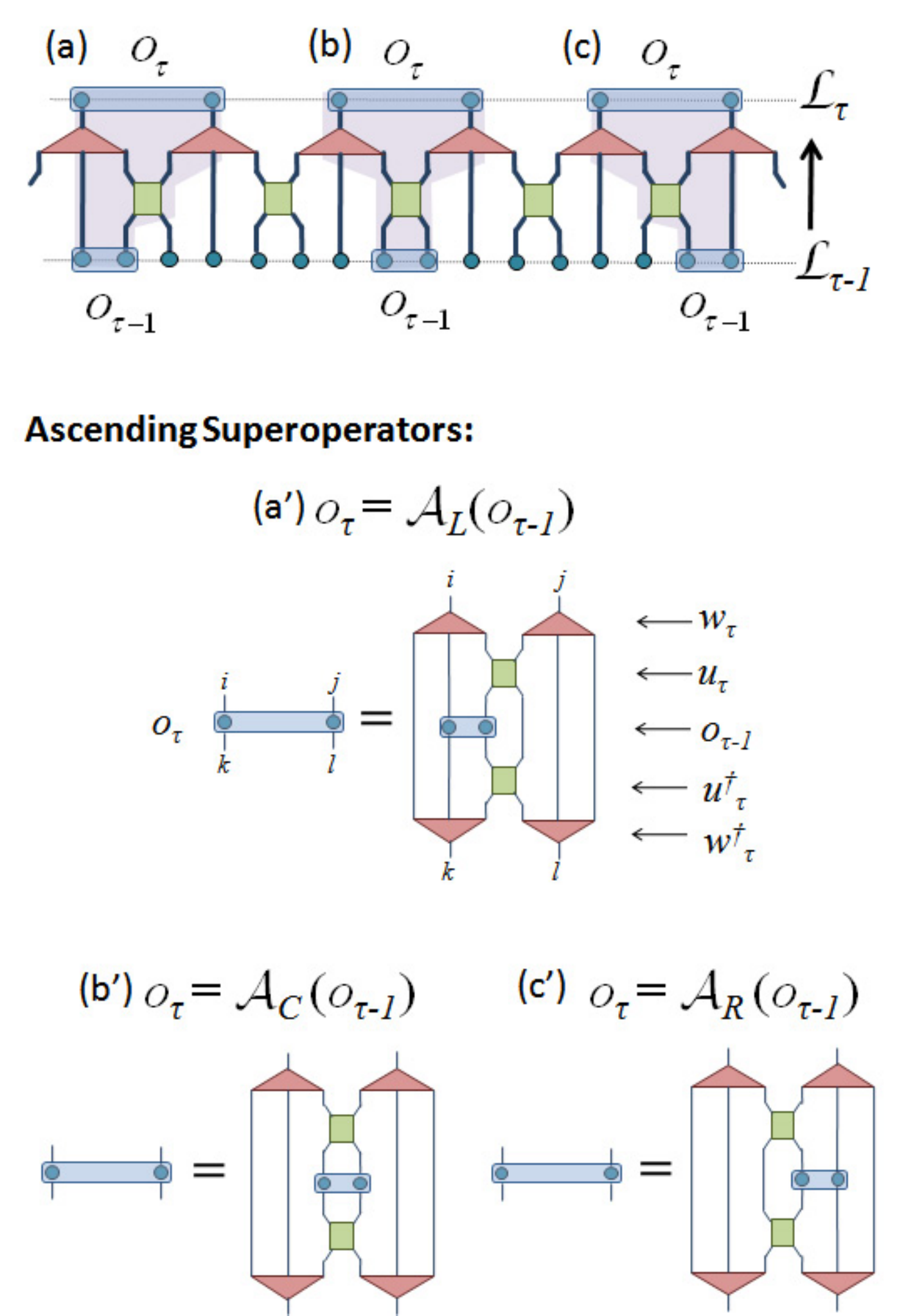}
\caption{(Colour online) The ascending superoperator $\mathcal{A}$ transforms a local operator $o_{\tau-1}$ of lattice $\mathcal{L}_{\tau-1}$ into a local operator $o_{\tau}$ of lattice $\mathcal{L}_{\tau}$ (for simplicity we omit the label $[r,r+1]$ that specifies the sites on which $o_{\tau-1}$ and $o_{\tau}$ are supported). Depending on the relative position between the support of $o_{\tau-1}$ and the closest disentangler, the operator can be lifted to lattice $\mathcal{L}_{\tau}$ in three different ways, indicated in the figure as (a), (b) and (c). Correspondingly, there are three structurally different forms of the ascending superoperator $\mathcal{A}$, namely left $\mathcal{A}_L$, center $\mathcal{A}_C$ and right $\mathcal{A}_R$, indicated as (a'), (b') and (c'). Notice that the figure completely specifies the tensor network representation of the superoperator, which is written in terms of the relevant disentanglers and isometries (and their Hermitian conjugates). An explicit form for the \emph{average ascending superoperator} $\bar{\mathcal{A}}$ of Eq. \ref{eq:avAsc} is obtained by averaging the above three tensor networks.}
\label{fig:AscendSuper}
\end{center}
\end{figure}

In the previous section we have seen that the MERA defines a sequence of increasingly coarser lattices $\{\mathcal{L}_0, \mathcal{L}_1, \cdots, \mathcal{L}_T\}$. Under the coarse-graining transformation $U_{\tau}^\dagger$ of Eq. \ref{eq:Utau}, a local operator $o^{[r,r+1]}_{\tau-1}$, supported on two consecutive sites $[r,r\!+\!1]$ of lattice $\mathcal{L}_{\tau-1}$, is mapped onto another local operator $o^{[r',r'+1]}_{\tau}$ supported on two consecutive sites $[r',r'\!+\!1]$ of lattice $\mathcal{L}_{\tau}$ (Fig. \ref{fig:3MERA}). This is so because in $U_{\tau}^\dagger o^{[r,r+1]}_{\tau-1} U_{\tau}$ most disentanglers and isometries of $U_{\tau}$ and $U_{\tau}^{\dagger}$ are annihilated in pairs according to Eq. \ref{eq:isometry}. The resulting transformation is implemented by means of the \emph{ascending} superoperator $\mathcal{A}$ described in Fig. \ref{fig:AscendSuper}, 
\begin{equation}
 o^{[r',r'+1]}_{\tau} = \mathcal{A}(o^{[r,r+1]}_{\tau-1}).
\label{eq:Ascending}
\end{equation}
In order to keep our notation simple, we do not specify on which lattice/sites the superoperator $\mathcal{A}$ is applied, even though $\mathcal{A}$ actually depends on $\tau$, $r$ and $r'$. Instead, when necessary we will simply indicate which of its three structurally different forms (namely, left $\mathcal{A}_L$, center $\mathcal{A}_C$ or right $\mathcal{A}_R$ in Fig. \ref{fig:AscendSuper}) is being used.

\begin{figure}[!htb]
\begin{center}
\includegraphics[width=8cm]{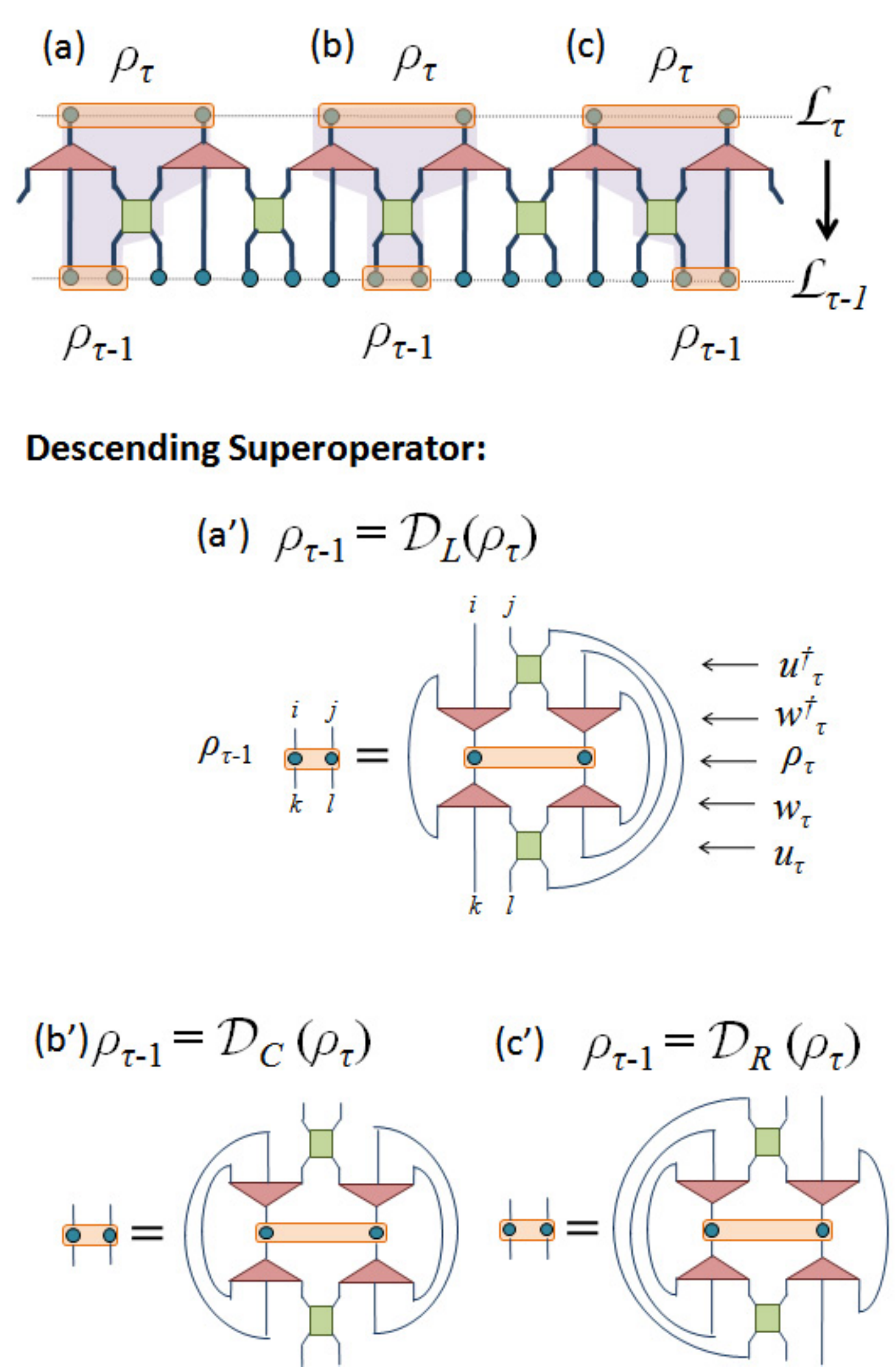}
\caption{(Colour online) The descending superoperator $\mathcal{D}$ transforms a local density matrix $\rho_{\tau}$ of lattice $\mathcal{L}_{\tau}$ into a local density matrix $\rho_{\tau-1}$ of lattice $\mathcal{L}_{\tau-1}$. Depending on the relative position between the support of $\rho_{\tau-1}$ and the closest disentangler, the density matrix $\rho_{\tau}$ canbe lowered to lattice $\mathcal{L}_{\tau-1}$ in three different ways, indicated in the figure as (a), (b) and (c). Correspondingly, there are three structurally different forms of the descending superoperator $\mathcal{D}$, namely left $\mathcal{D}_L$, center $\mathcal{D}_C$ and right $\mathcal{D}_R$, indicated as (a'), (b') and (c'). An explicit form for the \emph{average descending superoperator} $\bar{\mathcal{D}}$ of Eq. \ref{eq:avDesc} is obtained by averaging the above three tensor networks.}
\label{fig:DescendSuper}
\end{center}
\end{figure}

\begin{figure}[!tb]
\begin{center}
\includegraphics[width=7cm]{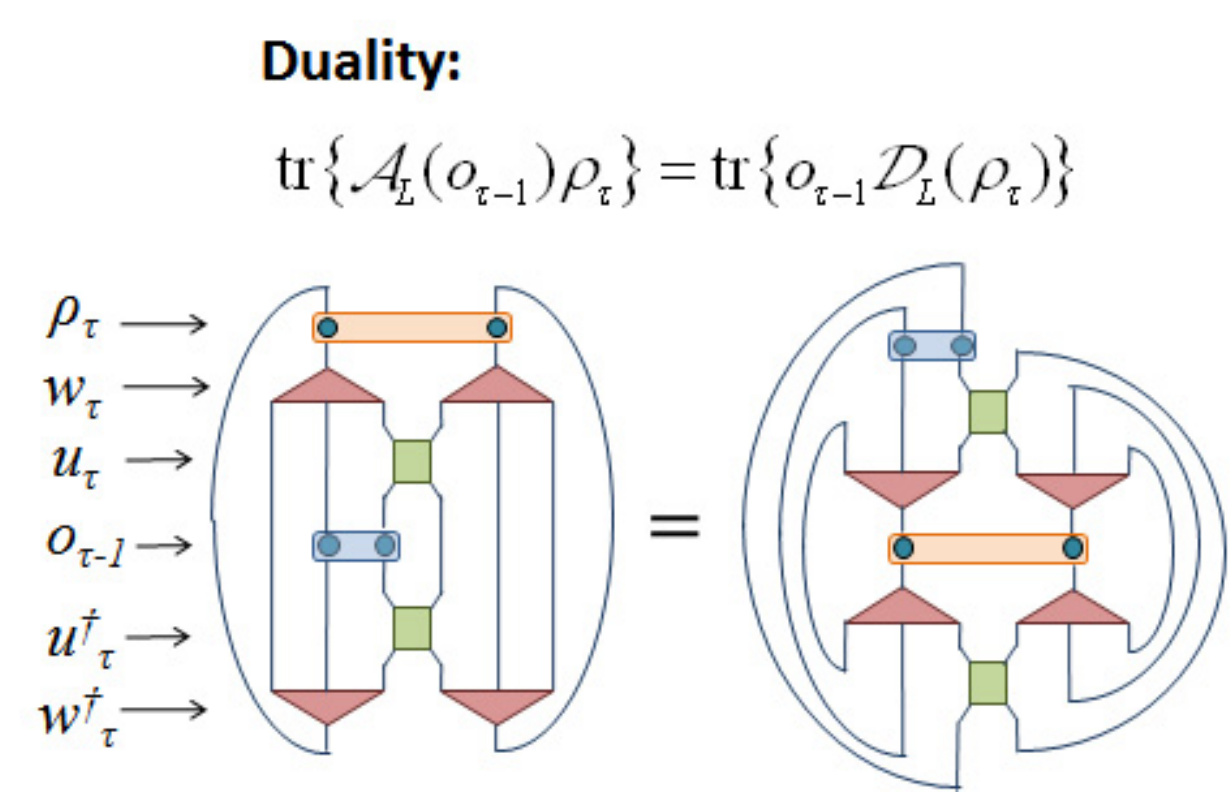}
\caption{(Colour online) The ascending and descending superoperators, $\mathcal{A}$ and $\mathcal{D}$, are dual to each other, see Eq. \ref{eq:Duality}. This becomes evident by inspecting the above figure, where the superoperators are explicitly decomposed in terms of disentanglers and isometries.} 
\label{fig:Duality}
\end{center}
\end{figure}

As above, let $[r,r\!+\!1]$ denote two consecutive sites of lattice $\mathcal{L}_{\tau-1}$ and let $[r',r'\!+\!1]$ denote two consecutive sites of lattice $\mathcal{L}_{\tau}$ that lay inside the past causal cone of $[r,r\!+\!1] \in \mathcal{L}_{\tau-1}$. Given a density matrix $\rho_{\tau}^{[r',r'+1]}$ in $\mathcal{L}_{\tau}$, the \emph{descending superoperator} $\mathcal{D}$ of Fig. \ref{fig:DescendSuper} produces a density matrix $\rho_{\tau-1}^{[r,r+1]}$ in $\mathcal{L}_{\tau-1}$,
\begin{equation}
 \rho^{[r,r+1]}_{\tau-1} = \mathcal{D}(\rho^{[r',r'+1]}_{\tau}).
\end{equation}
Notice that the descending superoperator $\mathcal{D}$ (which depends on $\tau$, $r$ and $r'$) is the dual of the ascending superoperator $\mathcal{A}$, $\mathcal{D} = \mathcal{A}^{\star}$. Indeed, as can be checked in Fig. \ref{fig:Duality}, by construction we have that, for any $o^{[r,r+1]}_{\tau-1}$ and $\rho^{[r',r'+1]}_{\tau}$,
\begin{equation}
	\tr \left( o^{[r,r+1]}_{\tau-1}\mathcal{D}(\rho^{[r',r'+1]}_{\tau}) \right) 
	= \tr \left( \mathcal{A}(o^{[r,r+1]}_{\tau-1}) \rho^{[r',r'+1]}_{\tau} \right).
	\label{eq:Duality}
\end{equation}
Correspondingly, there are also three structurally different forms of the descending superoperators, namely left $\mathcal{D}_L$, center $\mathcal{D}_C$ and right $\mathcal{D}_R$ in Fig. \ref{fig:DescendSuper}.

\begin{figure}[!tb]
\begin{center}
\includegraphics[width=7cm]{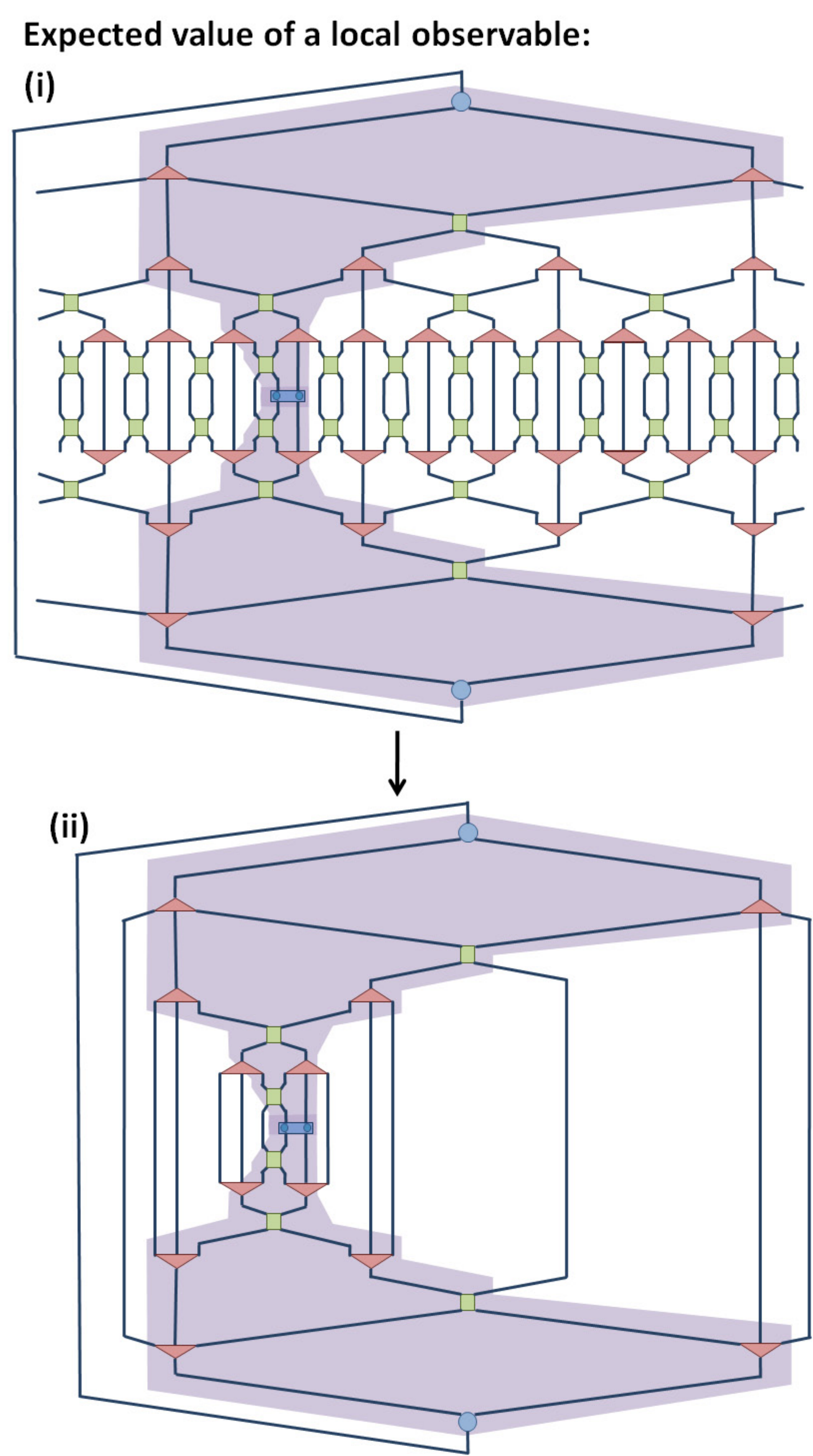}
\caption{(Colour online) (i) Tensor network corresponding to the expected value $\tr(o^{[r,r+1]}P)$ of Eq. \ref{eq:ev_o}. The two-site operator $o^{[r,r+1]}$ is represented by a four-legged rectangle in the middle of the tensor network. The shaded region represents the past causal cone of sites $r,r+1 \in \mathcal{L}$. (ii) All isometric tensors that lay outside the past causal cone of sites $r,r+1 \in \mathcal{L}$ annihilate and we are left with a simpler tensor network.} 
\label{fig:LocalObs1}
\end{center}
\end{figure}

\begin{figure}[!tb]
\begin{center}
\includegraphics[width=8cm]{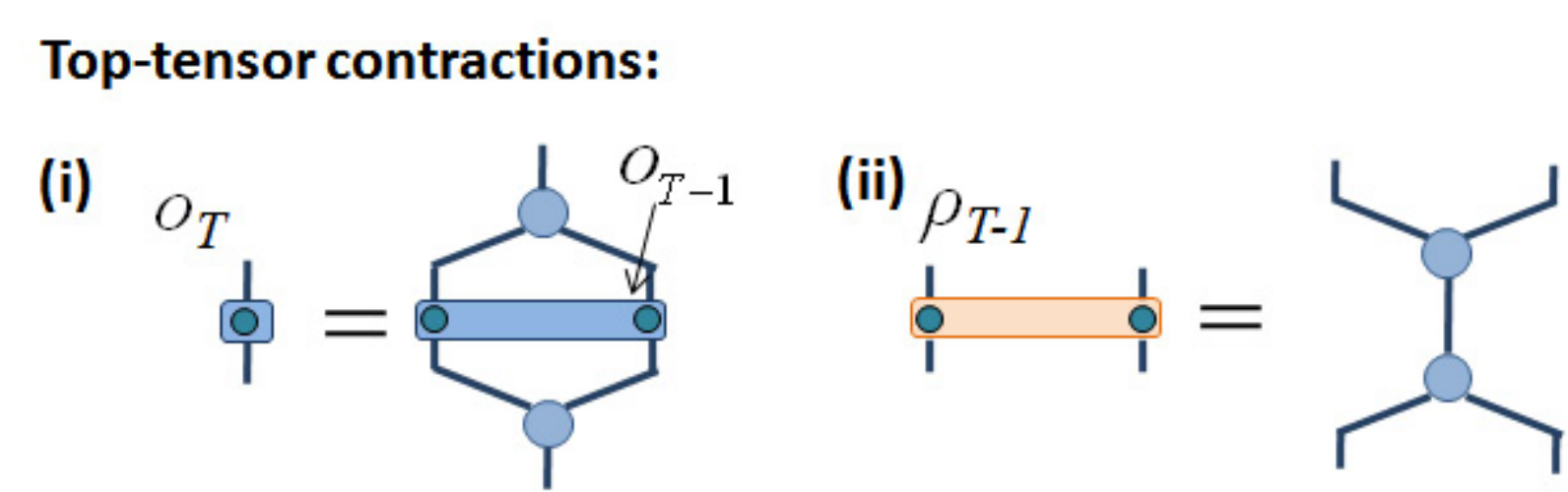}
\caption{(Colour online) (i) The \emph{top tensor} transforms a two-site operator $o_{T-1}$ defined on lattice $\mathcal{L}_{T-1}$ into a one-site operator (a $\chi_T \times \chi_T$ matrix) $o_{T}$ on the top of the MERA. (ii) The two-site density matrix $\rho_{T-1}$ on lattice $\mathcal{L}_{T-1}$ is obtained through contraction of the top tensor with its conjugate. Notice that $\rho_{T-1}$, as well as all $\rho_\tau$, are normalized to have trace $\tr(\rho_{\tau}) = \chi_T$.}
\label{fig:TopTensor}
\end{center}
\end{figure}

\subsection{Evaluation of a two-site operator}

We can now proceed to compute the expected value $\tr (o^{[r,r+1]}P)$ of Eq. \ref{eq:ev_o} from the MERA. This computation corresponds to contracting the tensor network depicted in the upper half of Fig. \ref{fig:LocalObs1}.

In a key first step, the contraction of the tensor network for $\tr (o^{[r,r+1]}P)$ is significantly simplified by the fact that, by virtue of Eq. \ref{eq:isometry}, each isometric tensor outside the past causal cone of sites $[r,r\!+\!1] \in \mathcal{L}$ is annihilated by its Hermitian conjugate. As a result, we are left with a new tensor network that contains only (two copies of) the tensors in the causal cone, as represented in the second half of Fig. \ref{fig:LocalObs1}. Because the past causal cones in the MERA have a bounded width, this tensor network can now be contracted with a computational effort that grows with $N$ just as O$(\log N)$. One can proceed in several ways:

\begin{figure}[!tb]
\begin{center}
\includegraphics[width=8cm]{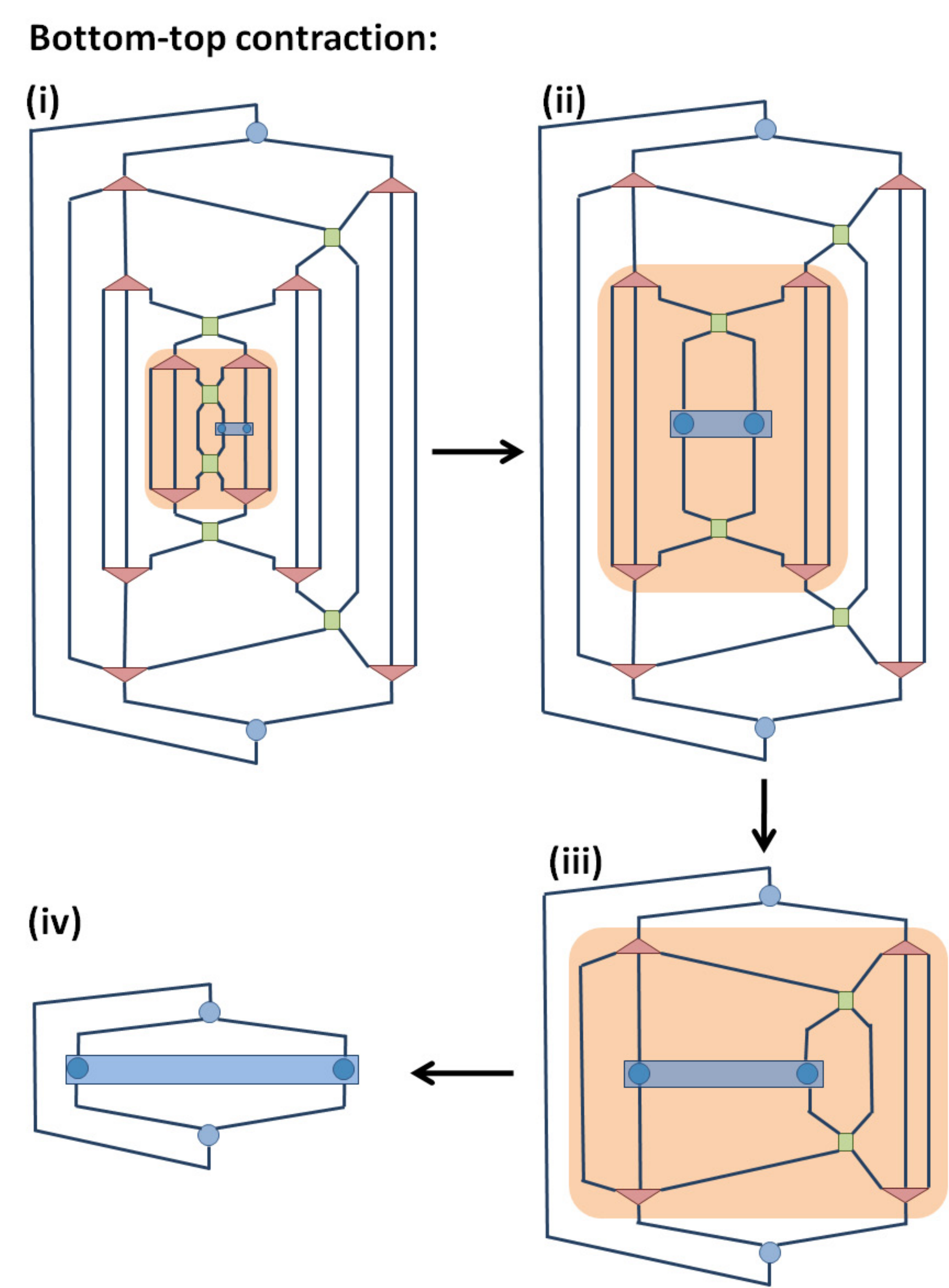}
\caption{(Colour online) The contraction of the tensor network in the lower half of Fig. \ref{fig:LocalObs1} using the bottom-top approach corresponds to employing the ascending super operator $\mathcal{A}$ a number of times. In this particular case, we first use (i) $\mathcal{A}_R$, then (ii) $\mathcal{A}_C$ and then (iii) $\mathcal{A}_L$, to bring the tensor network into a simple form whose contraction gives a complex number: the expected value of Eq. \ref{eq:ev_o}. } 
\label{fig:BottomTop}
\end{center}
\end{figure}

\begin{figure}[!tb]
\begin{center}
\includegraphics[width=8cm]{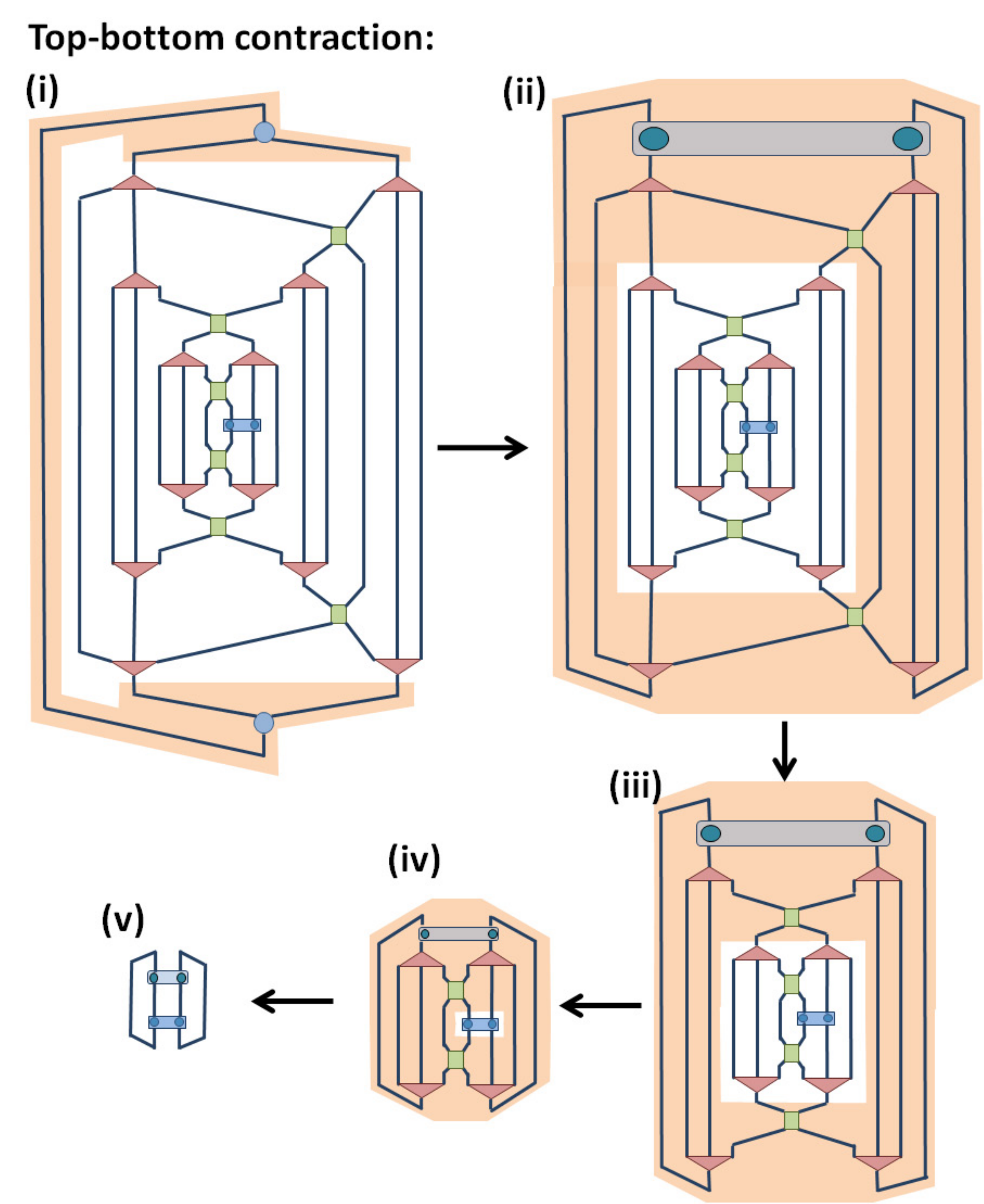}
\caption{(Colour online) The contraction of the tensor network in the lower half of Fig. \ref{fig:LocalObs1} using the top-bottom approach corresponds to first implementing (i) a \emph{top tensor} contraction followed by repeated application of the descending super operator $\mathcal{D}$. Specifically, here we first use (ii) $\mathcal{D}_L$, then (iii) $\mathcal{D}_C$ and then (iv) $\mathcal{D}_R$, in order to compute the appropriate density matrix $\rho^{[r,r+1]}$ for two sites $[r,r+1] \in \mathcal{L}$. With the density matrix $\rho^{[r,r+1]}$ we can finally compute (v) the expectation value $\tr (o^{[r,r+1]}P) = \tr(o^{[r,r+1]} \rho^{[r,r+1]})$.} 
\label{fig:TopBottom}
\end{center}
\end{figure}
 
\textbf{Bottom-top.---} In the bottom-top approach, we would start by contracting the indices of $o^{[r,r+1]}$ and the disentanglers and isometries of the first layer ($\tau=1$) of the causal cone; then we would contract the indices of disentanglers and isometries of the second layer ($\tau = 2$); and so on (Fig. \ref{fig:BottomTop}). However, this corresponds to repeatedly applying the ascending superoperator $\mathcal{A}$ on $o_0^{[r,r+1]} \equiv o^{[r,r+1]}$. Therefore this is precisely how we proceed, obtaining a sequence of increasingly coarse-grained operators
\begin{equation}
	o^{[r,r+1]}_0 ~ \stackrel{\mathcal{A}}{\rightarrow} ~ o^{[r_1,r_1+1]}_1 ~ \stackrel{\mathcal{A}}{\rightarrow} ~ o^{[r_2,r_2+1]}_2 ~ \stackrel{\mathcal{A}}{\rightarrow}~ \cdots ~ o_T
\end{equation}
supported on lattices $\mathcal{L}_0$, $\mathcal{L}_1$, $\mathcal{L}_2$, $\cdots$, and $\mathcal{L}_{T}$ respectively.
Here, the $\chi_T\times \chi_T$ matrix $o_T$ at the top of the MERA is obtained according to Fig. \ref{fig:TopTensor} and the expected value of Eq. \ref{eq:ev_o} corresponds to its trace,
\begin{equation}
	\tr(o^{[r,r+1]}P) = \tr(o_T).
\end{equation}
 
\textbf{Top-bottom.---} In the top-bottom approach, we would instead start by contracting the indices of the tensors in the top layer ($\tau=T$) of the causal cone; then we would contract the indices of the tensors in the layer right below ($\tau=T-1$); and so on (Fig. \ref{fig:TopBottom}). However, that corresponds to first computing a density matrix $\rho_{T-1}$ for the two sites of $\mathcal{L}_{T-1}$ according to Fig. \ref{fig:TopTensor}  and then repeatedly applying the descending superoperator $\mathcal{D}$. Therefore this is how we proceed, producing a sequence of two-site density matrices 
\begin{equation}
	\rho_{T-1} ~ \stackrel{\mathcal{D}}{\rightarrow} ~ \cdots ~ \rho^{[r_2,r_2+1]}_2 ~ \stackrel{\mathcal{D}} {\rightarrow} ~\rho^{[r_1,r_1+1]}_1 ~ \stackrel{\mathcal{D}} {\rightarrow} ~ \rho_0^{[r,r+1]}
	\label{eq:sequenceDM}
\end{equation}
supported on lattices $\mathcal{L}_{T-1}$, $\cdots$, $\mathcal{L}_2$, $\mathcal{L}_1$ and $\mathcal{L}_{0}$ respectively \cite{commentDM}. The last density matrix $\rho^{[r,r+1]} \equiv \rho^{[r,r+1]}_{0}$ describes the state of the two sites of $\mathcal{L}$ on which the local operator $o^{[r,r+1]}$ is supported. Therefore we can evaluate the expected value of $o^{[r,r+1]}$,
\begin{equation}
	\tr (o^{[r,r+1]}P) = \tr(o^{[r,r+1]} \rho^{[r,r+1]}).
\end{equation}

\textbf{Middle ground.---} More generally, one can also evaluate the expected value of Eq. \ref{eq:ev_o} through a mixed strategy where the ascending and descending superoperators are used to compute the operator $o^{[r_{\tau}, r_{\tau}+1]}_{\tau}$ and density matrix $\rho^{[r_{\tau}, r_{\tau}+1]}_{\tau}$ supported on lattice $\mathcal{L}_{\tau}$, which fulfill 
\begin{equation}
	\tr (o^{[r,r+1]}P) = \tr(o^{[r_{\tau}, r_{\tau}+1]}_{\tau} \rho^{[r_{\tau}, r_{\tau}+1]}_{\tau}).
\end{equation}

In all the cases above, one needs to use the ascending/descending superoperators about $T\approx \log N$ times, at a cost O$(\chi^8)$, so that the total computational cost is O$(\chi^8\log N)$.

\subsection{Evaluation of a sum of two-site operators}

In order to compute the expected value 
\begin{equation}
	\langle O \rangle_{\mathbb{V}_{U}} \equiv \tr (OP),~~~~~~~~~~O \equiv \sum_{r} o^{[r,r+1]}
\label{eq:ev_O1}
\end{equation}
of an operator $O$ on $\mathcal{L}$ that decomposes as the sum of two-site operators, we can write
\begin{equation}
\tr( O P) = \sum_{r} \tr( o^{[r,r+1]} P)
\label{eq:ev_O2}
\end{equation}
and individually evaluate each contribution $\tr( o^{[r,r+1]} P)$ by using e.g. the bottom-top strategy of the previous subsection, with a cost O$(\chi^8N\log N)$. However, by properly organizing the calculation, the cost of computing $\tr( O P)$ can be reduced to O$(\chi^8 N)$. We next describe how this is achieved. The strategy is closely related to the computation of expected values in the presence of translation invariance, as discussed later in this section. Again, there are several possible approaches:

\textbf{Bottom-top.---} We consider the sequence of operators
\begin{equation}
	O_0 ~ \stackrel{U_{1}^{\dagger}}{\rightarrow} ~ O_1 ~ \stackrel{U_{2}^{\dagger}}{\rightarrow} ~ O_2 ~ \stackrel{U_{3}^{\dagger}}{\rightarrow}~ \cdots ~ O_T, ~~~~~~~~O_0 \equiv O,
\label{eq:Os}
\end{equation}
where the operator $O_{\tau}$ is the sum of $N/3^{\tau}$ local operators,
\begin{equation}
	O_{\tau} = \sum_{r=1}^{N/3^{\tau}} o_{\tau}^{[r,r+1]}.
	\label{eq:O2}
\end{equation}
$O_{\tau-1}$ is obtained from $O_{\tau-1}$ by coarse-graining, $O_{\tau} = U_{\tau}^{\dagger} O_{\tau-1} U_{\tau}$. Each local operator $o^{[r,r+1]}_{\tau}$ in $O_{\tau}$ is the sum of three local operators from $O_{\tau-1}$ (see (a),(b) and (c) in Fig. \ref{fig:AscendSuper}), which are lifted to $\mathcal{L}_{\tau}$ by the three different forms of the ascending superoperator, $\mathcal{A}_L$, $\mathcal{A}_C$ and $\mathcal{A}_R$. Since $O_{\tau-1}$ has $N/3^{\tau-1}$ local operators, $O_{\tau}$ is obtained from $O_{\tau-1}$ by using the ascending superoperator $\mathcal{A}$ only $N/3^{\tau-1}$ times. Then, since $\sum_{\tau=0}^{T} 3^{-\tau} < 2$, this means that the entire sequence of Eq. \ref{eq:Os} requires using $\mathcal{A}$ only O$(N)$ times. Once $O_T$ is obtained, the expected value of $O$ follows from
\begin{equation}
	\tr(OP) = \tr(O_T).
\end{equation}

\textbf{Top-bottom.---} Here we consider instead the sequence of ensembles of density matrices
\begin{equation}
	E_{T-1} ~ \stackrel{U_{T-1}}{\rightarrow} ~ \cdots ~ E_2 ~ \stackrel{U_2} {\rightarrow} ~E_1 ~ \stackrel{U_1} {\rightarrow} ~ E_0,
	\label{eq:sequenceE}
\end{equation}
where $E_{\tau}$ is an ensemble of the $N/3^{\tau}$ two-site density matrices $\rho_{\tau}^{[r,r+1]}$ supported on nearest neighbor sites of $\mathcal{L}_{\tau}$,
\begin{equation}
	E_{\tau} \equiv \left\{ \rho_{\tau}^{[1,2]},  \rho_{\tau}^{[2,3]}, \cdots, \rho_{\tau}^{[N/3^{\tau},1]} \right\}
\end{equation}
From each density matrix in the ensemble $E_{\tau}$ we can generate three density matrices in the ensemble $E_{\tau-1}$ by applying the three different forms of the descending superoperator,  $\mathcal{D}_L$, $\mathcal{D}_C$ and $\mathcal{D}_R$ (see (a),(b) and (c) in Fig. \ref{fig:DescendSuper}). All the $N/3^{\tau-1}$ density matrices of ensemble $E_{\tau-1}$ can be obtained from density matrices of $E_{\tau}$ in this way. Since $\sum_{\tau=0}^{T} 3^{-\tau} < 2$, we see that by using the descending superoperator $\mathcal{D}$ only O$(N)$ times, we are able to compute all the density matrices in the sequence of ensembles of Eq. \ref{eq:sequenceE}. Once the ensemble $E_0\equiv E$ has been obtained, 
\begin{equation}
	E = \left\{ \rho^{[1,2]}, \rho^{[2,3]}, \cdots, \rho^{[N,1]} \right\},
\end{equation}
the expected value of $O$ follows from
\begin{equation}
	\tr(OP) = \sum_r \tr(o^{[r,r+1]}\rho^{[r,r+1]}).
\end{equation}

\textbf{Middle ground.---} More generally, we could build operator $O_\tau$ as well as ensemble $E_{\tau}$ and evaluate the expected value of $O$ from the equality
\begin{equation}
	\tr(OP) = \sum_{r} \tr(o_{\tau}^{[r,r+1]}\rho_{\tau}^{[r,r+1]}).
\end{equation}

Each of the strategies above require the use of the ascending/descending superoperators O$(N)$ times and therefore can indeed be accomplished with cost O$(\chi^8 N)$.

\begin{figure}[!tb]
\begin{center}
\includegraphics[width=8cm]{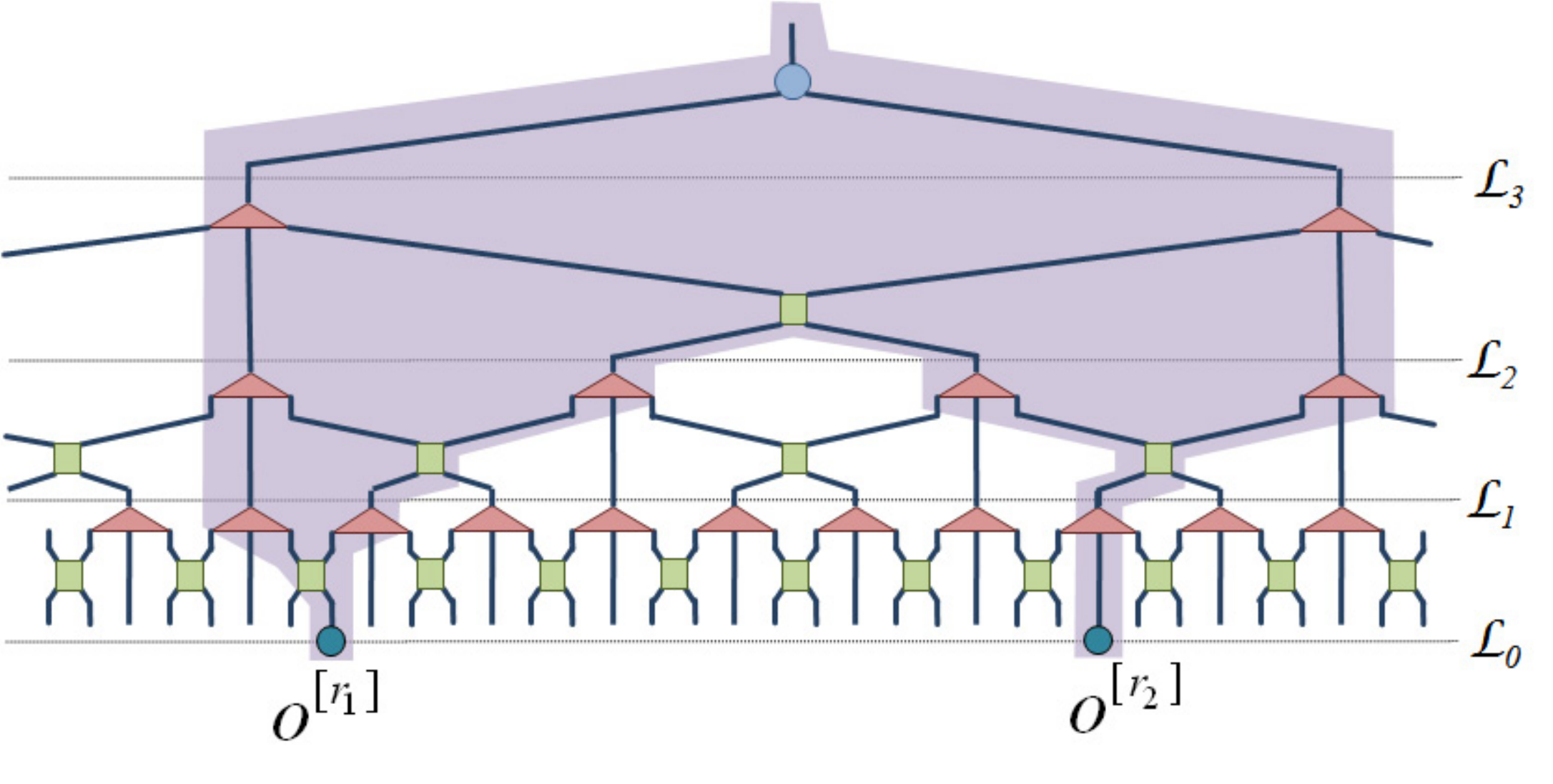}
\caption{(Colour online) In order to compute a two-point correlator $C_2(r_1,r_2)$ we need to consider the union of the past causal cones of sites $r_1$ and $r_2$. Notice that, in contrast with the case of a single local operator, the joint causal cone of two distant sites typically involves more than two contiguous sites of some lattice $\mathcal{L}_{\tau}$. This makes the computational cost scale as a power of $\chi$ larger than $\chi^8$.}
\label{fig:CorrDifficult}
\end{center}
\end{figure}

\begin{figure}[!tb]
\begin{center}
\includegraphics[width=7cm]{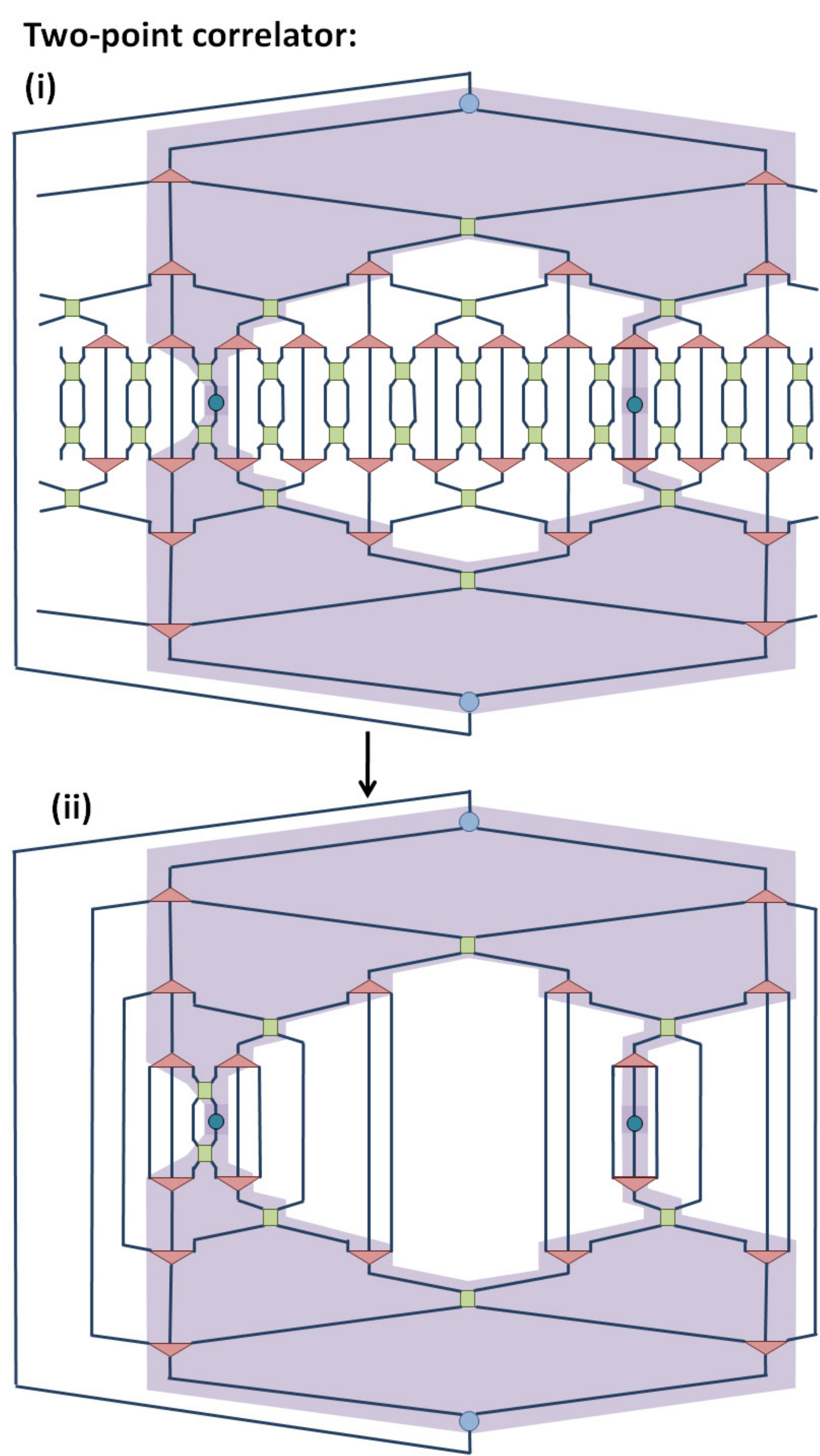}
\caption{(Colour online) (i) Tensor network to be contracted in order to evaluate a two-point correlator $C_2(r_1,r_2)$.
Similarly to the case of a local observable Fig. \ref{fig:3MERA}, the tensors outside of the casual cone annihilate in pairs due to their isometric character, Eq. \ref{eq:isometry}. The resulting tensor network (ii) is much simpler network. However, for a generic pair of sites $r_1,r_2\in\mathcal{L}$, the joint past causal cone will contain more than just two sites per layer, resulting in a computational cost that scales with $\chi$ as a power larger than $\chi^8$.} 
\label{fig:CorrDifficultFull}
\end{center}
\end{figure}

\subsection{Evaluation of two-point correlators}
 
Let us now consider the computation of a two-point correlator of the form
\begin{equation}
C_2(r_1,r_2) \equiv	\bra{\Psi}o^{[r_1]}\otimes o^{[r_2]}\ket{\Psi},
\label{eq:two-point}
\end{equation}
where $o^{[r]}$ and $o^{[s]}$ denote two one-site operators applied on two arbitrary sites $r$ and $s$ of $\mathcal{L}$, see Fig. \ref{fig:CorrDifficult}. Fig. \ref{fig:CorrDifficultFull} shows the tensor network to be contracted. Again, we can use Eq. \ref{eq:isometry} to remove all disentanglers and isometries that lay outside the joint past causal cone for sites $r$ and $s$. Then, we can proceed to contract the resulting tensor network, for instance through a bottom-top or top-bottom approach, with the help of the ascending and descending superoperators (and generalizations thereof). Notice that since at intermediate layers the two legs of the causal cone may contain two sites each one, in general we will need to compute operators/density matrices that span more than just two sites, and the cost of their computation will be larger than O$(\chi^8)$. 

\begin{figure}[!tb]
\begin{center}
\includegraphics[width=8cm]{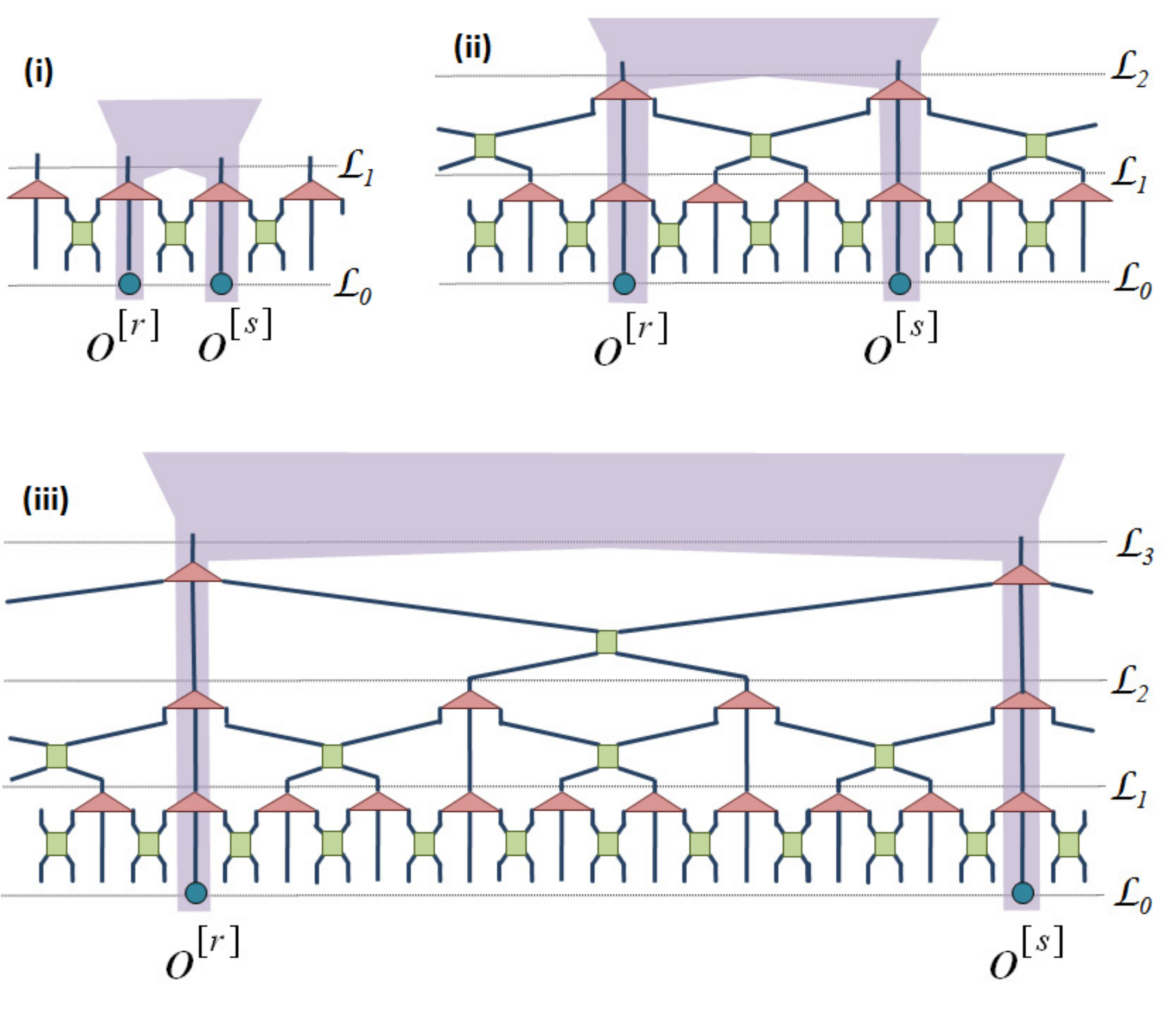}
\caption{(Colour online) Two-point correlators for specific pairs of sites $r,s$ [at distances of $3^q$ sites for $q=1,2,3...$] can be computed with cost $O(\chi^8\log N)$. This is due to the fact that the causal cones for each of $r,s$ contains only one site until they meet--- (i) at $\mathcal L_1$, (ii) at $\mathcal L_2$ or (iii) at $\mathcal L_3$.}
\label{fig:CorrEasy}
\end{center}
\end{figure}

\begin{figure}[!tb]
\begin{center}
\includegraphics[width=8cm]{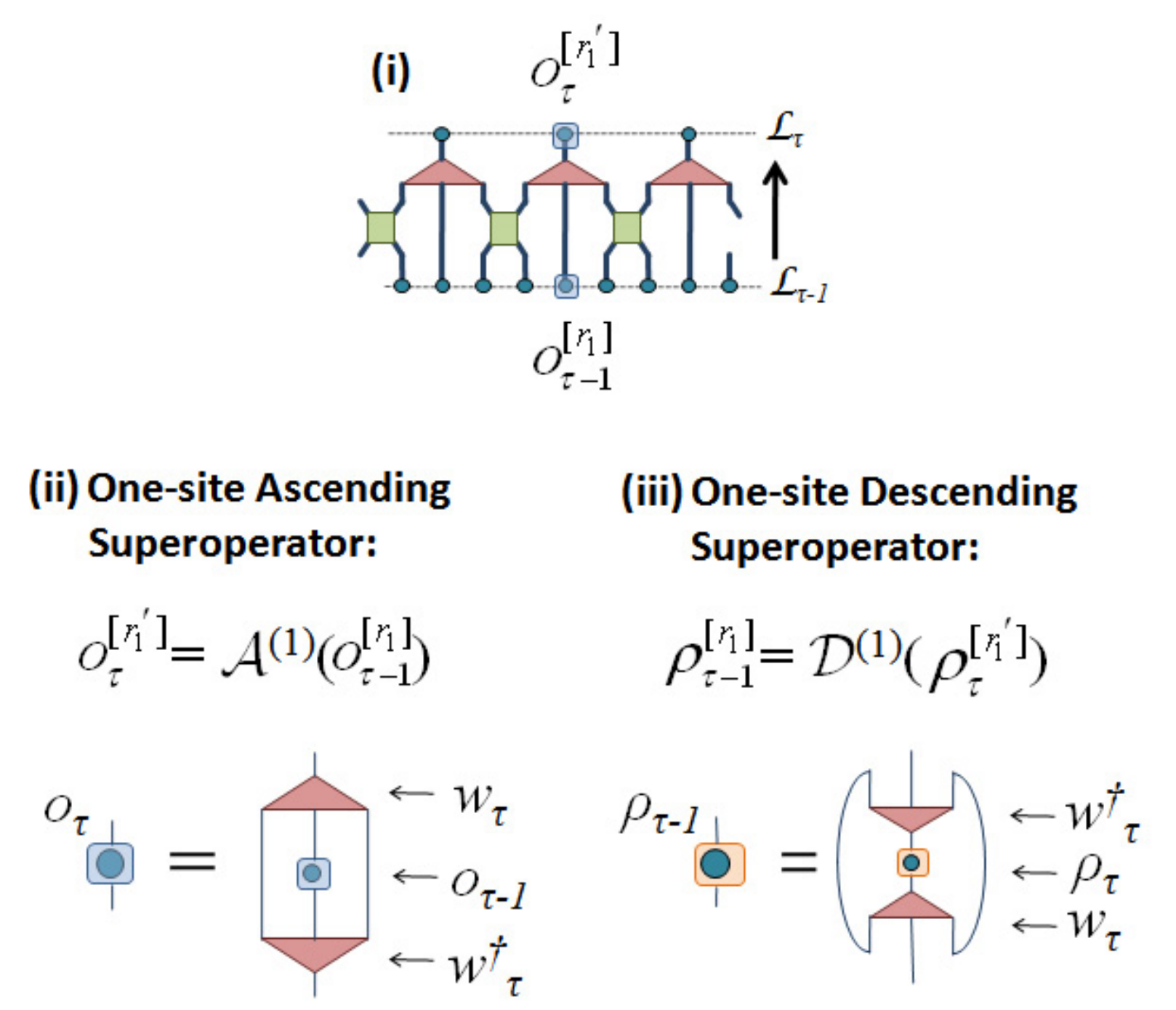}
\caption{(Colour online) A one-site operator $o_{\tau-1}$ supported on certain sites of $\mathcal{L}_{\tau-1}$ (corresponding to the central wire of an isometry $w_{\tau}$) is mapped onto a single-site operator on $\mathcal{L}_{\tau}$. In this case the (i) ascending and (ii) descending superoperators $\mathcal{A}^{(1)}$ and $\mathcal{D}^{(1)}$ have a very simple form.} 
\label{fig:OneSiteSuper}
\end{center}
\end{figure}

However, for specific choices of sites $r,s\in \mathcal{L}$, we are still able to compute $C_2(r,s)$ with overall cost $O(\chi^8\log N)$, as illustrated in Fig. \ref{fig:CorrEasy}. We emphasize that this was not possible in the binary 1D MERA and is one of the main reasons to work with the ternary 1D MERA. For such choices of sites $r$ and $s$, each of the two legs of the joint past causal cone contains just one site until, at some layer $\tau_0$, they fuse into a single two-site leg. We can introduce one-site ascending and descending superoperators $\mathcal{A}^{(1)}$ and $\mathcal{D}^{(1)}$ (Fig. \ref{fig:OneSiteSuper}), in terms of which we can express, for $\tau \leq \tau_0$, the transformation of a product operator $o_{\tau-1}^{[r]}\otimes o^{[s]}_{\tau-1}$ into a product operator
\begin{equation}
	o_{\tau}^{[r']}\otimes o^{[s']}_{\tau} = \mathcal{A}^{(1)}(o_{\tau-1}^{[r]})\otimes \mathcal{A}^{(1)}(o^{[s]}_{\tau-1}),
\end{equation}
or of a density matrix $\rho_{\tau}^{[r',s']}$ into a density matrix
\begin{equation}
	\rho_{\tau-1}^{[r,s]} = (\mathcal{D}^{(1)} \otimes \mathcal{D}^{(1)})(\rho_{\tau}^{[r',s']}),
\end{equation}
where $r,s \in \mathcal{L}_{\tau-1}$ and $r',s' \in \mathcal{L}_{\tau}$ are sites corresponding to single-site legs of the causal cone. In, say, the bottom-top approach we can compute the correlator of Eq. \ref{eq:two-point} by using the single-site ascending superoperator $\mathcal{A}^{(1)}$ for layers $\tau\leq\tau_0$ and then the two-site ascending super-operator $\mathcal{A}$ for layer $\tau > \tau_0$.

\subsection{Translation invariance} 

The computation of the expected value $\tr(o^{[r,r+1]}P)$ of a single local operator $o^{[r,r+1]}$ in the case of a translation invariant MERA can proceed as explained earlier in this section. In the present case one would expect the result to be independent of the sites $[r,r+1]\in\mathcal{L}$ on which the operator is supported, but a finite bond dimension $\chi$ typically introduces small space inhomogeneities in the reduced density matrix $\rho^{[r,r+1]}$ and therefore also in $\tr(o^{[r,r+1]}P) = \tr(o^{[r,r+1]}\rho^{[r,r+1]})$. 

Given a two-site operator $o$, an expected value that is independent of $[r,r+1]$ can be obtained by computing an average over sites,
\begin{eqnarray}
	\tr(o^{[r,r+1]}P) ~~\rightarrow && \frac{1}{N} \sum_r \tr(o^{[r,r+1]}P) \label{eq:av_ev1}\\ 
	&=& \frac{1}{N} \sum_r \tr(o^{[r,r+1]}\rho^{[r,r+1]}),
	\label{eq:av_ev2}
\end{eqnarray}
where the terms $o^{[r,r+1]}$ denote translations of the same operator $o$. This average can be computed e.g. by obtaining the $N$ density matrices $\rho^{[r,r+1]}$ individually and then adding them together, with an overall cost O$(\chi^8 N)$. However, with a better organization of the calculation the cost can be reduced to O$(\chi^8 \log N)$.

We first need to introduce average versions of the ascending and descending superoperators. Given a two-site operator $o_{\tau-1}$ in lattice $\mathcal{L}_{\tau-1}$, we can build a two-site operator $o_{\tau}$ by using an average of the three two-site operators resulting from lifting $o_{\tau-1}$ to lattice $\mathcal{L}_{\tau}$, namely $\mathcal{A}_{L}(o_{\tau-1})$, $\mathcal{A}_{L}(o_{\tau-1})$ and $\mathcal{A}_{L}(o_{\tau-1})$. In terms of the \emph{average ascending superoperator} $\bar{\mathcal{A}}$,
\begin{equation}
	\bar{\mathcal{A}} \equiv \frac{1}{3}(\mathcal{A}_L + \mathcal{A}_C + \mathcal{A}_R),
	\label{eq:avAsc}
\end{equation}
this transformation reads
\begin{equation}
	o_{\tau} = \bar{\mathcal{A}}(o_{\tau-1}).	
	\label{eq:avAsc2}
\end{equation}
Importantly, if we coarse-grain the translation invariant operator 
\begin{equation}
 \frac{1}{N_{\tau\!-\!1}} \sum_r o_{\tau\!-\!1}^{[r,r+1]},~~~~~~~~~~N_x\equiv N/3^{x},
 \label{eq:TIo}
\end{equation}
where $N_{\tau\!-\!1}$ is the number of sites of $\mathcal{L}_{\tau\!-\!1}$ and the terms $o_{\tau\!-\!1}^{[r,r+1]}$ denote translations of $o_{\tau\!-\!1}$, the resulting operator can be written as
\begin{equation}
 \frac{1}{N_{\tau}} \sum_r o_{\tau}^{[r,r+1]},
\end{equation}
where the terms $o_{\tau-1}^{[r,r+1]}$ denote translations of $o_{\tau}$ and where $o_{\tau}$ and $o_{\tau-1}$ are related through Eq. \ref{eq:avAsc2}. In other words, the average ascending superoperator $\bar{\mathcal{A}}$ can also be used to characterize the coarse-graining, in the translation invariant case, of operators of the form of Eq. \ref{eq:ev_O1}.

Let $\bar{\rho}_{\tau}$ denote the two-site density matrix obtained by averaging over all density matrices $\rho^{[r,r+1]}_{\tau}$ on different pairs $[r,r+1]$ of two contiguous sites of $\mathcal{L}_{\tau}$, 
\begin{equation}
	\bar{\rho}_{\tau} \equiv \frac{1}{N_{\tau}} \sum_r \rho_{\tau}^{[r,r+1]},
\end{equation}
and similarly for lattice $\mathcal{L}_{\tau-1}$,
\begin{equation}
	\bar{\rho}_{\tau-1} \equiv \frac{1}{N_{\tau-1}} \sum_r \rho_{\tau-1}^{[r,r+1]}.
\end{equation}
Recall that each density matrix $\rho^{[r,r+1]}_{\tau}$ on lattice $\mathcal{L}_{\tau}$ gives rise to three density matrices in $\mathcal{L}_{\tau-1}$ according to the three versions of the descending superoperator, namely $\mathcal{D}_{L}$, $\mathcal{D}_{C}$ and $\mathcal{D}_{R}$. It follows that the density matrix $\bar{\rho}_{\tau-1}$ can be obtained from the density matrix $\bar{\rho}_{\tau}$ by using the \emph{average descending superoperator},
\begin{equation}
	\bar{\mathcal{D}}  \equiv \frac{1}{3}(\mathcal{D}_L + \mathcal{D}_C + \mathcal{D}_R),
	\label{eq:avDesc}
\end{equation}
that is
\begin{equation}
	\bar{\rho}_{\tau-1} = \bar{\mathcal{A}}(\bar{\rho}_{\tau}).	
	\label{eq:avDesc2}
\end{equation}

We can now proceed to compute the average expected value of Eqs. \ref{eq:av_ev1}-\ref{eq:av_ev2}. This can be accomplished in several alternative ways.

\textbf{Bottom-top.---} Given a two-site operator $o$, we compute a sequence of increasingly coarse-grained operators 
\begin{equation}
	o_0 ~ \stackrel{\bar{\mathcal{A}}}{\rightarrow} ~ o_1 ~ \stackrel{\bar{\mathcal{A}}}{\rightarrow} ~ o_2 ~ \stackrel{\bar{\mathcal{A}}}{\rightarrow}~ \cdots ~ o_T,~~~~~~~o_0\equiv o,
\end{equation}
where $o_{\tau}$ is obtained from $o_{\tau-1}$ by means of the average ascending superoperator $\bar{\mathcal{A}}$. Then we simply have
\begin{equation}
	\frac{1}{N} \sum_r \tr(o^{[r,r+1]}P) = \tr(o_T).
\end{equation}

\textbf{Top-bottom.---} Alternatively, we can compute the sequence of average density matrices
\begin{equation}
	\bar{\rho}_{T-1} ~ \stackrel{\bar{\mathcal{D}}}{\rightarrow} ~ \cdots ~ \bar{\rho}_2 ~ \stackrel{\bar{\mathcal{D}}} {\rightarrow} ~\bar{\rho}_1 ~ \stackrel{\bar{\mathcal{D}}} {\rightarrow} ~ \bar{\rho}_0,
\end{equation}
where $\bar{\rho}_{\tau-1}$ is obtained from $\bar{\rho}_{\tau}$ by means of the average descending superoperator $\bar{\mathcal{D}}$ and where $\bar{\rho} \equiv \bar{\rho}_{0}$ corresponds to
the average density matrix on lattice $\mathcal{L}$, 
\begin{equation}
	\bar{\rho} \equiv \frac{1}{N} \sum_r \rho^{[r,r+1]},
\end{equation}
in terms of which we can express the average expected value as
\begin{equation}
	\frac{1}{N} \sum_r \tr(o^{[r,r+1]}P) = \tr(o\bar{\rho}).
\end{equation}

\textbf{Middle ground.---} As costumary, we can also use both $\bar{\mathcal{A}}$ and $\bar{\mathcal{D}}$ to compute $o_{\tau}$ and $\bar{\rho}_{\tau}$, and evaluate the average expected value as
\begin{equation}
	\frac{1}{N} \sum_r \tr(o^{[r,r+1]}P) = \tr(o_\tau\bar{\rho}_{\tau}).
\end{equation}

In all the above strategies the average ascending and descending superoperators $\bar{\mathcal{A}}$ and $\bar{\mathcal{D}}$ are used O$(\log(N))$ times and therefore the computational cost scales as O$(\chi^8 \log N)$.

To summarize, with a translation invariant MERA we can coarse-grain a single two-site operator $o$ (with a transformation that involves averaging over all possible causal cones) or compute the average density matrix $\bar{\rho}$ by using the average ascending/descending superoperators. This leads to a sequence of operators $o_{\tau}$ and density matrices $\bar{\rho}_{\tau}$,
\begin{eqnarray}
	&&o_0  \stackrel{\bar{\mathcal{A}}}{\rightarrow}  o_1 \stackrel{\bar{\mathcal{A}}}{\rightarrow} ~\cdots~ \stackrel{\bar{\mathcal{A}}}{\rightarrow} o_T, ~~~~~~~~~ o_0 \equiv o, \label{eq:ATI}\\
	&&\bar{\rho}_0  \stackrel{\bar{\mathcal{D}}}{\leftarrow}  \bar{\rho}_1 \stackrel{\bar{\mathcal{D}}}{\leftarrow} ~\cdots~ \stackrel{\bar{\mathcal{D}}}{\leftarrow} \bar{\rho}_T, ~~~~~~~~~ \bar{\rho}_0 \equiv \bar{\rho}, \label{eq:DTI}
\end{eqnarray}
from which the expected value of $o$ is obtained as $\tr(o\bar{\rho})$, as $\tr(o_T)$ or, more generally, as $\tr(o_{\tau}\bar{\rho}_{\tau})$.

\subsection{Scale invariance}

In the case of a translation invariant MERA that is also scale invariant, the average ascending superoperators $\bar{\mathcal{A}}$ is identical on each layer $\tau$, since it is always made of the same disentangler $u$ and isometry $w$. We then refer to it as the \emph{scaling superoperator} $\mathcal{S}$ \cite{CFT}. Its dual $\mathcal{S}^{*}$ corresponds to the descending superoperator $\bar{\mathcal{D}}$. 

As derived in Ref. \cite{Transfer}, the expected value of a local observable $o$ in the thermodynamic limit can be obtained by analyzing the spectral decomposition of the scaling superoperator $\mathcal{S}$,
\begin{equation}
	\mathcal{S}(\bullet) = \sum_{\alpha} \lambda_{\alpha} \phi_{\alpha} \tr(\hat{\phi}_{\alpha} \bullet),~~~~~~\tr(\hat{\phi}_{\alpha} \phi_{\beta}) = \delta_{\alpha\beta}.
	\label{eq:spectral}
\end{equation}
We refer to the eigenoperators $\phi_{\alpha}$ of $\mathcal{S}$, 
\begin{equation}
	\mathcal{S}(\phi_{\alpha}) = \lambda_{\alpha} \phi_{\alpha},
\end{equation}
as the \emph{scaling operators}. Notice that the operators $\hat{\phi}_{\alpha}$, which are bi-orthonormal to the operators $\phi_{\alpha}$, are eigenoperators of $\mathcal{S}^{*}$,
\begin{equation}
	\mathcal{S}^{*}(\hat{\phi}_{\alpha}) = \lambda_{\alpha} \hat{\phi}_{\alpha}.
\end{equation}

We recall that the scaling operator $\mathcal{S}$ is made of isometric tensors (cf. Eq. \ref{eq:isometry}) and therefore the identity operator $\mathbb{I}$ is an eigenoperator of $\mathcal{S}$ with eigenvalue 1 (that is, $\mathcal{S}$ is unital),
\begin{equation}
	\mathcal{S}(\mathbb{I}) = \mathbb{I}.
\end{equation}
On the other hand, since the MERA is built as a quantum circuit ---and descending through the causal cone corresponds to advancing in the time of a quantum evolution--- it is obvious that the descending superoperator $\mathcal{D}$ is a quantum channel, and so are $\bar{\mathcal{D}}$ and $\mathcal{S}^{*}$ (see also \cite{Transfer}). In particular, $\mathcal{S}^{*}$ is a contractive superoperator \cite{Super}, which means that the eigenvalues $\lambda_{\alpha}$ in Eq. \ref{eq:spectral} are constrained to fulfill  $|\lambda_{\alpha}| \leq 1$. In practical simulations \cite{CFT} one finds that the identity operator $\mathbb{I}$ is the only eigenoperator of $\mathcal{S}$ with eigenvalue one, $\lambda_{\mathbb{I}} = 1$, and that $|\lambda_{\alpha}| < 1$ for $\alpha \neq \mathbb{I}$. Let $\hat{\rho}$ denote the corresponding unique fixed point of $\mathcal{S}^{*}$, 
\begin{equation}
	\mathcal{S}^{*}(\hat{\rho}) = \hat{\rho}.
	\label{eq:hat_rho}
\end{equation}
In an infinite system, the local density matrix of any lattice $\mathcal{L}_{\tau}$ (with finite $\tau$) results from applying $\mathcal{S}^{*}$ on $\rho^{T}$ an infinite number of times, and it is therefore equal to the fixed point $\hat{\rho}$ \cite{Transfer}. Consequently, Eqs. \ref{eq:ATI} and \ref{eq:DTI} are then replaced with
\begin{eqnarray}
	&&o_0  \stackrel{\mathcal{S}}{\rightarrow}  o_1 \stackrel{\mathcal{S}}{\rightarrow} o_2 \stackrel{\mathcal{S}}{\rightarrow} o_3 ~\cdots , ~~~~~~~~~ o_0 \equiv o, \label{eq:ASI}\\
	&&\hat{\rho}~  \stackrel{~\mathcal{S}^{*}}{\leftarrow}  \hat{\rho} \stackrel{~\mathcal{S}^{*}}{\leftarrow}  \hat{\rho} \stackrel{~\mathcal{S}^{*}}{\leftarrow}  \hat{\rho} ~~\cdots, ~~~~~~~~~  \label{eq:DSI}
\end{eqnarray}
where in addition, by decomposing $o$ in terms of the scaling operators $\phi_{\alpha}$,
\begin{equation}
	o = \sum_{\alpha} c_{\alpha} \phi_{\alpha}, ~~~c_{\alpha} \equiv \tr(\hat{\phi}_{\alpha} o),
\label{eq:decompose_o}
\end{equation}
we can explicitly compute $o_{\tau}$:
\begin{equation}
	o_{\tau} = \big(\underbrace{\mathcal{S}\circ \cdots \circ \mathcal{S}}_{\tau \mbox{ \scriptsize{times}}}\big)   (o) = \sum_{\alpha} c_{\alpha} (\lambda_{\alpha})^{\tau}\phi_{\alpha}.
\end{equation}
This expression shows that, unless $c_{\mathbb{I}} \neq 0$, the operator $o_{\tau}$ decreases exponentially with $\tau$ (recall that $|\lambda_{\alpha}| < 1$ for $\alpha \neq \mathbb{I}$) and its expected value must vanish. The average expected value of $o$ then reads:
\begin{equation}
	\lim_{N\rightarrow \infty} \frac{1}{N} \sum_r \tr(o^{[r,r+1]}P) = \tr(o\hat{\rho}).
\label{eq:SI_o}
\end{equation}

Two-point correlators for selected positions can also be expressed in a simple way, by considering the \emph{one-site scaling superoperator} $\mathcal{S}^{(1)}$, which is how we refer to the superoperator $\mathcal{A}^{(1)}$ of Fig. \ref{fig:OneSiteSuper} in the case of a scale invariant MERA. Its spectral decomposition, 
\begin{equation}
	\mathcal{S}^{(1)}(\bullet) = \sum_{\alpha} \mu_{\alpha} \psi_{\alpha} \tr(\hat{\psi}_{\alpha} \bullet),~~~~~~\tr(\hat{\psi}_{\alpha} \psi_{\beta}) = \delta_{\alpha\beta},
	\label{eq:spectraltwo}
\end{equation}
provides us with a new set of (one-site) scaling operators $\psi_{\alpha}$. Given two arbitrary one-site operators $o$ and $o'$, we can always decompose them in terms of these $\psi_{\alpha}$ (similarly as in Eq. \ref{eq:decompose_o}). Thus we can focus directly on a correlator of the form	$\langle \psi_{\alpha}^{[r]} \psi_{\beta}^{[s]}\rangle$. Here $r$ and $s$ are restricted to selected positions as in Fig. \ref{fig:CorrEasy}. Then we have
\begin{equation}
	\langle \psi_{\alpha}^{[r]} \psi_{\beta}^{[s]}\rangle = \frac{C_{\alpha\beta}} {|r-s|^{\Delta_{\alpha}+\Delta_{\beta}}},
\label{eq:SICorr}
\end{equation}
where $\Delta_{\alpha} \equiv \log_{3} \mu_{\alpha}$ is the \emph{scaling dimension} of the scaling operator $\psi_{\alpha}$, whereas $C_{\alpha\beta}$ is given by
\begin{equation}
	C_{\alpha\beta} \equiv \tr\left((\psi_{\alpha}\otimes \psi_{\beta})\hat{\rho}\right),
\label{eq:SICorr2}
\end{equation}
with $\hat{\rho}$ from Eq. \ref{eq:hat_rho}. 

In deriving Eq. \ref{eq:SICorr} we have used that, by construction,  $|r-s| = 3^q$ for some $q=1,2,3,\cdots$. Coarse-graining $\psi_{\alpha}^{[r]} \psi_{\beta}^{[s]}$ a number $q$ of times produces a multiplicative factor $(\mu_{\alpha}\mu_{\beta})^q$ and the residual two-site operator $\psi_{\alpha}^{[0]} \psi_{\beta}^{[1]}$, whose expected value gives $C_{\alpha\beta}$. On the other hand, by noting that $\mu^{q} = \mu^{\log_3 |r-s|} = |r-s|^{\log_3 \mu} = |r-s|^{\log_3 \Delta}$, we arrive at $(\mu_{\alpha}\mu_{\beta})^q = |r-s|^{\Delta_{\alpha}+\Delta_{\beta}}$, which explains the denominator in Eq. \ref{eq:SICorr}. 

We note that the polynomial decay of correlators in the scale invariant MERA was established in Ref. \cite{MERA}. Their connection with the eigenvalues of the scaling superoperator, Eq. \ref{eq:spectral}, was formalized in Ref. \cite{Transfer}. Its closed expression in Eq. \ref{eq:SICorr}, including the coefficients $C_{\alpha\beta}$, was derived in Ref. \cite{CFT}, where also three-point correlators were considered. [Ref. \cite{CFT} also unveiled a connection between the scale invariant MERA and conformal field theory. The coefficients $C_{\alpha\beta}$ of two-point correlators and the analogous for three-point correlators are the key to identify the operator algebra of primary fields and their towers of descendant fields.]

In conclusion, from the scale invariant MERA we can characterize the expected value of local observables and two-point correlators, as expressed in Eqs. \ref{eq:SI_o} and \ref{eq:SICorr}-\ref{eq:SICorr2}. All critical exponents of the theory can be extracted from the scaling dimensions $\Delta_{\alpha}$. 

The required manipulations include computing $\bar{\rho}$ from $\mathcal{S}$ (using sparse diagonalization techniques) and diagonalizing $\mathcal{S}^{(1)}$, all of which can be accomplished with the ternary 1D MERA with cost O$(\chi^8)$.

 

\section{Optimization of a Disentangler/Isometry}
\label{sect:optim}

In preparation for the algorithms to be described in the next section, here we explain how to optimize a single tensor of the MERA.

Let $H$ be a Hamiltonian made of nearest neighbor, two-site interactions $h^{[r,r+1]}$,
\begin{equation}
	H = \sum_{r} h^{[r,r+1]}.
	\label{eq:H}
\end{equation}
For purposes of the optimization below, we choose each term $h^{[r,r+1]}$ so that it has no positive eigenvalues, $h^{[r,r+1]} \leq 0$. [This can be achieved with the simple replacement $h^{[r,r+1]} \rightarrow h^{[r,r+1]} - \lambda_{\max} I$, where $\lambda_{\max}$ is the largest eigenvalue of $h^{[r,r+1]}$.]

\begin{figure}[!tb]
\begin{center}
\includegraphics[width=7cm]{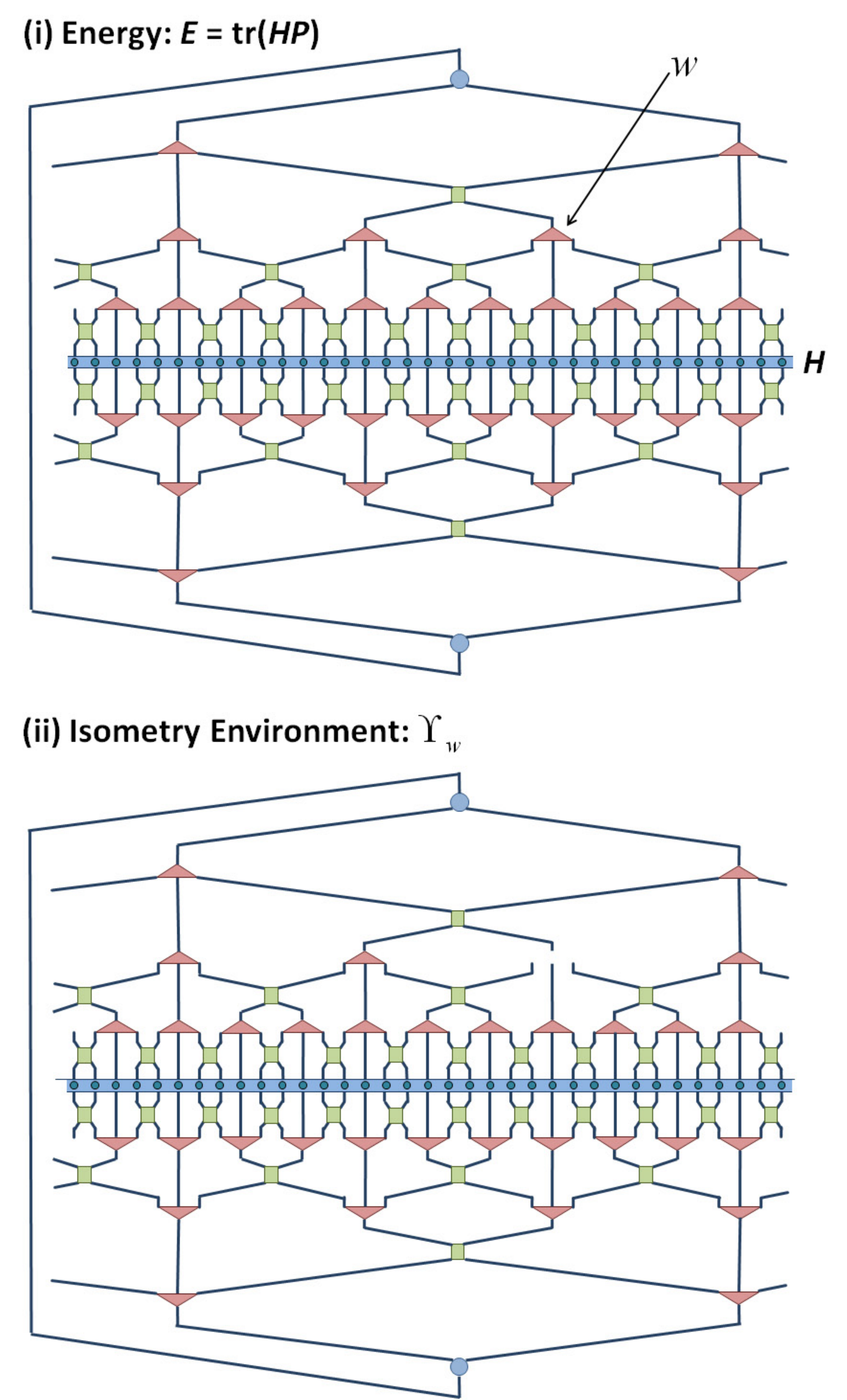}
\caption{(Colour online) (i) The energy of a MERA, defined $E \equiv \tr(HP)$, is represented explicitly in terms of a tensor network. The removal of an isometry $w$ from this network gives (ii) the environment $\Upsilon_w$ for $w$ (and similarly for disentanglers $u$). By construction we have that $E = \textrm{tr} (w \Upsilon_w) $.}
\label{fig:IsoLargeEnviro}
\end{center}
\end{figure}

Our goal for the time being will be to minimize the energy (Fig. \ref{fig:IsoLargeEnviro}.i)
\begin{equation}
	E \equiv \tr(HP),
\label{eq:E}
\end{equation}
where $P$ is a projector onto the $\chi_T$-dimensional subspace $\mathbb{V}_{U} \in \mathbb{V}_{\mathcal{L}}$, see Eq. \ref{eq:P}, by modifying only one of the tensors of the MERA. The optimization of a disentangler $u$ is very similar to that of an isometry $w$, and we can focus on describing the latter in more detail. 

Suppose then that, given a MERA, we want to optimize an isometry $w$ while keeping the rest of the tensors fixed. The cost function $E$ is quadratic in $w$ (more specifically, it depends bi-linearly on $w$ and $w^\dagger$),
\begin{equation}
	E(w) = \tr (\sum_s w M_s w^{\dagger} N_s) + c_1,
\end{equation}
where $M_s$ and $N_s$ are two sets of matrices and $c_1$ is a constant (that originates in all the Hamiltonian terms of Eq. \ref{eq:H} outside the future causal cone of $w$). Unfortunately there is no known algorithm to solve a quadratic problem subject to the additional isometric constraint of Eq. \ref{eq:isometry}. One can, however, attempt several approximate strategies (see Ref. \cite{OldAlgorithm} for some possibilities). Here we describe an iterative approach based on linearizing the cost function $E(w)$. 
 
In this approach, we temporarily regard $w$ and $w^{\dagger}$ as independent tensors, and optimize $w$ while keeping $w^{\dagger}$ fixed. The cost function reads, up to the irrelevant constant, simply
\begin{equation}
	E^{\star}(w) \equiv \tr( w \Upsilon_w ), ~~~~~~\Upsilon_w \equiv \sum_s M_s ~w^{\dagger}N_s,
\end{equation}
where we call the matrix $\Upsilon_w$ the \emph{environment} of the isometry $w$ and we treat it as if it was indepedent of $w$. $E^{\star}(w)$ is then minimized by the choice $w = -WV^{\dagger}$, where $V$ and $W$ are the unitary transformations in the singular value decomposition of the environment, $\Upsilon_w = VSW^{\dagger}$,
\begin{equation}
	\min_w E^{\star}(w) = \min_w (w VSW^{\dagger}) =  -\tr(S) = -\sum_\alpha s_\alpha
\end{equation}
(here $s_{\alpha}\geq 0$ are the singular values of $\Upsilon_w$). 

\begin{figure}[!htb]
\begin{center}
\includegraphics[width=7cm]{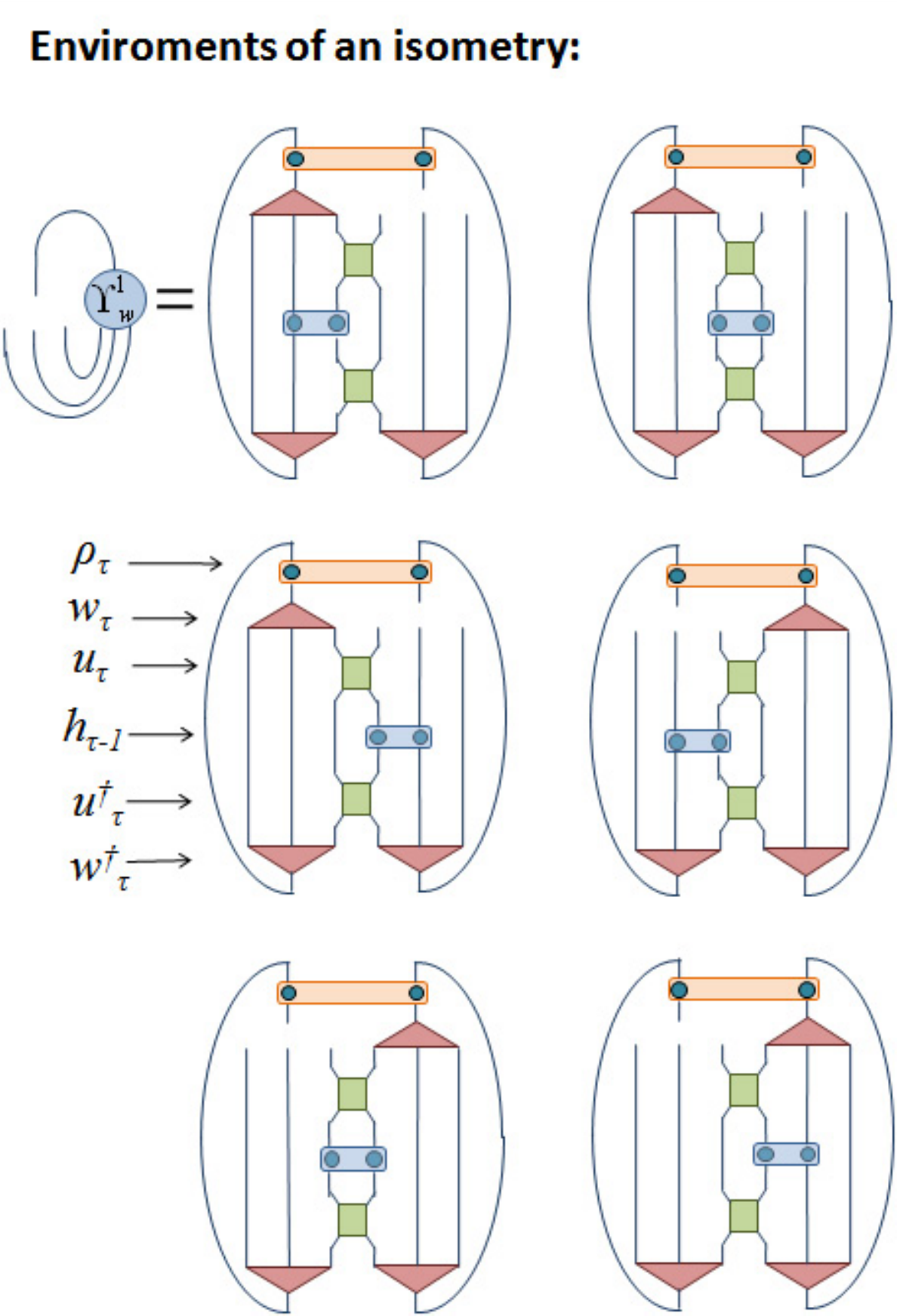}
\caption{(Colour online) Tensor network corresponding to the 6 different contributions $\Upsilon_{w}^{i}$ to the environment $\Upsilon_w= \sum\nolimits_{i=1}^6 \Upsilon_{w}^{i}$ of the isometry $w$. Notice that at each iteration of L1-L4 we need to recompute each $\Upsilon_{w}^i$ since it depends on the updated $w^{\dagger}$. Nevertheless, the Hamiltonian term and density matrix that appears in $\Upsilon_u$ remain the same throughout the optimization and only need to be computed once.}
\label{fig:IsoEnviro}
\end{center}
\end{figure}

Accordingly, given an initial isometry $w$, the optimization is performed by iterating the following four steps $q_{\mbox{\tiny{one}}}$ times:

\newcounter{Lcount} 
\begin{list}{L\arabic{Lcount}.}  {\usecounter{Lcount}}\setcounter{Lcount}{0} 
	\item Compute the environment $\Upsilon_w$ with the newest version of $w^{\dagger}$ (as explained below, see also Fig \ref{fig:IsoEnviro}).
	\item Compute the singular value decomposition $\Upsilon_w = VSW^{\dagger}$.
	\item Compute the new isometry $w' = -WV^{\dagger}$.
	\item Replace $w^{\dagger}$ with $w'^{\dagger}$.
\end{list}

The environment $\Upsilon_w$ of an isometry $w$ (at layer $\tau$) can be decomposed as the sum of 6 contributions $\Upsilon_w^i$ ($i=1,\cdots,6$), each one expressed as a tensor network that involves neighboring isometric tensors of the same layer $\tau$ (disentanglers and isometries) as well as one Hamiltonian term $h^{[r,r+1]}_{\tau-1}$ and one density matrix $\rho_{\tau}^{[r',r'+1]}$, see Fig. \ref{fig:IsoEnviro}. The two-site Hamiltonian term $h^{[r,r+1]}_{\tau-1}$ collects contributions from all the Hamiltonian terms in Eq. \ref{eq:H} included in the future causal cone of the sites $r,r+1$ of $\mathcal{L}_{\tau-1}$ and is computed with the help of the ascending superoperator $\mathcal{A}$. Similarly, the two-site density matrix $\rho_{\tau}^{[r',r'+1]}$ is computed with the help of the descending superoperator $\mathcal{D}$. The computation of $h^{[r,r+1]}_{\tau-1}$ and $\rho_{\tau}^{[r',r'+1]}$, which only needs to be performed once during the optimization of $w$, has a cost $O(\chi^8 \log N)$. 

On the other hand, once we have $h^{[r,r+1]}_{\tau-1}$ and $\rho_{\tau}^{[r',r'+1]}$, computing $\Upsilon_{\omega}$ has a cost $O(\chi^8)$ and needs to be repeated at each iteration of the steps L1-L4, with a total cost $O(\chi^8 q_{\mbox{\tiny{one}}})$. In actual MERA simulations we find that the cost function $E_{w}$ typically drops very close to the eventual minimum already after a small number of iterations $q_{\mbox{\tiny{one}}}$ of the order of 10.

\begin{figure}[!htb]
\begin{center}
\includegraphics[width=6cm]{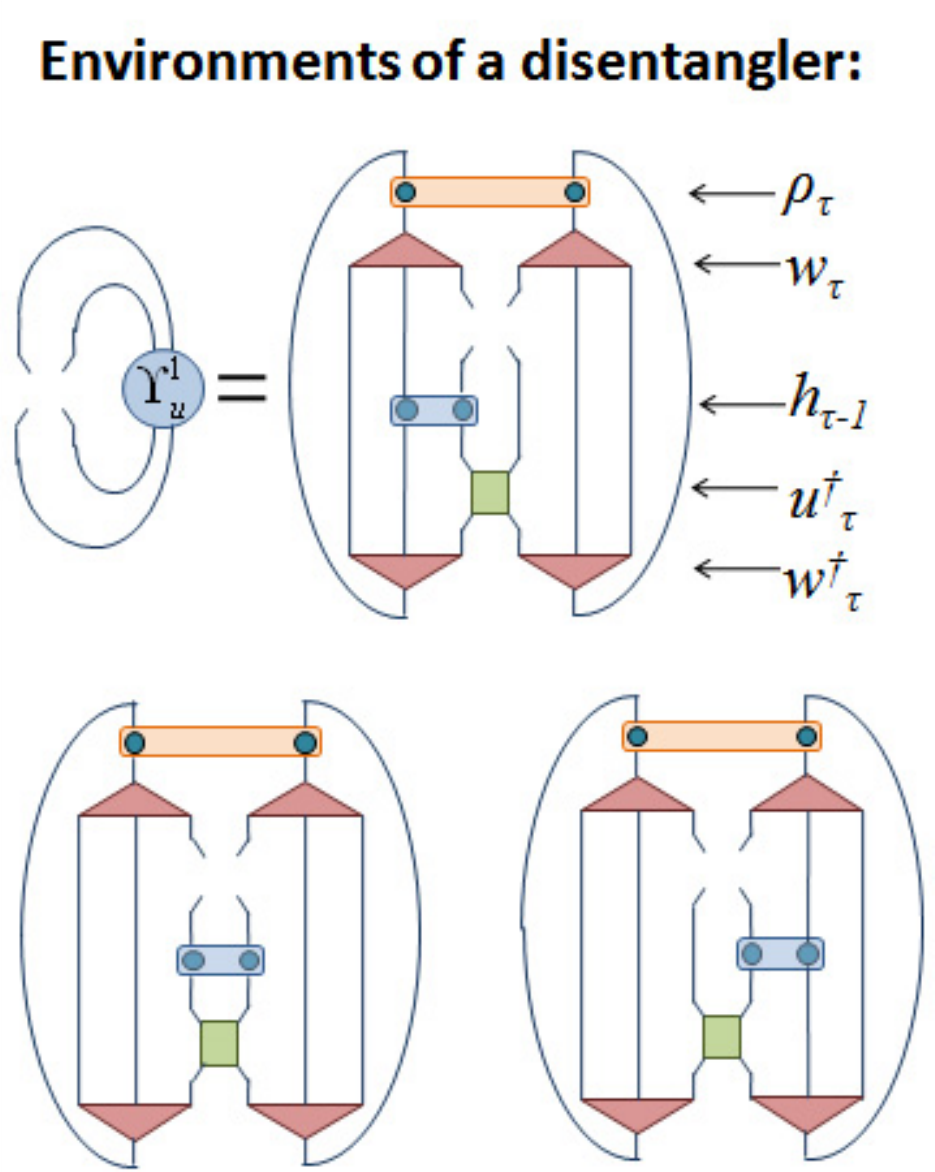}
\caption{(Colour online) Tensor networks corresponding to the 3 different contributions $\Upsilon_{u}^{i}$ to the environment $\Upsilon_u= \sum\nolimits_{i=1}^3 \Upsilon_{u}^{i}$ of the disentangler $u$. Notice that at each iteration of L1-L4 we need to recompute each $\Upsilon_{u}^i$ since it depends on the updated $u^{\dagger}$. Nevertheless, the Hamiltonian term and density matrix that appears in $\Upsilon_u$ remain the same throughout the optimization and only need to be computed once.}
\label{fig:DisEnviro}
\end{center}
\end{figure}

The optimization of a disentangler $u$ is achieved analogously, but in this case the environment $\Upsilon_u$ decomposes into three contributions $\Upsilon_u^i$ ($i=1, 2, 3$), see Fig. \ref{fig:DisEnviro}. The required Hamiltonian terms and density matrices can be computed at a cost $O(\chi^8 \log N)$, while the optimization of $u$ following steps L1-L4 has a cost $O(\chi^8 q_{one})$.

\section{Optimization of the MERA}
\label{sect:algorithm}

In this section we explain a simple algorithm to optimize the MERA so that it minimizes the energy of a local Hamiltonian of the form Eq. \ref{eq:H}. We first describe the algorithm for a generic system, and then discuss a number of specialized variations. These are directed to exploit translation invariance, scale invariance and to simulate systems where there is a finite range of correlations.

\subsection{The algorithm}

The basic idea of the algorithm is to attempt to minimize the cost function of Eq. \ref{eq:E} by sequentially optimizing individual tensors of the MERA, where each tensor is optimized as explained in the previous section.

By choosing to sweep the MERA in an organized way, we are able to update all its $O(N)$ tensors once with cost $O(\chi^8 N)$. Here we describe a bottom-top approach  where the MERA is updated layer by layer, starting with the bottom layer $\tau=1$ and progressing upwards all the way to the top layer (top-bottom and combined approaches are also possible). 

Given a starting MERA and the Hamiltonian of Eq. \ref{eq:H}, a bottom-top sweep is organized as follows:
 
\begin{list}{A\arabic{Lcount}.}  {\usecounter{Lcount}}\setcounter{Lcount}{0} 
\item Compute all two-site density matrices $\rho^{[r,r+1]}_{\tau}$ for all layers $\tau$ and sites $r \in \mathcal{L}_{\tau}$.
\end{list}
Starting from the lowest layer and for growing values of $\tau=1,2,\cdots, T-1$, repeat the following two steps:
\begin{list}{A\arabic{Lcount}.}  {\usecounter{Lcount}}\setcounter{Lcount}{1} 
	\item Update all disentanglers $u$ and isometries $w$ of layer $\tau$.
	\item Compute all two-site Hamiltonian terms $h^{[r,r+1]}_{\tau}$ for layer $\tau$.
\end{list}
Then, finally,
\begin{list}{A\arabic{Lcount}.}  {\usecounter{Lcount}}\setcounter{Lcount}{3} 
\item Update the top tensor of the MERA.
\end{list}

In step A1, we compute all nearest neighbor reduced density matrices $\rho^{[r,r+1]}_{\tau}$ for all the lattices $\mathcal{L}_{\tau}$, so that they can be used in step A2. We first compute the density matrix for the two sites of $\mathcal{L}_{T-1}$ as explained in Fig. \ref{fig:TopTensor}. Then we use the descending superoperator $\mathcal{D}$ to compute the 6 possible nearest neighbor, two-site density matrices of $\mathcal{L}_{T-2}$. More generally, given all the relevant density matrices of $\mathcal{L}_{\tau}$, we use $\mathcal{D}$ to obtain all the relevant density matrices of $\mathcal{L}_{\tau-1}$. In this way, the number of operations is proportional to the number of computed density matrices, namely $O(N)$, and the total cost is $O(\chi^8 N)$.

Step A2 breaks into a sequence of single-tensor optimizations that sweeps a given layer $\tau$ of the MERA. Each individual optimization in that layer is performed as explained in the previous section. Note that in order to optimize, say, an isometry $w$, we build its environment $\Upsilon_w$ by using (i) the density matrices computed in step A1; (ii) the Hamiltonian terms that were either given at the start for $\tau=1$ or have been computed in A3 for $\tau>1$; (iii) the neighboring disentanglers and isometries within the layer $\tau$. We can proceed, for instance, by updating all disentanglers of the layer from left to right, and then update all the isometries. This can be repeated a number $q_{\mbox{\tiny{lay}}}$ of times until the cost function does not change significantly.

In step A3, the new disentanglers and isometries of layer $\tau$ are used to build the ascending superoperator $\mathcal{A}$, which we then apply to the Hamiltonian terms of layer $\tau-1$ to compute the Hamiltonian terms for layer $\tau$. As explained after Eq. \ref{eq:O2}, each Hamiltonian term in layer $\tau$ is built from three contributions from layer $\tau-1$.

In step A4, the optimized top tensor corresponds to the $\chi_T$ eigenvectors with smaller energy eigenvalues of the Hamiltonian of the two-site lattice $\mathcal{L}_{T-1}$, obtained by exact diagonalization.

The overall optimization of the MERA consists of iterating steps A1-A4 until some pre-established degree of convergence in the energy $E$ is achieved. Suppose this occurs after $q_{\mbox{\tiny{iter}}}$ iterations. Then the cost of the optimization scales as $O(\chi^8 N q_{\mbox{\tiny{one}}}q_{\mbox{\tiny{lay}}}q_{\mbox{\tiny{iter}}})$. We observe that it is often convenient to keep $q_{\mbox{\tiny{one}}}$ and $q_{\mbox{\tiny{lay}}}$ relatively small (say between 1 and 5), since it is not worth spending much effort optimizing a single tensor/layer that will have to be optimized again later on with a modified cost function.

\subsection{Translation invariant systems}
\label{sect:trans}

When the Hamiltonian $H$ is invariant under translations, we can use a translation invariant MERA \cite{TI_MERA}.

In this case, each layer $\tau$ is characterized by a disentangler $u_\tau$ and an isometry $w_{\tau}$. In addition, on each lattice $\mathcal{L}_{\tau}$ we have one two-site hamiltonian $h_\tau$ and one average density matrix $\bar{\rho}_{\tau}$. Then a bottom-top sweep of the MERA breaks into the steps A1-A4 for the inhomogeneous case above, but with the following simplifications:

In step A1, we compute $\bar{\rho}_{\tau-1}$ from $\bar{\rho}_{\tau}$ using the average descending superoperator $\bar{\mathcal{D}}$ of Eq. \ref{eq:avDesc},
\begin{equation}
	\bar{\rho}_{\tau} \rightarrow \bar{\rho}_{\tau-1} = \bar{\mathcal{D}}(\bar{\rho}_{\tau}).
\end{equation}
Then the whole sequence $\{\bar{\rho}_{T-1}, \cdots, \bar{\rho}_1, \bar{\rho}_0 \}$, with $T\approx \log N$, is computed with cost $O(\chi^8 \log N)$.

In step A2, the minimization of the energy $E$ by optimizing, say, the isometry $w_{\tau}$ is no longer a quadratic problem (since a larger power of $w_{\tau}$ appears now in the cost function). Nevertheless, we still linearize the cost function $E$ and optimize $w_{\tau}$ according to the steps L1-L4 of the previous section. Namely, we build the environment $\Upsilon_{w}$ (which now contains copies of $w_{\tau}$ and $w_{\tau}^{\dagger}$, all of them treated as frozen), compute its singular value decomposition to build the optimal $w'_{\tau}$, and then replace $w_{\tau}$ and $w_{\tau}^{\dagger}$ with $w'_{\tau}$ and $w'^{\dagger}_{\tau}$ in the tensor network for $\Upsilon_{w}$ before starting the next iteration. 

In step A3, the new hamiltonian term $h_{\tau}$ is obtained from $h_{\tau-1}$ using the average ascending superoperator $\bar{\mathcal{A}}$,
\begin{equation}
	h_{\tau-1} \rightarrow h_{\tau} = \bar{\mathcal{A}}(h_{\tau-1}).
\end{equation}

Step A4 proceeds as in the inhomogeneous case.

The overall cost of optimizing the MERA is in this case  $O(\chi^8 \log (N) q_{\mbox{\tiny{one}}}q_{\mbox{\tiny{lay}}}q_{\mbox{\tiny{iter}}})$.


\subsection{Scale invariant systems}
\label{sect:critical}

Given the Hamiltonian $H$ for an infinite lattice at a quantum critical point, where we expect the system to be invariant under rescaling, we can use a scale invariant MERA to represent its ground state \cite{ER,MERA,FreeFermions,FreeBosons,Transfer,CFT}. [A scale invariant MERA is also relevant in the context of topological order in the infra-red limit of the RG flow \cite{QuantumDouble,StringNet}, both for finite and infinite systems; we will not consider such systems here].

Let us assume that all disentanglers and isometries are copies of a unique pair $(u,w)$. Then, as explained in Ref. \cite{CFT}, the optimization algorithm can be specialized to take advantage of scale invariance as follows:

In step A1, we apply sparse diagonalization techniques to compute the fixed point density matrix $\hat{\rho}$ from the superoperator $\mathcal{S}^{*}$. This amount to applying $\mathcal{S}^{*}$ a number of times and therefore can be accomplished with cost $O(\chi^8)$. 

In step A2, the environment for e.g. the isometry $w$, $\Upsilon_w$, is computed as a weighted sum of environments for different layers $\tau=1,2,\cdots$. In a translation invariant MERA the environment for layer $\tau$ is a function $f(u_{\tau},w_{\tau},\rho_\tau,h_{\tau-1})$ of the pair $(u_{\tau},w_{\tau})$, the density matrix $\rho_{\tau}$ and the Hamiltonian term $h_{\tau-1}$ (specifically, $f$ is the sum of the diagrams in Fig. \ref{fig:IsoEnviro}). A scale invariant MERA corresponds to the replacements
\begin{equation}
	(u_{\tau},w_{\tau}) \rightarrow (u,w),~~~~~~ \rho_{\tau} \rightarrow \hat{\rho},
\end{equation}
so that only $h_{\tau-1}$ retains dependence on $\tau$. We then choose the average environment
\begin{equation}
	\Upsilon_{w} \equiv \sum_{\tau=1}^{\infty} \frac{1}{3^{\tau}} f(u,w,\hat{\rho},h_{\tau-1}),
\end{equation}
where the weight $1/3^{\tau}$ reflects the fact that for each isometry at layer $\tau$ there are $3$ isometries at layer $\tau-1$. Using linearity of the $f$ in its fourth argument we arrive at
\begin{equation}
	\Upsilon_{w} = f(u,w,\hat{\rho},\bar{h}), ~~~~~\bar{h} \equiv \sum_{\tau=1}^{\infty} \frac{1}{3^{\tau}} h_{\tau-1}.
\end{equation}
In practice, only a few terms of the expansion of $\bar{h}$ (say $\tau=1,2,3,4$) seem to be necessary. Given $\Upsilon_{w}$, the optimization proceeds as usual with a singular value decomposition.

Steps A3 and A4 are not necessary. 

That is, the algorithm to minimize the expected value of $H$ consists simply in iterating the following two steps:

\begin{list}{ScInv\arabic{Lcount}.}  {\usecounter{Lcount}}\setcounter{Lcount}{0} 
	\item Given a pair ($u,w$), compute a pair $(\hat{\rho},\bar{h})$.
	\item Given the pairs ($u,w$) and $(\hat{\rho},\bar{h})$, update the pair ($u,w$).
\end{list}

In practical simulations it is convenient to include a few (say one or two) transitional layers at the bottom of the MERA, each one characterized by a different pair ($u_{\tau}$, $w_{\tau}$). In this way the bond dimension $\chi$ of the MERA can be made independent of the dimension $d$ of the sites of $\mathcal{L}$. These transitional layers are optimized using the algorithm for translation invariant systems.


\begin{figure}[!tb]
\begin{center}
\includegraphics[width=8cm]{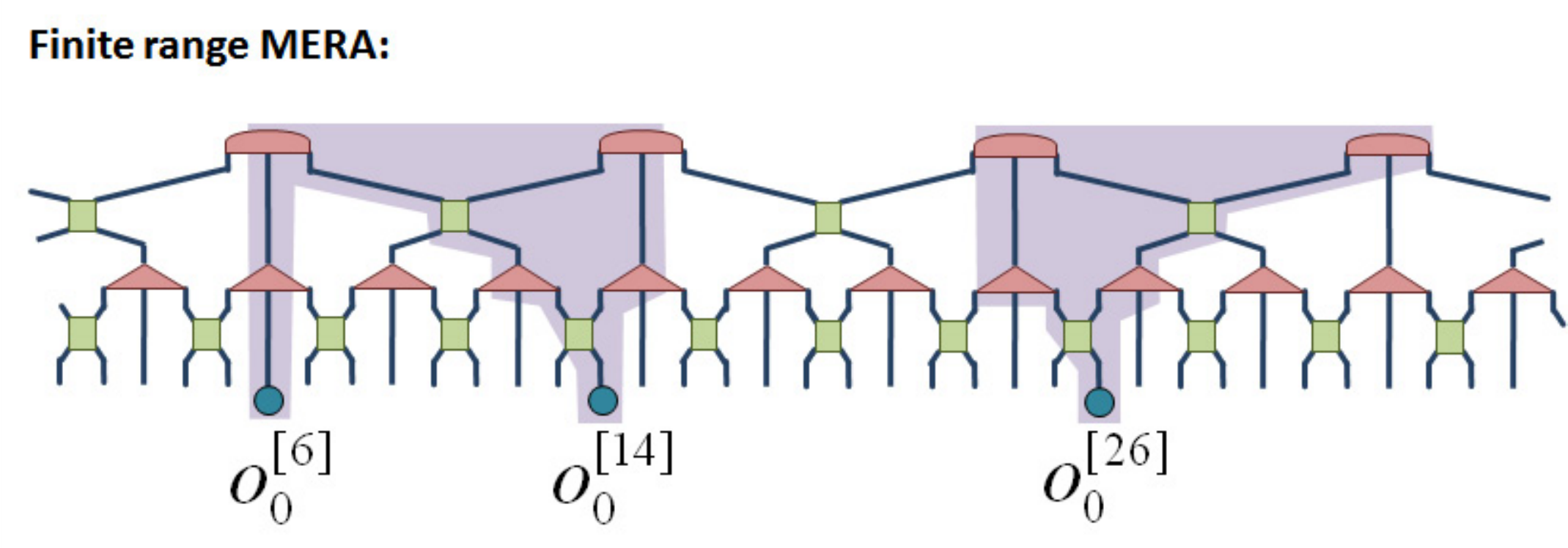}
\caption{(Colour online) A \emph{finite correlation range} MERA with $T'=2$ layers is used to represent a state of $N=36$ sites. Since it lacks the uppermost layers, only sites within a finite distance or range $\zeta \approx 3^{T'}$ of each other may be correlated. More precisely, only pairs of sites $(r_1,r_2)$ whose past casual cones intersect can be correlated, as it is the case with the pair of sites $(6,14)$ but not with $(14,26)$, for which we have $\langle {o^{[14]} o^{[26]} } \rangle  = \langle {o^{[14]} } \rangle \langle {o^{[26]} } \rangle$.}
\label{fig:FiniteCorr}
\end{center}
\end{figure}

\subsection{Finite range of correlations}
\label{sect:constant}

A third variation of the basic algorithm consists in setting the number $T'$ of layers in the MERA to a value smaller than its usual one $T \approx \log_3 N$ (in such a way that the number $N_{T'}$ of sites on the top lattice $\mathcal{L}_{T'}$ may still be quite large) and to consider a state $\ket{\Psi}$ of the lattice $\mathcal{L}$ such that after $T'$ coarse-graining transformations it has become a product state,
\begin{equation}
	\ket{\Psi_0} \rightarrow \ket{\Psi_1} \rightarrow \cdots \rightarrow \ket{\Psi_{T'}},
\end{equation}
where $\ket{\Psi_{T'}} = \ket{0}^{\otimes N_{T'}}$.
For instance, Fig. \ref{fig:FiniteCorr} shows a MERA for $N=36$ and $T'=2$. The four top tensors of this MERA are of type $(0,3)$, where the lack of upper index indicates that the top lattice $\mathcal{L}_{T'}$ is in a product state of its $N_{T'} = 4$ sites.

We refer to this ansatz as the \emph{finite range} MERA, since it is such that correlations in $\ket{\Psi}$ are restricted to a finite range $\zeta$, roughly $\zeta\approx 3^T$ sites, in the sense that regions separated by more than $\zeta$ sites display no correlations. This is due to the fact that the past causal cones of distance regions of $\mathcal{L}$ have zero intersection, see Fig. \ref{fig:FiniteCorr}.

Given a ground state $\ket{\Psi}$ with a finite correlation length $\xi$, the finite range MERA with $\zeta \approx\xi$ turns out to be a better option to represent $\ket{\Psi}$ than the standard MERA with $T \approx \log_3 N$ layers, in that it offers a more compact description and the cost of the simulations is also lower since there are less tensors to be optimized. The algorithm is adapted in a straightforward way. The only significant difference is that the top isometries, being of type $(0,3)$, do not require any density matrix in their optimization (their environment is only a function of neighboring disentanglers and isometries, and of Hamiltonian terms).

A clear advantage of the finite range MERA is in a translation invariant system, where the cost of a simulation with range $\zeta = 3^{T'}$ is O$(\chi^8 \log_3 \zeta)$, that is, independent of $N$. This allows us to take the limit of an infinite system. We find that, given a translation invariant Hamiltonian $H = \sum_{r=1}^N h^{[r,r+1]}$, where $h^{[r,r+1]}$ is the same for all $r\in\mathcal{L}$, the optimization of a finite range MERA will lead to the same collection of optimal disentanglers and isometries $\{(u_1,w_1),(u_2,w_2), \cdots, (u_{T'},w_{T'})\}$, for different lattice sizes $N, N',N''\cdots$ larger than $\zeta$. This is due to the existence of disconnected causal cones, which imply that the cost functions  for the optimization are not sensitive to the total system size provided it is larger than $\zeta$. As a result, $\{(u_1,w_1),(u_2,w_2), \cdots, (u_{T'},w_{T'})\}$ can be used to define not just one but a whole collection of states $\ket{\Psi(N)}$, $\ket{\Psi(N')}$, $\ket{\Psi(N'')}$, $\cdots$, for lattices of different sizes $N,N',N'', \cdots$, such that they all have the same two-site density matrix $\rho$ and therefore also the same expected value of the energy per link,
\begin{equation}
	\bra{\Psi(N)}h\ket{\Psi(N)} = \bra{\Psi(N')}h\ket{\Psi(N')} = \cdots
\end{equation}
In particular, we can use the finite range MERA algorithm to obtain an upper bond for the ground state energy of an infinite system, even though only $T'$ pairs $(u_{\tau},w_{\tau})$ are optimized.

\begin{figure}[!tb]
\begin{center}
\includegraphics[width=8cm]{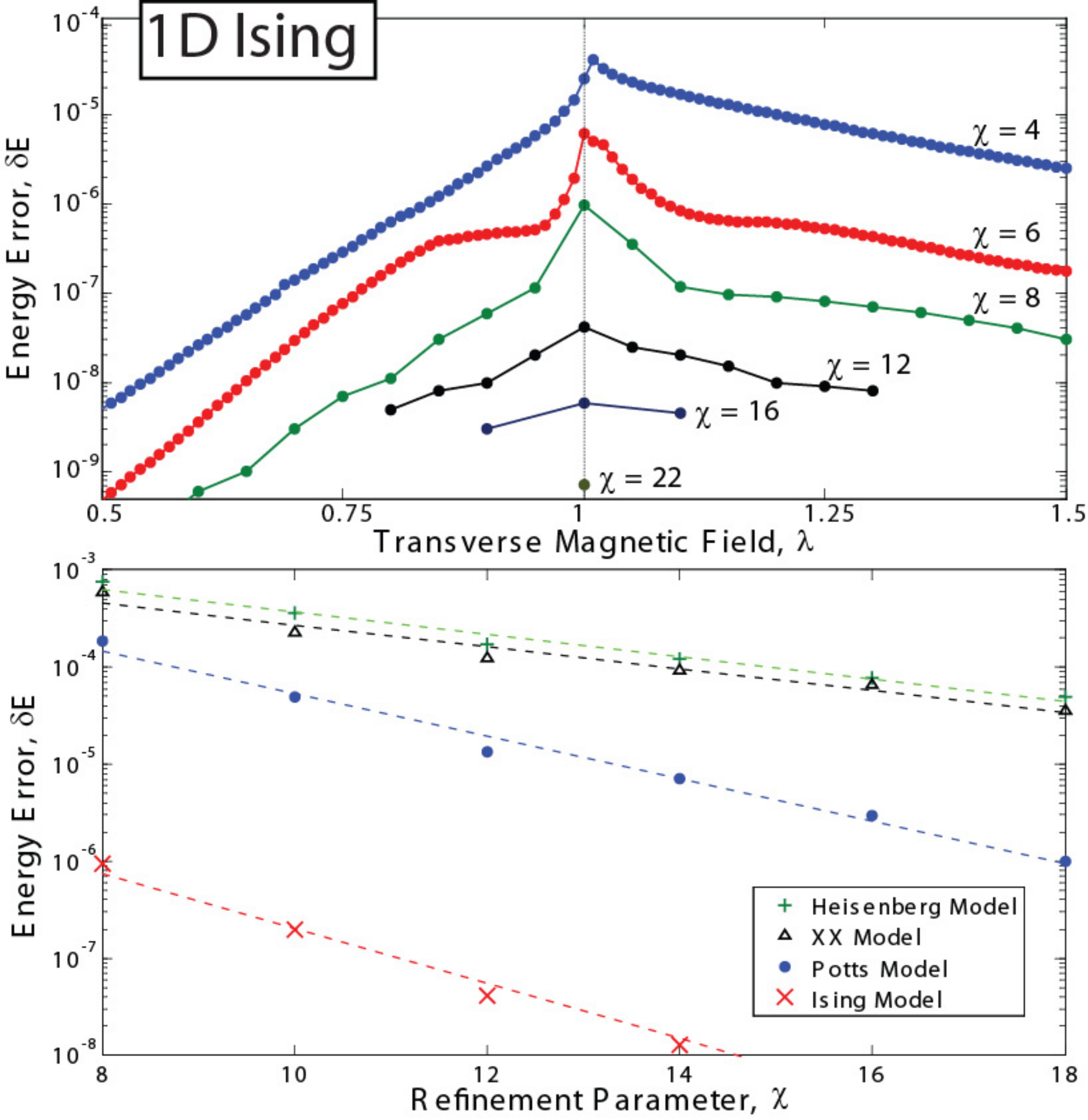}
\caption{(Colour online) \emph{Top:} The energy error of the MERA approximations to the ground-state of the infinite Ising model, as compared against exact analytic values, is plotted both for different transverse magnetic field strengths and different values of the MERA refinement parameter $\chi$. The finite correlation range algorithm (with at most $T=5$ levels) was used for non-critical ground states, whilst the scale invariant MERA was used for simulations at the critical point. It is seen that representing the ground-state is most computationally demanding at the critical point, although even at criticality the MERA approximates the ground-state to between 5 digits of accuracy ($\chi=4$) and 10 digits of accuracy ($\chi=22$). \emph{Bottom:} Scale-invariant MERA are used to compute the ground-states of infinite, critical, 1D spin chains of Eqs. \ref{eq:hamising}-\ref{eq:hams} for several values of $\chi$. In all instances one observes a roughly exponential convergence in energy over a wide range of values for $\chi$ as indicated by trend lines (dashed). Energy errors for Ising, XX and Heisenberg models are taken relative to the analytic values for ground energy whilst energy errors presented for the Potts model are taken relative to the energy of a $\chi=22$ simulation.} \label{IsingPottsEnergyError}
\end{center}
\end{figure}

\section{Benchmark Calculations for 1D systems}
\label{sect:benchmark}

In order to benchmark the algorithms of the previous section, we have analyzed zero temperature, low energy properties of a number of 1D quantum spin systems. Specifically, we have considered the Ising model \cite{Ising}, the 3-state Potts model \cite{Potts}, the XX model \cite{XX} and the Heisenberg models \cite{Heisenberg}, with Hamiltonians
\begin{eqnarray}
H_{{\rm{Ising}}}  &=& \sum_r \left(\lambda \sigma^{[r]}_z  + \sigma^{[r]}_x \sigma^{[r+1]}_x \right) \label{eq:hamising}\\
H_{{\rm{Potts}}}  &=&  \sum_r \left(\lambda M^{[r]}_{z}  + \sum_{a=1,2} M^{[r]}_{x,a} M^{[r+1]}_{x,3-a} \right)\label{hampotts}
\end{eqnarray}
\begin{eqnarray}
H_{{\rm{XX}}}   &=&  \sum_r \left(\sigma^{[r]}_x \sigma^{[r+1]}_x + \sigma^{[r]}_y \sigma^{[r+1]}_y \right) \\
H_{{\rm{Heisenberg}}}  &=&  \sum_r \left(\sigma^{[r]}_x \sigma^{[r+1]}_x + \sigma^{[r]}_y \sigma^{[r+1]}_y + \sigma^{[r]}_z \sigma^{[r+1]}_z \right)\nonumber\\ 
 \label{eq:hams}
\end{eqnarray}
where $\sigma_x$, $\sigma_y$ and $\sigma_z$ are the spin $1/2$ Pauli matrices and $M_{x,1}$, $M_{x,2}$ and $M_z$ are the matrices
\begin{eqnarray}
M_z  &\equiv& \left( {\begin{array}{*{20}c}
   2 & 0 & 0  \\
   0 & { - 1} & 0  \\
   0 & 0 & { - 1}  \\
\end{array}} \right),\\
M_{x,1}  &\equiv& \left( {\begin{array}{*{20}c}
   0 & 1 & 0  \\
   0 & 0 & 1  \\
   1 & 0 & 0  \\
\end{array}} \right),\; M_{x,2}  \equiv \left( {\begin{array}{*{20}c}
   0 & 0 & 1  \\
   1 & 0 & 0  \\
   0 & 1 & 0  \\
\end{array}} \right).
\end{eqnarray}
We assume periodic boundary conditions in all instances and use a translation invariant MERA to represent an approximation to the ground state and, in some models, also the first excited state. For Ising and Potts models the parameter $\lambda$ is the strength of the transverse magnetic field applied along the $z$-axis, with $\lambda_c=1$ corresponding to a quantum phase transition. Both the XX model and Heisenberg model are quantum critical as written.

\begin{figure}[!tb]
\begin{center}
\includegraphics[width=8cm]{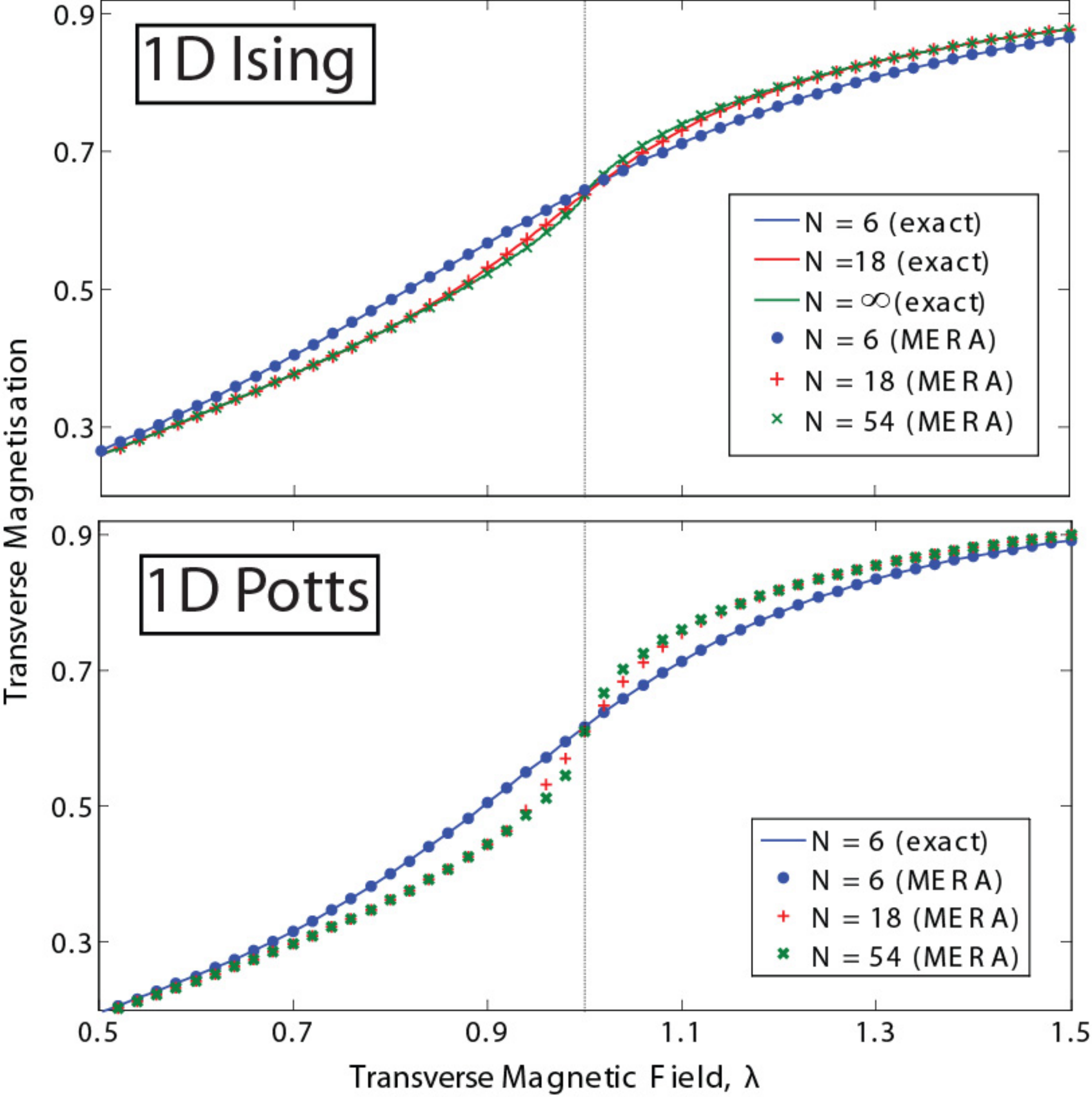}
\caption{(Colour online) The transverse magnetization $\left\langle {\sigma _z } \right\rangle$ for Ising and $\frac{1}{2}\left\langle {M _z } \right\rangle$ for Potts, is plotted for translation invariant chains of several sizes $N$. \emph{Top:} For the Ising model, the magnetization given from $\chi=8$ MERA matches those from exact diagonalization for small system sizes ($N=6,18$), whilst the magnetisation from the $N=54$ MERA approximates that from the thermodynamic limit (known analytically). \emph{Bottom:} Equivalent magnetisations for the Potts model, here computed with a $\chi=12$ MERA. Simulations with larger $N$ systems show little change from the $N=54$ data, again indicating that $N=54$ is already close to the thermodynamic limit.} \label{IsingPottsExpectZ}
\end{center}
\end{figure}

Fig.~\ref{IsingPottsEnergyError} shows the accuracy obtained for ground-state energies of the above models in the limit of an infinite chain, as a function of the refinement parameter $\chi$. Simulations were performed with either the finite correlation range algorithm (for the non-critical Ising) or the scale invariant algorithm (for critical systems). In all cases one observes roughly exponential convergence to the exact energy with increasing $\chi$. For any fixed value of $\chi$, the MERA consistently yields more accuracy for some models than for others. For the Ising model, the cheapest simulation considered ($\chi=4$) produced 5 digits of accuracy, whilst the most computationally expensive simulation ($\chi=22$) produced 10 digits of accuracy \cite{accuracyIsing}. The time taken for the MERA to converge, running on a 3GHz dual-core desktop PC with 8Gb of RAM, is approximately a few minutes/hours/days/weeks for $\chi=4,8,16,22$ respectively. We stress that these simulations were performed on single desktop computers; a parallel implementation of the code running on a computer cluster might bring significantly larger values of $\chi$ within computational reach. 

\begin{figure}
\begin{center}
\includegraphics[width=8cm]{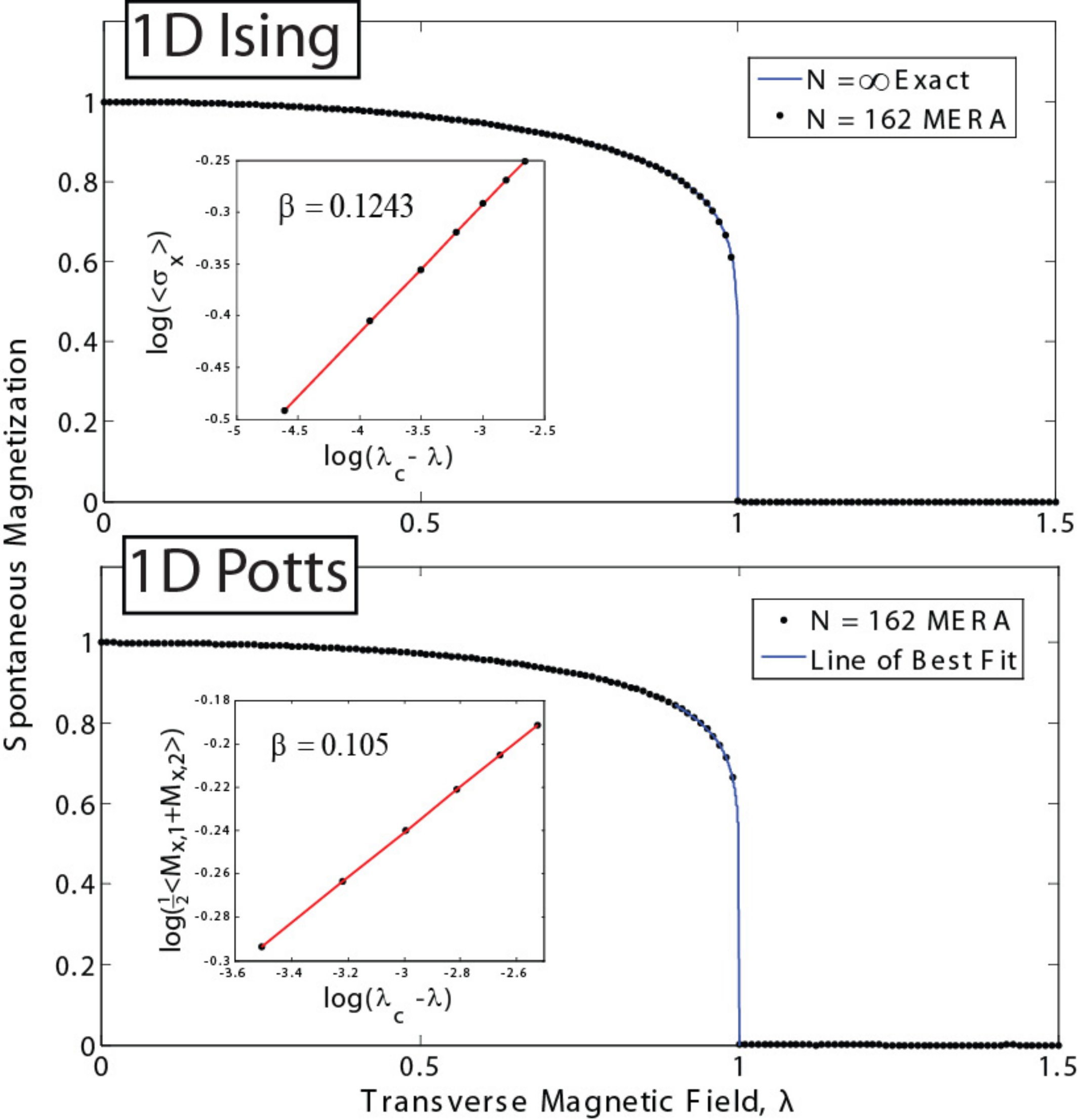}
\caption{(Colour online) \emph{Top:} Spontaneous magnetization $\left\langle {\sigma _x } \right\rangle$ computed with a $\chi=8$ MERA for a periodic Ising system of $N=162$ sites. The results closely approximate the analytic values of magnetization known for the thermodynamic limit. A fit of the data near the critical point yeilds a critical exponent $\beta_\textrm{MERA} = 0.1243$, with the exact exponent known as $\beta_\textrm{ex} = 1/8$. \emph{Bottom:} An equivalent phase portrait of the Potts model, here with spontaneous magnetization $\frac{1}{2}\left\langle {M_{x,1}  + M_{x,2} } \right\rangle$, is computed with a $\chi=14$ MERA and is plotted with a fit of the data near the critical point. The fit yields a critical exponent $\beta_\textrm{MERA} = 0.105$ with the exact exponent known to be $\beta_\textrm{ex} = 1/9$.} \label{IsingPottsSpont}
\end{center}
\end{figure}

Fig. ~\ref{IsingPottsExpectZ} demonstrates the ability of the MERA to reproduce finite size effects. It shows the transverse magnetization as a function of the transverse magnetic field for several system sizes. The results smoothly interpolate between those for small system sizes and those for an infinite chain, and match the available exact solutions. On the other hand, the MERA can also be used to explore the phase diagram of a system. Fig.~\ref{IsingPottsSpont} shows the spontaneous magnetization, which is the system's order parameter, for a $1D$ chain of $N=162$ sites for both Ising and Potts models, where $N$ has been chosen large enough that the results under consideration do not change singnificantly with the system size (thermodynamic limit). A fit for the critical exponent of the Ising model gives $\beta_\textrm{MERA}=0.1243$ whilst the fit for the Potts model produces $\beta_\textrm{MERA}=0.105$. These values are within less than $1\%$ and $6\%$ of the exact exponents $\beta=1/8$ and $\beta=1/9$ for the Ising and Potts models respectively. Obtaining an accurate value for this critical exponent through a fit of the data near the critical point is difficult due to the steepness of the curve near the critical point. Through an alternative method involving the scaling super-operator $\mathcal{S}$ (Sect. \ref{sect:local}), more accurate critical exponents have been obtained in Ref. \cite{CFT}. 

\begin{figure}[!tb]
\begin{center}
\includegraphics[width=8cm]{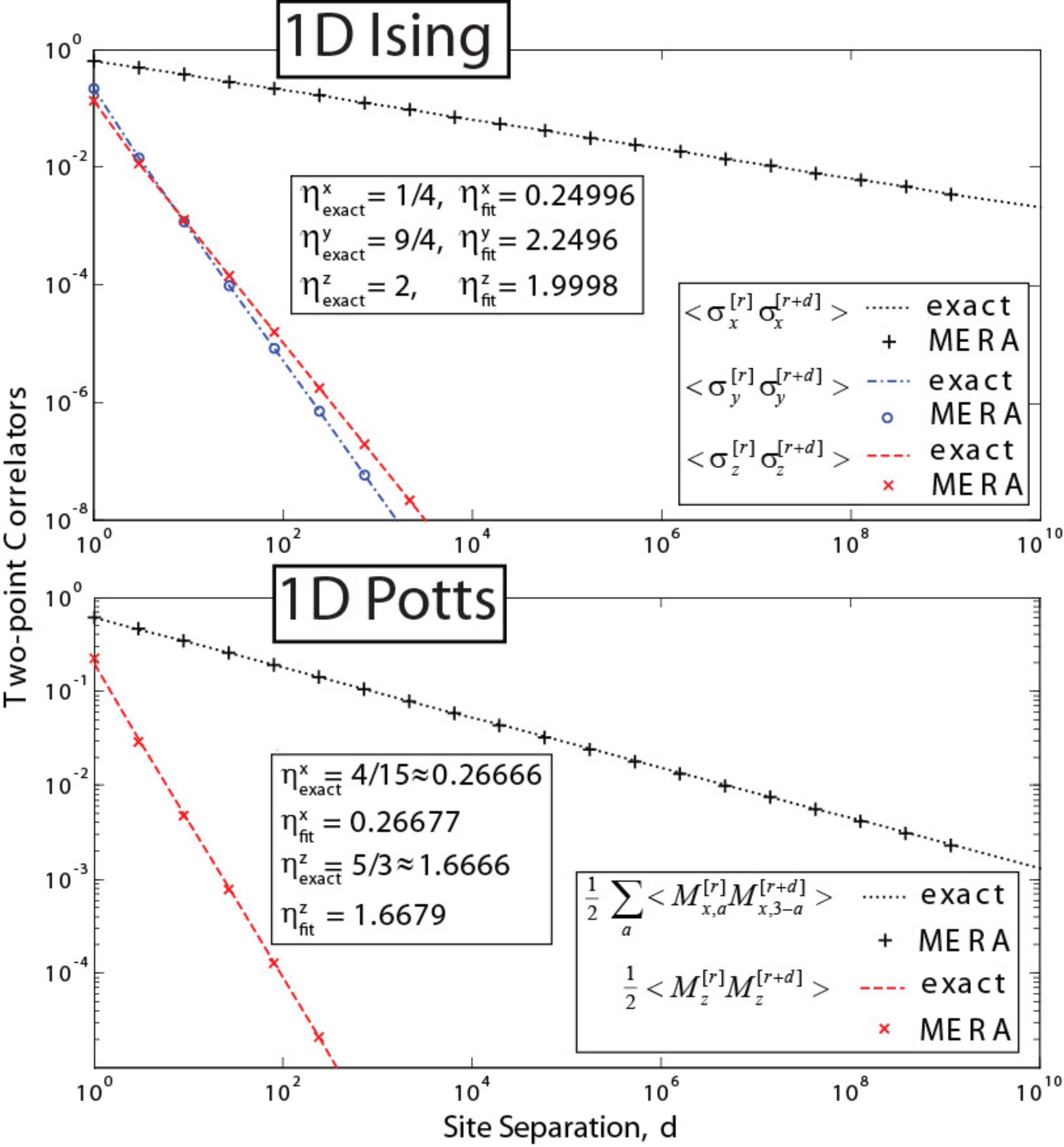}
\caption{(Colour online) Two-point correlators for infinite 1D Ising and Potts chains at criticality ($\lambda=1$), as computed with $\chi=22$ scale-invariant MERA. Correlators for the Ising model are compared against analytic solutions \cite{Ising} whilst those for the Potts model are plotted against the polynomial decay predicted from CFT \cite{CFTbook}. A scale-invariant MERA produces polynomial decay of correlators at all length scales; a fit of the form $\langle{\sigma _x^{[r]} \sigma _x^{[r + d]} } \rangle  \propto d^{ - \eta ^x }$ for Ising correlators generated by the MERA gives the decay exponent $\eta ^x=0.24996$, close to the known analytic value $1/4$ and similarly for the fits on other correlators. Indeed the MERA here reproduces exact $\langle{\sigma _x^{[r]} \sigma _x^{[r + d]} } \rangle$ correlators for the Ising model at a distance up to $d=10^9$ sites within $0.6\%$ accuracy. Critcal exponents for the Potts are also reproduced very accurately.  
  } \label{IsingPottsCorr}
\end{center}
\end{figure}

\begin{figure}[!tb]
\begin{center}
\includegraphics[width=8cm]{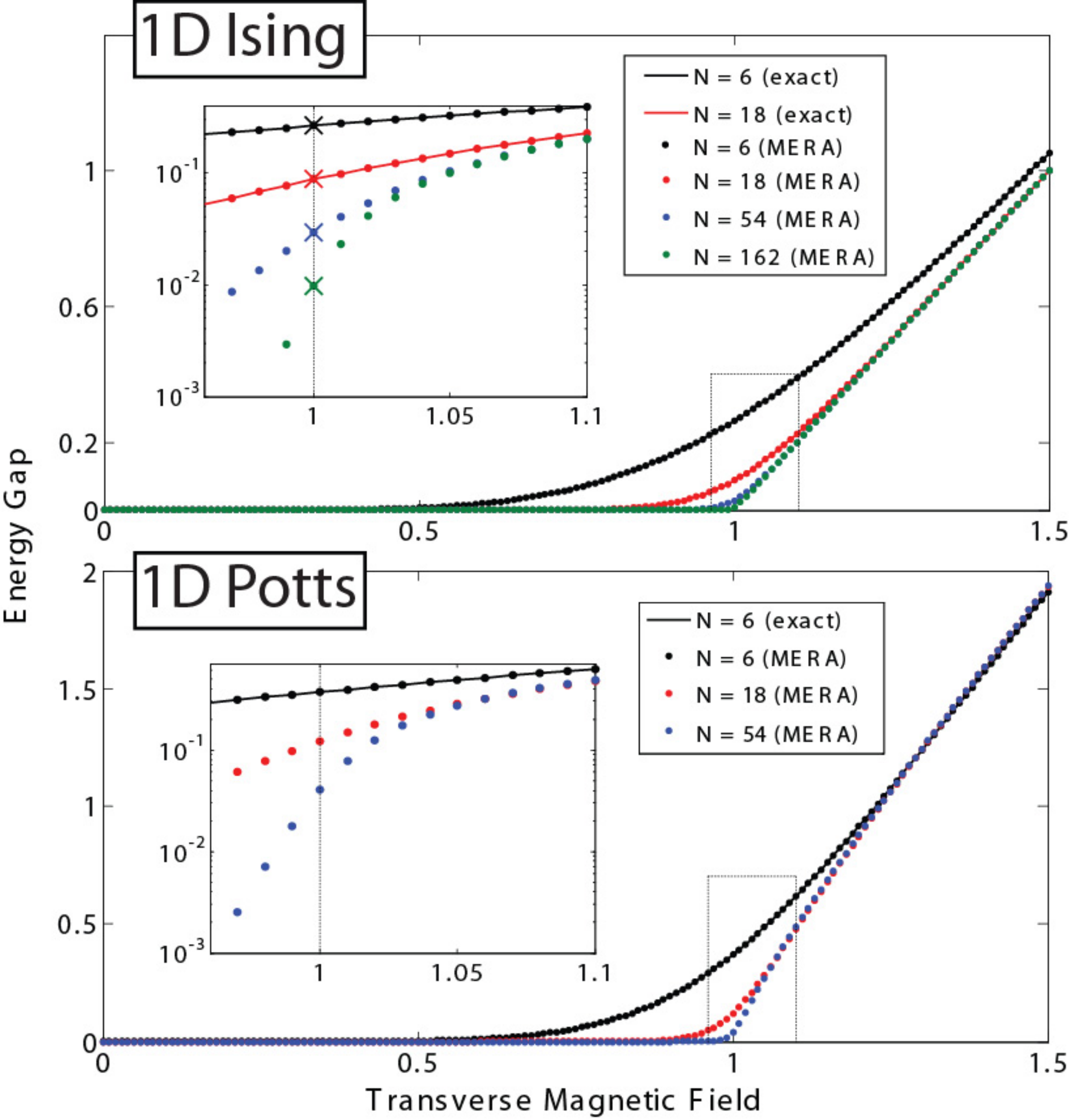}
\caption{(Colour online) \emph{Top:} A $\chi=8$ MERA is used to compute the energy gap $\Delta E$ (the energy difference between the ground and $1^{\textrm{st}}$ excited state) of the Ising chains as a function of the transverse magnetic field. The gap computed with MERA for $N=6,18$ sites is in good agreement with that computed through exact diagonalization of the system. \emph{Inset:} Crosses show analytic values of energy-gaps at the critical point for $N=\{6,18,54,162 \}$. Even for the largest system considered, $N=162$, the gap computed with MERA $\Delta E_{\textrm{MERA}}=9.67\times 10^{-3}$ compares well with the exact value $\Delta E_{\textrm{ex}}=9.69\times 10^{-3}$. \emph{Bottom:} Equivalent data for the Potts Model where simulations have been performed with a $\chi=14$ MERA to account for the increased computational difficulty of this model.} \label{IsingPottsGap}
\end{center}
\end{figure}

The previous results refer to local observables. Let us now consider correlators. A scale-invariant MERA, useful for the representation of critical systems, gives polynomial correlators at all length scales, as shown in Fig.~\ref{IsingPottsCorr} for the critical Ising and Potts models. Note that Fig. \ref{IsingPottsCorr} displays the correlators that are most convenient to compute (as per Fig.~\ref{fig:CorrEasy}). These occur at distances $d=3^q$ for $q=1,2,3\ldots$ and are evaluated with cost $O(\chi^8)$. Evaluation of arbitrary correlators is possible (see Fig.~\ref{fig:CorrDifficult}) but its cost is several orders of $\chi$ more expensive. The precision with which correlators are obtained is remarkable. A $\chi=22$ MERA for the Ising model gives $\langle{\sigma _x^{[r]} \sigma _x^{[r + d]} } \rangle$ correlators, at distances up to $d=10^9$ sites, accurate to within $0.6\%$ of exact correlators. Critical exponents $\eta$ are obtained through a fit of the form $C(r,r+d) \propto d^{ - \eta }$ with $C$ as correlators of $x,y$ or $z$ magnetization. For the Ising model, the exponents for $x,y$ and $z$ magnetizations are obtained with less than $0.02\%$ error each. For the Potts model exponents are obtained with less than $0.04\%$ and $0.08\%$ error for $x$ and $z$ magnetization respectively. 

Finally, we also demonstrate the ability of MERA to investigate low-energy excited states by computing the energy gap in the Ising and Potts models. Fig.~\ref{IsingPottsGap} shows that the gap grows linearly with the magnetic field $\lambda$ and independent of $N$ in the disordered phase $\lambda >> \lambda_c$, whilst at criticality it closes as $1/N$. Even a relatively cheap $\chi=8$ MERA reproduces the known critical energy gaps to within $0.2\%$ for systems as large as $N=162$ sites. The expected value of arbitrary local Hamiltonians (besides the energy) can also be easily evaluated for the excited state.

\section{Conclusions}

After reviewing the conceptual foundations of the MERA \cite{ER,MERA} (Sect. II-III), in this paper we have provided a rather self-contained description of an algorithm to explore low energy properties of lattice models (Sect. IV-V), and benchmark calculations addressing 1D quantum spin chains (Sect. VI).

Many of the features of the MERA algorithm highlighted by the present results can also be observed by investigating systems of free fermions \cite{FreeFermions} and free bosons \cite{FreeBosons} in $D=1,2$ dimensions. These include (i) the ability to consider arbitrarily large systems, (ii) the ability to compute the low-energy subspace of a Hamiltonian, (iii) the ability to disentangle non-critical systems completely, (iv) the ability to find a scale-invariant representation of critical systems and finally (v) the reproduction of accurate polynomial correlators for critical systems. However, the algorithms of Refs. \cite{FreeFermions,FreeBosons} exploit the formalism of Gaussian states that is characteristic of free fermions and bosons and cannot be easily generalized to interacting systems. Instead, the algorithms discussed in this paper can be used to address arbitrary lattice models with local Hamiltonians.

An alternative method to optimise the MERA is with a time evolution algorithm as described in Ref. \cite{TimeEvolution}. The time evolution algorithm has a clear advantage: it can be used both to compute the ground state of a local Hamiltonian (by simulating an evolution in imaginary time) and to study lattice dynamics (by simulating an evolution in real time). We find, however, that the algorithm described in the present paper are a better choice when it comes to computing ground states. On the one hand, the time-evolution algorithm has a time step $\delta t$ that needs to be sequentially reduced in order to diminish the error in the Suzuki-Trotter decomposition of the (euclidean) time evolution operator. In the present algorithm, convergence is faster and there is no need to fine tune a time step $\delta t$. In addition, the present algorithm allows to compute not only the ground state but also low energy excited states. It is unclear how to use the time evolution algorithm to achieve the same.


The benchmark calculations presented in this manuscript refer to 1D systems. For such systems, however, DMRG \cite{DMRG} already offers an extraordinarily successful approach. The strength of entanglement renormalization and the MERA relies on the fact that the present algorithms can also address large 2D lattices, as discussed in Ref. \cite{Finite2D,Scalable2D}.

The authors thank Frank Verstraete for key suggestions that lead to some of the optimization methods described in Sect. \ref{sect:optim}, and thank R. N. C. Pfeifer for useful discussions and comments. Support from the Australian Research Council (APA, FF0668731, DP0878830) is acknowledged.


\begin{thebibliography}{99}
\bibitem{TI_MERA} A translation invariant MERA, characterized by one disentangler and one isometry at each layer, need not represent a translation invariant state $\ket{\Psi}$. Hence the need to consider the average density matrix $\bar{\rho}$.
\bibitem{commentDM} Each operator $\rho_{\tau}$ in Eq. {\ref{eq:sequenceDM}} is both Hermitian ($\rho_{\tau}^{\dagger} = \rho_{\tau}$) and non-negative ($\bra{\phi}\rho_{\tau}\ket{\phi} \geq 0, \forall \phi$) but its trace is $\tr(\rho_{\tau})=\chi_T$. For simplicity, we call $\rho_{\tau}$ a density matrix for any $\chi_T \geq 1$, even though it is only a proper density matrix for $\chi_T=1$.
\bibitem{accuracyIsing} We note that the $\chi=22$ simulation for the Ising model gives less accurate results than the trend line  observed for smaller $\chi$ in Fig. \ref{IsingPottsEnergyError} would suggest. It is possible that numerical errors (such as errors in the sparse eigenvalue decomposition used) may have been significant at this level of accuracy.
\end{thebibliography}
\end{document}